\documentclass[letterpaper,onecolumn,11pt,accepted=2021-12-09]{quantumarticle}
\pdfoutput=1
\usepackage[utf8]{inputenc}
\usepackage[english]{babel}
\usepackage[T1]{fontenc}
\usepackage{amsmath}
\usepackage{hyperref}

\usepackage{tikz}
\usepackage{lipsum}

\numberwithin{equation}{section}

\newcounter{resultcounter}[section]

\newtheorem{thm}[resultcounter]{Theorem}
\newtheorem{lem}[resultcounter]{Lemma}
\newtheorem{prop}[resultcounter]{Proposition}
\newtheorem{cor}[resultcounter]{Corollary}

\newcommand{\sym}{ {\mathfrak S}_{\phi\leftrightarrow\psi}}

\newcommand{\m}{{\rm M}}

\newcommand{\aes}{a_e^{(s)}}
\newcommand{\Qes}{Q_e^{(s)}}

\newcommand{\Ees}{E_e^{(s)}(\lambda)}

\usepackage{amsmath}
\usepackage{amssymb}
\usepackage{graphicx}
\usepackage{color}

\renewcommand{\r}{{\rm R}}
\newcommand{\s}{{\rm S}}

\renewcommand{\i}{{\rm i}}

\def\qed{\hfill $\Box$\medskip}

\newcommand{\scalprod}[2]{\left\langle {#1}, {#2}\right\rangle}

\newcommand{\bbbone}{\mathchoice {\rm 1\mskip-4mu l} {\rm 1\mskip-4mu l}
	{\rm 1\mskip-4.5mu l} {\rm 1\mskip-5mu l}}

\newcommand{\R}{{\rm R}}

\begin{document}
	
\title{Dynamics of 	
		Open Quantum Systems I,  
		
		Oscillation and Decay}	
\author{Marco Merkli}
\affiliation{Department of Mathematics and Statistics, Memorial University of Newfoundland, St. \!John's, A1C 5S7, Canada}
\orcid{0000-0002-3990-6155}
	\email{merkli@mun.ca}
\homepage{https://www.math.mun.ca/~merkli/}
\maketitle

\centerline{\small Dedication: {\em Per Bap.}}

\begin{abstract}
We develop a framework to analyze the dynamics of a finite-dimensional quantum system $\s$ in contact with a reservoir $\r$. The full, interacting $\s\r$ dynamics is unitary. The reservoir has a stationary state but otherwise dissipative dynamics. We identify a main part of the full dynamics, which approximates it for small values of the $\s\r$ coupling constant, uniformly for all times $t\ge 0$. The main part consists of explicit oscillating and decaying parts. We show that the reduced system evolution is Markovian for all times. The technical novelty is a detailed analysis of the link between the dynamics and the spectral properties of the generator of the $\s\r$ dynamics, based on Mourre theory. We allow for $\s\r$ interactions with little regularity, meaning that the decay of the reservoir correlation function only needs to be polynomial in time, improving on the previously required exponential decay.

In this work we distill the structural and technical ingredients causing the characteristic features of oscillation and decay of the $\s\r$ dynamics. In the companion paper \cite{Markov2} we apply the formalism to the concrete case of an $N$-level system linearly coupled to a spatially infinitely extended thermal bath of non-interacting Bosons. 
\end{abstract}

\section{Introduction}

The fundamental evolution equation of quantum theory is the Schr\"odinger equation, which governs the dynamics of {\em closed} quantum systems,  isolated from their surroundings. In contrast, when a system is interacting with its surroundings, then the evolution equation for the degrees of freedom of the system alone have to be deduced from other fundamental principles. One approach is to view the system plus its surroundings {\em together} as a closed system, meaning that they evolve according to the Schr\"odinger equation involving both, the degrees of freedom of the system and those of the surroundings. This amounts to a huge number of variables, even if the system itself is small. The evolution equation for the system is obtained by ``tracing out'' the degrees of freedom of the environment, and its solution is called the reduced system dynamics. Generically, this situation is described as follows. The Hilbert space of states is 
\begin{equation}
\label{i4}
{\mathcal H}={\mathcal H}_\s\otimes{\mathcal H}_\r,
\end{equation}
describing the full complex into a partition of a system part $\s$ and a reservoir part $\r$. Traditionally, one calls the degrees of freedom making up the surroundings of $\s$ a {\em reservoir} (a thermal reservoir if one has in mind a big system in thermal equilibrium). The total system ($\s+\r$) being closed, the dynamics of an initial (time $t=0$) $\s\r$ density matrix $\rho(0)$ is given by 
$$
\rho(t) = e^{-\i t H}\rho(0) e^{\i t H},
$$
in accordance with the Schr\"odinger equation -- also called the von-Neumann equation in the case of density matrices. Here, $H$ is the total Hamiltonian, which is the sum 
\begin{equation}
\label{i5}
H= H_\s+ H_\r+ \lambda V
\end{equation}
of individual system and reservoir Hamiltonians plus an interaction term $\lambda V$, where $\lambda$ is a {\em coupling constant}. The operators $H_\s$ and $H_\r$ act non-trivially only on the system and reservoir factors of $\mathcal H$, while $V$ mixes up those factors. Typically $V=A\otimes B$ (or a sum of such terms), where $A$ and $B$ are system and reservoir operators, respectively. The reduced system density matrix at time $t$ is given by
\begin{equation}
\label{i0}
\rho_\s(t) = {\rm tr}_\r \big( e^{-\i t H}\rho(0)e^{\i t H}\big),
\end{equation}
where ${\rm tr}_\r$ denotes the partial trace over the reservoir part of $\mathcal H$. For $\lambda=0$, the dynamics is uncoupled, $e^{-\i t H}=e^{-\i t H_\s}\otimes e^{-\i t H_\r}$ and each factor (subsystem) evolves individually and independently. The dynamics is usually known explicitly in the uncoupled case, and the challenge is to find out what happens when $\lambda\neq 0$. The idea then is, usually, to carry out perturbation theory for small $\lambda$. If the initial state is not correlated, $\rho(0)=\rho_\s(0)\otimes\rho_\r(0)$, then the map
$$
V_t : \rho_\s(0) \mapsto \rho_\s(t)
$$
is well defined. It is the flow, mapping an arbitrary system initial state to its state at time $t\ge0$.  For $\lambda=0$, we have $V_t=e^{t {\mathcal L}_\s}$, where ${\mathcal L}_\s\rho = -\i [H_\s,\rho]$. However, for $\lambda\neq 0$, generically, $\s$ and $\r$ become entangled during the evolution, and this prevents $V_t$ from having the above simple structure. Namely, $V_t$ cannot be written as $e^{t {\mathcal L}}$ for any ${\mathcal L}$; this means that $V_t$ is not a semigroup in $t$ any longer (we call $e^{t{\mathcal L}}$ a semigroup instead of a group, if we restrict the values of $t$ to $t\ge0$). Apart from encoding the Markovian characteristics, any dynamics of the form $e^{t\mathcal L}$ has the practical advantage that the eigenvalues and eigenvectors of the operator $\mathcal L$ encode the dynamical behaviour. Spectral theory thus becoms a  main tool in the analysis of such a Markovian dynamics.

\bigskip

{\bf Markovian approximation.\ } Even though the flow $V_t$ is not of exponential (semigroup) form, one expects that under suitable conditions, $V_t$ can be {\em approximated} by a semigroup,
$$
V_t\approx e^{t {\mathcal L}(\lambda)},
$$
with ${\mathcal L}(0)={\mathcal L}_\s$. This is called the {\em Markovian approximation} \cite{BP, RivasHuelga}. The approximate evolution equation, which reads in differential form
\begin{equation}
\label{i2}
\dot\rho_\m(t) = {\mathcal L}(\lambda)\rho_\m(t),
\end{equation}
is the ubiquitous {\em Markovian master equation}. On physical grounds, its validity is plausible under two main assumptions: Firstly, the reservoir loses its memory quickly and secondly, the reservoir is `large' and the interaction is weak, so as to guarantee that the reservoir dynamics is not affected much by the inteaction with the system. These assumptions, also called the Markov- and Born approximations, respectively, are quantified by the decay of the reservoir correlation function (quick memory loss) and a smallness condition on $\lambda$ (weak coupling). The main two issues are 
\begin{itemize}
	\item[(a)] How to construct ${\mathcal L}(\lambda)$ and 
	\item[(b)] How to estimate the error made by the Markovian approximation.
\end{itemize} 
The generator ${\mathcal L}(\lambda)$ has to be constructed starting from the full $\s\r$ dynamics (interacting Hamiltonian $H$ as above) and reducing the evolution to the system component alone by tracing out the reservoir degrees of freedom. The literature on this topic is enourmous, and there have been many proposals for ${\mathcal L}(\lambda)$. Eventually, the so-called Davies generator \cite{Da1,Da2} 
\begin{equation}
\label{Dgen}
{\mathcal L} = {\mathcal L}_\s+\lambda^2K
\end{equation}
emerged as the `right' generator. Here, $K$ is an operator describing the effect of $\r$ on $\s$, which is explicitly determined by the interaction operator $V$ and the reservoir correlation function. The approximate Markovian dynamics $e^{t {\mathcal L}}$ is a so-called CPTP semigroup \cite{AL,CP} (completely positive, trace preserving), which guarantees, in particular, that the approximate solution $\rho_\m(t)$ is a (positive definite, normalized) density matrix for all $t$. For other generators proposed in the literature, this is not the case \cite{DS}.  Of course, the Davies generator is not the remedy for all ailments. There are plenty of interesting situations where the usual Davies generator \eqref{Dgen} does not lead to a good approximation. An important class of examples are systems with small energy level spacings, for which a different generator has to be used \cite{ML,Majetal, Trush, NR,  Rivas,M3level,Mtransport}.

\subsection{Goals}

We have two goals:
\begin{itemize}
	\item[G1] Identify a general framework for the perturbation theory of the dynamics of a unitarily evolving total system-reservoir complex. Isolate the dominant term of the total dynamics for small coupling. Derive from it the open system dynamics. Do this with weak regularity requirements on the interaction between the parts.

	\item[G2] Show how a widely used model of open quantum systems, an $N$-level system coupled linearly to a spatially infinitely extended system of non-interacting Bosons in thermal equilibrium, fits the general framework. Show the validity of the Markovian master equation for this model, under weak regularity conditions.  
\end{itemize}

We achieve the first goal in the current manuscript, while the second one is addressed in the companion paper \cite{Markov2}. The setup for G1 is more general than the usual setup found in the open quantum systems literature. This is motivated by us trying to distill the essential ingredients that imply characteristic oscillation and decay features of the total dynamics, as presented in our main result, Theorem \ref{motherthm}. The companion paper \cite{Markov2} can be read independently of the current work, and the concrete model considered there will be more familiar to many researchers in the field. We hope that a reader finding the current manuscript somewhat abstract may still consult \cite{Markov2} for the application part, which presents new results on this classical family of models.

\bigskip

We consider a bipartite Hilbert space $\mathcal H$, of the form \eqref{i4}, carrying a unitary dynamics $e^{\i t L_\lambda}$, with a self-adjoint generator (compare to \eqref{i5})
$$
L_\lambda=L_\s+L_\r+\lambda I,
$$
were $L_\s$ and $L_\r$ are called the system and reservoir Liouville operators, acting non-trivially only on the respective factors of $\mathcal H$, $\lambda$ is a coupling constant and $I$ is an interaction operator. Our assuptions are expressed mathematically precisely in (A1)-(A5), Section \ref{sec:setup}. The main two assumptions are described as follows. 
\begin{itemize}
	\item We assume that the reservoir dynamics has a {\em unique stationary state}. In other words, the reservoir Liouvillean $L_\r$ has a unique eigenvalue. This eigenvalue is zero and non-degen\-erate. The unique stationary state (eigenvector) is denoted by $\Omega_\r$. 
	
	\item We assume that away from the reservoir stationary state, the total dynamics is {\em dissipative}. By this we mean that for vectors $\phi,\psi\in\mathcal H$ satisfying $P_\r\phi=P_\r\psi=0$, where $P_\r =\bbbone_{{\mathcal H}_\s}\otimes |\Omega_\r\rangle\langle\Omega_\r|$ is the projection onto the reservoir stationary state, we have
	\begin{equation}
	\label{i6}
	|\langle\phi, e^{\i t \bar L_\lambda}\psi\rangle|\sim t^{-k}\mbox{\qquad  for large $t$, for some $k\ge 2$.}
	\end{equation}
	Here, $\bar L_\lambda$ denotes the operator $L_\lambda$ restricted to the range of $P_\r^\perp=\bbbone-P_\r$ (the space of vectors $\psi$ that have a vanishing overlap with the reservoir stationary state $\Omega_\r$).
\end{itemize}
How do we phrase \eqref{i6} precisely? This is done  by representing $e^{\i t \bar L_\lambda}$ by the Fourier-Laplace transform of the resolvent operator $(\bar L_\lambda-z)^{-1}$,
$e^{\i t \bar L_\lambda}= -\tfrac{1}{2\pi\i}\int_{\mathbb R} e^{\i t x}(\bar L_\lambda-x-\i 0_+)^{-1} dx$.  Using $e^{\i t x}=  \frac{1}{\ i t}\tfrac{d}{dx} e^{\i t x}$ and integrating by parts with respect to $x$ gives 
$$
e^{\i t \bar L_\lambda} = \frac 1t \frac{1}{2\pi }\int_{\mathbb R} e^{\i t x}(\bar L_\lambda-x-\i 0_+)^{-2}dx.
$$ 
Stronger decay in $t$ involves the Fourier-Laplace transform of higher powers $(\bar L_\lambda-x-\i 0_+)^{-1}$ (repeat the integration by parts).  To estimate such integrals, we require some powers of the resolvent $(\bar L_\lambda-z)^{-1}$ to be bounded in $z\in{\mathbb C}_-=\{z: {\rm Im}z<0\}$ (weakly on a suitable set of vectors). We give a rigorous discussion of this point below, after \eqref{m3.1}. Our method requires decay $\sim t^{-2}$, which results in the assumption that $\partial^j_z (\bar L_\lambda-z)^{-1}$ is bounded in $z\in {\mathbb C}_-$, for $j=0,1,2$, see \eqref{m03}.
\smallskip

\begin{itemize}
	\item Further conditions: We assume some non-degeneracy of the eigenvalues of $L_\lambda$ and some suitable relative boundedness conditions on the interaction operator $I$ (see (A1)-(A5)). 
\end{itemize}
These further conditions, given in detail in Section \ref{sec:setup}, are less fundamental (and can be weakened, or entirely removed). 
\medskip



We now explain what we mean by {\em weak regularity requirement} on the interaction, stated in the Goals G1, G2 above. A result similar to Theorem \ref{motherthm} (which implies the validity of the Markovian approximation) was proven in \cite{KM2,MAOP} under an {\em analyticity condition} on the reduced resolvent. Instead of requiring the first three derivatives of $(\bar L_\lambda-z)^{-1}$ to be (weakly) bounded in $z\in{\mathbb C}_-$, it was assumed in those works that $(\bar L_\lambda-z)^{-1}$ has an analytic  extension as $z$ moves from the lower complex plane across the real line into the upper complex plane. The difference between the two cases is not merely a technical issue, as the regularity determines physical features. For instance, the stronger analyticity condition implies that overlaps \eqref{i6}, as well as the reservoir correlation function, decay {\em exponentially quickly} in $t$.  Improving the `analytic deformation theory' used in \cite{KM1,KM2,MAOP} to the weaker regularity regime considered in the current paper and in \cite{Markov2}, is not totally easy. It is done using `Mourre theory', which is more delicate and technically trickier than the analytic theory. This improvement is worth pursuing as it sheds light on the question of how quickly the reservoir has to lose its memory in order for the Markovian approximation still to hold. The current work and \cite{Markov2} show that polynomial decay of the memory (reservoir correlation function) suffices. Incidentally, this polynomial memory loss still drives an exponentially quick approach of the system to its final state (c.f. Theorem \ref{motherthm}). It is well known that if the memory loss is not quick enough, then the Markovian approximation is not valid \cite{Rivas,RivasMME,Li}. We plan on extending our methods to this case in the future.

\begin{itemize}
\item[] Remarks	in view of applications.
\begin{itemize} 
\item[(1)] In applications \cite{Markov2,Mcorr}, we need, in general, to describe mixed states (such as equilibrium states). The Hilbert space $\mathcal H$ considered in the current paper does not, in general, describe pure states. Rather, $\mathcal H$ here is the space of purification (or, vectorization) of mixed states in question. Similarly, $L_\lambda$ is the {\em Liouville} operator, describing the dynamics expressed in the purification Hilbert space. This setup allows us to treat mixed and pure states on the same footing.
	
{\em Example. } The system equilibrium state of a qubit with Hamiltonian $H_\s=\tfrac{\omega_0}{2}\sigma_z =\tfrac{\omega_0}{2}|\!\uparrow\rangle -\tfrac{\omega_0}{2}|\!\downarrow\rangle$ is the mixed state $\rho_{\s,\beta}\propto e^{-\beta H_\s}$. Its purification is given by $\Psi_{\beta,\s}\propto e^{-\beta \omega_0/4} |
\!\uparrow\rangle\otimes|\!\uparrow\rangle +e^{\beta \omega_0/4} |
\!\downarrow\rangle\otimes|\!\downarrow\rangle$,  a normalized vector in the {\em purification space} ${\mathcal H}_\s={\mathbb C}^2\otimes{\mathbb C}^2$. 
The Liouvillean is $L_\s = H_\s\otimes\bbbone_{{\mathbb C}^2} - \bbbone_{{\mathbb C}^2}\otimes H_\s$. The key link between $\rho_{\s,\beta}$ and its purification $\Psi_{\s,\beta}$ is that for all observables $A$ and times $t$, the expectations can be expressed as,  
$$
{\rm tr}_{{\mathbb C^2}}(e^{-\i t H_\s}\rho_{\s,\beta}e^{\i t H_\s} A)=\langle e^{-\i t L_\s}\Psi_{\s,\beta}, (A\otimes \bbbone_{{\mathbb C^2}})e^{-\i t L_\s}\Psi_{\s,\beta}\rangle_{{\mathcal H}_\s}.
$$ 
Given $\rho_{\s,\beta}$ and $H_\s$, neither $\Psi_{\s,\beta}$ nor $L_\s$ are unique -- but there are standard choices for them \cite{BR,MAOP,BFS,JPrte,DJP}. The reservoir we consider in applications \cite{Markov2,Mcorr} consists of non-interacting Bose particles in infinite position space ${\mathbb R}^3$. They are in a state of thermal equilibrium at temperature $1/\beta>0$, determined by Planck's black body momentum density distribution: there are  $(e^{-\beta |k|}-1)^{-1} d^3k$ particles having momentum $k\in{\mathbb R}^3$ in a volume $d^3k$ centered around $k$, for each unit volume in position space. Usual Bosonic Fock space ${\mathcal F}(L^2({\mathbb R}^3,d^3k))$ does not accommodate this state, as any state $\psi\in {\mathcal F}(L^2({\mathbb R}^3,d^3k))$ (or denisty matrix) describes finitely many particles, which results in a vanishing particle density at infinite volume! However, one can find a Hilbert space ${\mathcal H}_\r$ (turning out to be the tensor product of two copies of Fock space) and a vector $\Omega_\r$ (turning out to be the tensor product of two vacua) which do represent the equilibrium state at positive density. The details of this construction are not the topic here, they can be found in textbook style in \cite{Mlnotes} (see also \cite{MAOP}, or the original paper \cite{AW}). In the mathematical literature, representing a mixed state as a normalized vector in a Hilbert space  is known as the Gelfand-Naimark-Segal construction (GNS); in the physics and chemistry literature it is also sometimes called the {\em thermofield} method.
	
\item[(2)] We assume that $\dim{\mathcal H}_\s<\infty$. However, the decay \eqref{i6} necessitates that $\bar L_\lambda$ does not have any eigenvalues, for otherwise clearly the overlap in \eqref{i6} would be independent of time on eigenvectors. This means that the spectrum of $\bar L_\lambda$ must be continuous. In particular, the Hilbert space ${\mathcal H}_\r$ has to be infinite dimensional.
	
{\em Example.\ } In the case of a reservoir of infinitely extended Bose particles, continuous energy spectrum arises due to the infinite volume limit. We prove in Theorem 3.1 of \cite{Markov2} that \eqref{i6} holds for this reservoir. The proof is based on the fact that the estimate holds for $\bar L_\lambda$ replaced by $\bar L_0$, together with a suitable perturbation theory in $\lambda$. 
	
\end{itemize}

\end{itemize}

\subsection{Explanation of the main result}

Our main result is Theorem \ref{motherthm}. It states that for coupling constants $\lambda$ small enough, we have the  expansion 
\begin{equation}
e^{\i tL_\lambda} = e^{\i t M(\lambda)}\otimes P_\r + P^\perp_\r e^{\i t \bar L_\lambda } P^\perp_\r +R(\lambda,t),
\label{i9}
\end{equation}
valid for all $t\ge 0$. Here, $P_\r =\bbbone_{{\mathcal H}_\s}\otimes|\Omega_\r\rangle\langle\Omega_\r|\equiv |\Omega_\r\rangle\langle\Omega_\r|$ is the projection onto the reservoir stationary state $\Omega_\r$ and $P_\r^\perp = \bbbone-P_\r$. The operator on the left side, $e^{\i t L_\lambda}$, is the propagator of the full, unitary $\s\r$ evolution. The first term on the right side describes a non-trivial evolution of $\s$, generated by an operator $M(\lambda)$, while $\r$ is projected onto the stationary state $\Omega_\r$. This part of the dynamics is Markovian. The second term on the right side describes the dynamics of states (vectors) in the range of $P^\perp_\r$, so states with vanishing overlaps with the reservoir stationary state $\Omega_\r$. Finally, $R(\lambda,t)$ is a remainder term. The equality \eqref{i9} is understood in the weak sense, that is, when $\langle\phi, \cdot\, \psi\rangle$ is applied to both sides, for suitable vectors $\psi,\phi\in\mathcal H$ belonging to a dense set. This set of vectors includes all uncorrelated $\s\r$ states in which the system is in any state and the reservoir is in the stationary state $\Omega_\r$. However, the set also includes a wide class of correlated initial $\s\r$ states; this aspect is exploited in \cite{Mcorr} to show the validity of the Markovian approximation even for correlated initial states.  In Theorem \ref{motherthm}, 
\begin{itemize}
	\item We find the detailed structure of $e^{\i t M(\lambda)}$. It consists of explicit oscillating terms and explicit decaying terms with decay rates $\propto 1/\lambda^2$.
	
	\item We show that the dissipative term decays polynomially in time, 
$$
|\langle\phi, e^{\i t \bar L_\lambda}\psi\rangle |\le C/(1+t^2),\qquad \mbox{for $\phi,\psi\in{\rm Ran} P^\perp_\r$.}
$$ 
	
	\item We show that the remainder satisfies $|\langle\phi, R(\lambda,t)\psi\rangle|\le C|\lambda|^{1/4}$ for all $t\ge 0$.  
\end{itemize}
We apply our main result \eqref{i9} to the concrete class of open quantum systems explained in goal G2 above in \cite{Markov2,Mcorr}.

\subsection{New results in the theory of open quantum systems}

We verify in \cite{Markov2} that the assumptions (A1)-(A5) of Section \ref{sec:setup} below are satisfied, and hence that Theorem \ref{motherthm} holds, for a standard class of open quantum systems described in goal G2 above. For this model, the generator $M(\lambda)$ is just the Davies generator $\mathcal L$, \eqref{Dgen}, as shown in \cite{MAOP}, and Theorem \ref{motherthm} yields the following new results.

\begin{itemize}
	\item[1.] Proof of the validity of the Markov approximation for all times \cite{Markov2}.
	
	Davies \cite{Da1,Da2} showed the validity of the approximation in the {\em ultra-weak coupling limit}.\footnote{In the mathematical literature, the term `weak coupling limit', or `Van-Hove limit' is used for this regime.} Namely, for any $a>0$,
	$$
	\lim_{\lambda\rightarrow 0}\ \sup_{0\le \lambda^2t<a}\|V_t -e^{t{\mathcal L}} \| =0,
	$$
	where $\|\cdot\|$ is the norm of super operators and $\mathcal L$ is the Davies generator \eqref{Dgen}.  This means that $e^{t{\mathcal L}}$ is a good approximation of the true system dynamics $V_t$ for small values of $\lambda$, {\em but only for times up to $t\sim \lambda^{-2}$}. In particular, Davies' result does not guarantee that the asymptotic system dynamics is approximated by the Markovian master equation. Our results overcomes this defect. Namely, we show in \cite{Markov2} that there is a $\lambda_0>0$ such that if $|\lambda|<\lambda_0$, then
	\begin{equation}
	\label{i1}
	\sup_{t\ge 0}\|V_t -e^{t{\mathcal L}} \| \le C|\lambda|^{1/4},
	\end{equation}
	for a constant $C$ independent of $\lambda, t$.
	
	We can rephrase the result \eqref{i1}. Let $\rho_\s(t)$ be the system density matrix \eqref{i0} and let $\rho_\m(t)$ be the solution of the Markovian master equation \eqref{i2}, with ${\mathcal L}(\lambda) = {\mathcal L}$ the Davies generator, and with equal initial conditions, $\rho_\s(0)=\rho_\m(0)$. Then \eqref{i1} asserts that
	$$
	\|\rho_\s(t)-\rho_\m(t)\|_1 \le C|\lambda|^{1/4},\qquad \forall t\ge 0,
	$$ 
	where $\|\sigma\|_1={\rm tr}_\s(|\sigma|)$ is the trace norm and $C$ is a constant independent of the initial condition.

	\item[2.] Proof of the validity of the Born approximation \cite{Mcorr}.
	
	Our main result \eqref{i9} gives an expansion of the full, interacting $\s\r$ dynamics. When reduced to the system dynamics alone, it results in \eqref{i1}. However, when analyzed in full, \eqref{i9} is shown in \cite{Mcorr} to yield the following results.
	\begin{itemize} 
		\item For uncorrelated initial states $\rho(0) = \rho_\s(0)\otimes|\Omega_\r\rangle\langle\Omega_\r|$, where $\s$ is in any state and $\r$ is in equilibrium, the reservoir stays in equilibrium  during the coupled evolution up to an error of $O({|\lambda|^{1/4}})$, for all times $t\ge 0$. This is a proof of the validity of the Born approximation, for all times. 
		
		\item For a large class of correlated initial $\s\r$ states, the correlations decay polynomially in time. After this decay has happened, the reservoir is in thermal equilibrium and the system evolves according to the Markovian dynamics generated by the Davies generator, up to errors $O(|\lambda|^{1/4})$, uniformly in times $t\ge 0$. 
	\end{itemize}
\end{itemize}

\subsection{Relation to earlier work}
\label{sect:newresults}

It is not our aim to present a detailed discussion of the huge literature on the dynamics of open quantum systems, as the goal of the current manuscript is the construction of a general mathematical framework. Let us rather discuss some mathematically rigorous works related to ours.

The spectral methods developed here have a certain similarity with the theory of metastable states in many-body quantum theory. There, the Hilbert space does not have the structure \eqref{H}, but rather, $L_0$ is the kinetic energy operator of $N$ particles and $I$ is an (interaction) potential. In this Schr\"odinger operator setup, initial states close to an eigenstate of $L_0$ stay bound (spatially localized)  for a long time under the evolution $e^{-\i t L_\lambda}$, but eventually decay for large times \cite{Simon, Hunziker1990, HS}. It is intuitively plausible, quite generally, that eigenvectors of $L_0$ associated to an unstable eigenvalue $e$, describing a bound states of a quantum system for $\lambda=0$, turn into `almost-bound' states for small $\lambda\neq 0$. This phenomenon, and some related mathematical tools (complex spectral deformation and Mourre theory, resolvent representation of the propagator), are common to the many-body Schr\"odinger and the open system setups.
\medskip

The dynamics of an $N$-level system coupled linearly to a spatially infinitely extended reservoir of non-interacting Bose particles (as we will consider as an application of the current results in \cite{Markov2, Mcorr}) has been investigated in detail before. It is based on the representation of the infinite volume equilibrium state as a vector in a purification Hilbert space, which was first constructed in the early sixties in \cite{AW}. However, it was not exploited to analyze the dynamics of open systems until the pioneering papers \cite{JPrte,BFS} developed the spectral approach. In these works, the phenomenon of {\em Return to Equilibrium (RtE)} was proven: States `close' to the (coupled) $\s\r$ equilibrium state are driven to equilibrium in the long time limit. In contrast with usual open system dynamics results, RtE is a result about the full system-reservoir dynamics, not only about the reduced system dynamics. Averages of reservoir observables also converge to the equilibrium values. While \cite{JPrte,BFS} based their analysis on analytic deformation methods, a Mourre theory version for RtE was developed in \cite{FM,M}. A bit later, in \cite{MBS}, the formalism used to show RtE was refined and a detailed description of the dynamics of each system density matrix element was obtained.  Then in \cite{KM1,KM2,MAOP} is was shown that the main term of the reduced system dynamics is a completely positive trace preserving semigroup. Under general assumptions, the generator of the semigroup is the Davies generator, as is shown in \cite{MAOP}.

In terms of technique, the paper \cite{KM1}, where Mourre theory is used to analyze the dynamics, is closest to the present work. However, as explained in discussion point (v) after Theorem \ref{motherthm}, the approach of \cite{KM1} is not suited to show the Markovian approximation. In the present work, we make substantial changes to the method of \cite{ KM1}, changes which we explain in Section~\ref{proofsect}.
\medskip

{\em Further rigorous work.} Establishing the validity of Markovian master equations is an active research field. In \cite{ML,Majetal}, the authors develop a so-called coarse-grained Master equation (CGME) and they compare its validity to that of the Davies and the Redfield master equations. They show that the CGME combines the advantages of the other two, but without incorporating their disadvantages. Namely, the CGME is a good approximation of the true system dynamics regardless of the system level spacing (Davies cannot do this) and the equation is CPTP (Redfield is not). However, the error bounds for neither of the three Master equations are uniform in time; the error of the CGME is $\propto O(|\lambda| e^{c\lambda^2 t})$ for some constant $c$ (linked explicitly to the fastest system decoherence time scale). In a similar vein, \cite{Trush,NR} develop a Master equation, again a CPTP equation, which describes the approximation of the true system dynamics for arbitrarily small system level spacings, but again with remainders guaranteed to be small only for finite times. In \cite{NR} the authors also provide a bound on the error generated by the Born approximation, for finite times. We have applied our resonance theory to systems with almost degenerate system levels in \cite{M3level,Mtransport} and shown that almost degeneracy leads to a separation of time scales in the system dynamics. The papers \cite{M3level,Mtransport} assume the stronger regularity (analyticity), but we do not see any obstacle to adapting the general theory for weak regularity developed here to the case of degenerate or almost degenerate system levels.

\section{Approach and assumptions}
\label{sec:setup}

The purpose of this section is to define the setup and state precisely the assumptions we make. The core of our overall strategy is to link the dynamics to the spectral data of the generator $L_\lambda=L_0+\lambda I$. A basic mechanism producing irreversible dynamics, or decay, is the disappearence (instability) of eigenvalues of $L_0$ under the perturbation $\lambda I$. The instability occurs because the eigenvalues of $L_0$ are embedded in continuous spectrum. Their fate after perturbation is analyzed using `singular perturbation theory'. We explain the main ideas of it in Section \ref{sub:instability}. In Section \ref{sec:link} we explain how stable and unstable eigenvalues of $L_\lambda$ affect the propagator $e^{\i tL_\lambda}$. 

Instead of making a list of assumptions, we are trying to proceed in an {\em in}ductive manner: We explain the main strategy of our approach, and while doing so, the assumptions should emerge naturally.

\subsection{Basic properties of the Liouville operator $L_\lambda$}

We consider a  bipartite Hilbert space
\begin{equation}
	{\mathcal H}={\mathcal H}_\s\otimes{\mathcal H}_\r,
	\label{H}
\end{equation} 
where $\dim{\mathcal H}_\s<\infty$, and a family of self-adjoint operators 
\begin{equation}
\label{m1}
L_\lambda = L_0 +\lambda I,
\end{equation}
where both $L_0$ and $I$ are self-adjoint, $\lambda\in\mathbb R$ and $L_0$ is of the form
\begin{equation}
L_0 = L_\s\otimes\bbbone_\r + \bbbone_\s\otimes L_\r.
\label{3}
\end{equation}
This is the setup describing  the composition of a system ($\s$)  plus reservoir ($\r$) arrangement, in which $L_\s$ and $L_\r$ generate the free (non interacting) dynamics of the individual components, $I$ is an interaction operator and $\lambda$ is a coupling constant.

\begin{figure}[b]
	\centering
	\includegraphics[width=12cm]{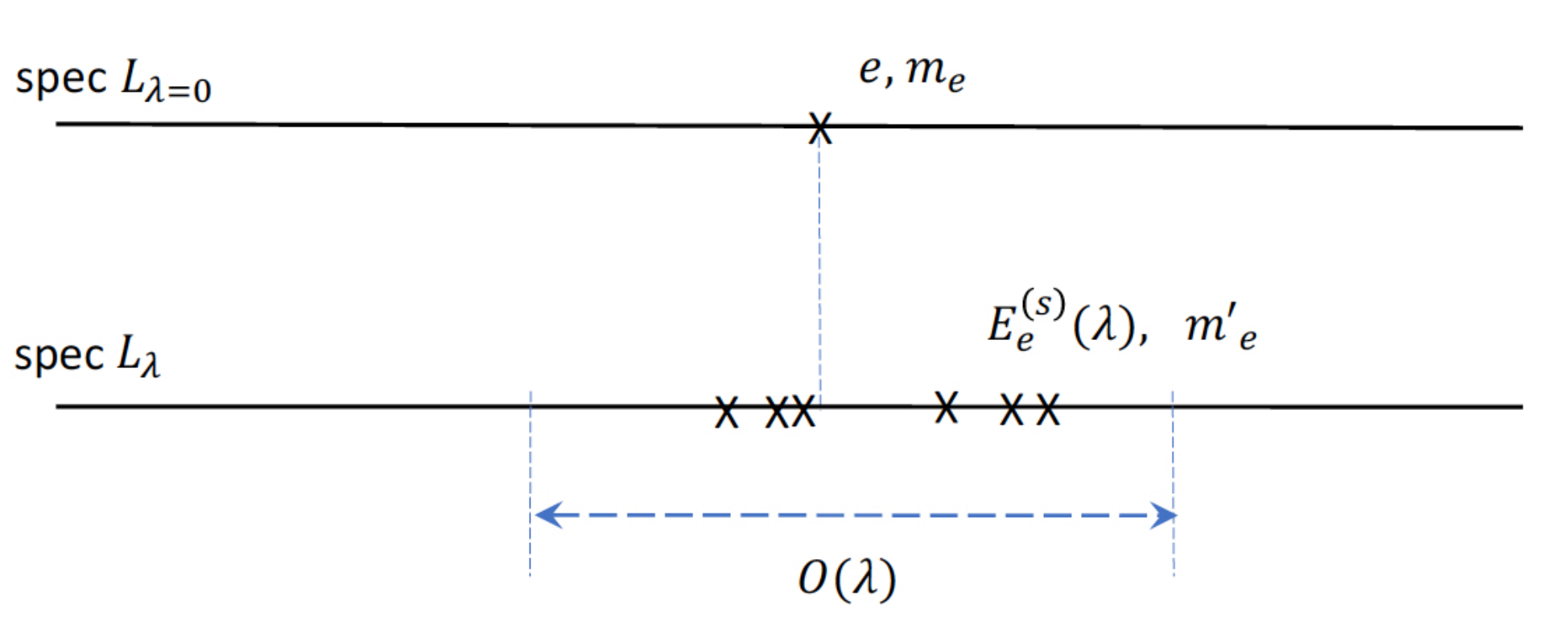}
	\caption{An eigenvalue $e$ of $L_0$ splits into eigenvalues $E_e^{(s)}\!(\lambda)$, $s=1,\ldots,m_e'$, of $L_\lambda$.}
	\label{Fig1}
\end{figure}

We suppose that $L_0$ has finitely many eigenvalues $e$, of finite multiplicity $m_e$, possibly embedded in continuous spectrum, which can cover parts or all of $\mathbb R$. The set of eigenvalues $e$ of $L_0$ is denoted by ${\mathcal E}_0$ and the associated orthogonal eigenprojection is $P_e$. We are going to impose a regularity condition ({\em c.f.} (A1) below) which implies the following picture for small $\lambda$: All eigenvalues of $L_\lambda$  lie inside an $O(\lambda)$ neighbourhood of ${\mathcal E}_0$. Moreover, within such a neighbourhood around any given $e\in{\mathcal E}_0$, either $L_\lambda$ does not have any eigenvalues for $\lambda\neq 0$ (we say $e$ is unstable), or $L_\lambda$ does have some eigenvalues, with summed multiplicity $m'_e$ not exceeding $m_e$ (we say $e$ is stable if $m'_e=m_e$, and  partially stable if $0<m'_e<m_e$). In the analytic perturbation theory of isolated eigenvalues, the summed multiplicity $m'_e$ would always equal $m_e$, but for embedded eigenvalues, it is generically strictly less than $m_e$.  Fig.~\ref{Fig1} gives a graphical depiction of the situation. 
\bigskip

Let us now introduce the assumptions and discuss their meaning. We start with an assumption to simplify the  bookkeeping. It is not essential for our method to work and could be removed at the cost of a more cumbersome presentation.
\begin{itemize}
\item[{\bf (A1)}] For $\lambda\neq 0$ small enough, all eigenvalues of $L_\lambda$ are simple.
\end{itemize}
We call these eigenvalues $E^{(s)}_e(\lambda)$, with  $s=1,\ldots,m_e'\le m_e$, and 
\begin{equation}
\lim_{\lambda\rightarrow 0}E_e^{(s)}(\lambda)=e,\qquad s=1,\ldots,m_e'.
\label{0m2}
\end{equation}

The next assumption is a key characteristic for the physical situation we want to describe. We assume  that the reservoir dynamics has a single stationary state and that the coupled dynamics is dissipative on the orthogonal complement of this state. 
\begin{itemize}
\item[{\bf (A2)}] The reservoir dynamics has a unique stationary state $\Omega_\r\in{\mathcal H}_\r$, that is, ${\rm Ker} L_\r={\mathbb C}\Omega_\r$. We denote 
\begin{equation}
\label{POmega}
P_\r=\bbbone_\s\otimes|\Omega_\r\rangle\langle\Omega_\r| \qquad \mbox{and} \qquad P^\perp_\r=\bbbone_{\mathcal H}- P_\r.
\end{equation}
On the orthogonal complement the full, coupled dynamics is dissipative in the sense that there exists a $\lambda_*>0$ and a dense set ${\mathcal D}\subset{\mathcal H}$ such that $\forall \lambda$ with $0\leq |\lambda|<\lambda_*$ and $\forall \phi,\psi\in{\mathcal D}$, 
\begin{eqnarray}
	\max_{0\le j\le 2}\sup_{z\in{\mathbb C}_-}\Big| \partial_z^j \langle\phi, R_z^{P_\r}(\lambda) \psi\rangle\Big| &\le& C_1(\phi,\psi)<\infty,
	\label{m03}\\
	\sup_{z\in{\mathbb C}_-}\Big| \partial_\lambda\langle\phi, R_z^{P_\r}(\lambda) \psi\rangle\Big| &\le& C_1(\phi,\psi)<\infty.
	\label{m03.1}
\end{eqnarray}
Here ${\mathbb C}_- =\{ z\in {\mathbb C} : {\rm Im} z<0\}$ is the open lower complex half plane and 
$$
R_z^{P_\r}(\lambda) = (P^\perp_\r L_\lambda P^\perp_\r-z)^{-1}\upharpoonright_{{\rm Ran}P^\perp_\r}
$$  
is the reduced resolvent. Here, $\upharpoonright_{{\rm Ran}P^\perp_\r}$ denotes the restriction of an operator to the subspace ${\rm Ran}P^\perp_\r$.  In \eqref{m03}, \eqref{m03.1},
 $C_1$ is well defined (finite) on ${\mathcal D}\times{\mathcal D}$.
\end{itemize}

\noindent
Discussion of Assumption (A2).
\begin{itemize} 
\item[(i)] The estimate \eqref{m03.1} is of technical nature, but the estimate \eqref{m03} is key as it implies that the dynamics generated by $\bar L_\lambda\equiv P^\perp_\r L_\lambda P^\perp_\r\upharpoonright_{{\rm Ran}P^\perp_\r}$  is {\em dissipative}, meaning that  $\lim_{t\rightarrow\infty}\langle \phi, e^{\i t \bar L_\lambda}\psi\rangle=0$. More precisely, it follows from \eqref{m03}, with $0\le j\le 2$ replaced by $0\le j\le k$, that 
\begin{equation}
	\big|\langle \phi, e^{\i t \bar L_\lambda} \psi\rangle\big| \le \frac{C}{(1+t^2)^{k/2}},
	\label{m4.1}
\end{equation}
provided $(\bar L_\lambda+\i)\phi\in {\mathcal D}$ and $(\bar L_\lambda+\i)\psi\in{\mathcal D}$. Specifically, \eqref{m03} implies time decay $\propto t^{-2}$ ($k=2$) of overlaps \eqref{m4.1}. We prove \eqref{m4.1} at the end of this section. 

\item[(ii)] We argue that \eqref{m03} is a natural assumption. Namely, \eqref{m03} for $\lambda=0$ is implied by
\begin{equation}	
\max_{0\le j\le 2}\sup_{z\in{\mathbb C}_-}\Big| \partial_z^j \langle\phi, (L_\r -z )^{-1} P_\r^\perp  \psi\rangle_{{\mathcal H}_\r}\Big| \le C_1(\phi,\psi)<\infty,
\label{m03.1.1}
\end{equation}
where we understand here ({\em c.f.} \eqref{POmega}) $P_\r=|\Omega_\r\rangle\langle\Omega_\r|$, $P_\r^\perp=\bbbone_\r-P_\r$. This is readily seen by writing 
$$
R_z^{P_\r}(0) = (L_0-z)^{-1} P^\perp_\r = \sum_{e\in{\mathcal E}_0} P_{\s,e}\otimes (L_\r-e-z)^{-1} P^\perp_\r,
$$
where $P_{\s,e}$ is the spectral projection of $L_\s$ associated to $e$, so that $\sum_{e\in{\mathcal E}_0}P_{\s,e}=\bbbone_\s$. In turn, \eqref{m03.1.1} is a natural assumption on the dynamics of a reservoir, since that dynamics should be dissipative away from the stationary state. In concrete applications, one starts with \eqref{m03} for $\lambda=0$ and then proves its validity for small $\lambda\neq 0$ by perturbation theory \cite{Markov2}.

\item[(iii)] In some recent works on the dynamics on open quantum systems \cite{KM2,MAOP}, the assumption \eqref{m03} is replaced by the condition that $z\mapsto \langle\phi, R^{P_\r}_z(\lambda) \psi\rangle$ have a {\em meromorphic continuation} from $z$ in the lower complex plane across the real axis into the upper plane. This is a much stronger condition than \eqref{m03}. In applications to open quantum systems, this difference means that the reservoir correlation function has to decay exponentially quickly in time for the  meromorphic situation, while under the present assumption, the decay only needs to be polynomial, {\em c.f.} \cite{Markov2}. 

\item[(iv)] Without loss of generality, we may assume that $C_1(\phi,\psi)=C_1(\psi,\phi)$ and that  $C_1(\phi,\psi)=C_1(P^\perp_\r\phi, P^\perp_\r\psi)$ in \eqref{m03}, \eqref{m03.1}.
\end{itemize}

As is well known (see e.g. Proposition 4.1 of \cite{Cycon}) if $A$ is a self-adjoint operator and  for each vector $\phi$ in some dense set, there exists a constant $C(\phi)$ such that  
$$
\liminf_{\epsilon\rightarrow 0_+} \sup_{x\in(a,b)}|\langle\phi, (A-x+\i\epsilon)^{-1}\phi\rangle|\le C(\phi),
$$
then the spectrum of $A$ in the interval $(a,b)$ is purely absolutely continuous. Thus the estimate \eqref{m03} with $j=0$ implies that the spectrum of $P^\perp_\r L_\lambda P^\perp_\r$ acting on ${\rm Ran}P_\r^\perp$ is purely absolutely continuous and for $\lambda=0$, this implies that the spectrum of $L_0$ reduced to ${\rm Ran}P_\r^\perp$ is purely absolutely continuous. On the finite dimensional part ${\rm Ran}P_\r\cong {\mathcal H}_\s$, the operator $L_0$ is (identified with) $L_\s$ which has pure point spectrum ${\mathcal E}_0$. Therefore $L_0$ has absolutely continuous spectrum except for the eigenvalues ${\mathcal E}_0$, the same as those of $L_\s$ and the eigenprojection $P_e$ of $L_0$ associated to $e$  is given by
\begin{equation}
P_e = ( P_{\s,e}\otimes \bbbone_\r) P_\r \equiv P_{\s,e}\otimes P_\r,
\label{m7.1}
\end{equation}
where $P_{\s,e}$ is the eigenprojection associated to $e$ as an eigenvalue of $L_\s$.

\subsection{Instability of embedded eigenvalues and how to track them}
\label{sub:instability} 

If $e$ is an {\em isolated} eigenvalue of  $L_0$ then by standard analytic perturbation theory \cite{Kato} $L_\lambda$ has eigenvalues $E^{(s)}_e(\lambda)$ close to $e$, for $\lambda$ small. Those eigenvalues coincide with the eigenvalues of the operator
\begin{equation}
e P_e  +\lambda P_eIP_e -\lambda^2 P_e I (L_0-e)^{-1}P^\perp_eIP_e +O(\lambda^3).
\label{m7}
\end{equation}
Each term in the expansion \eqref{m7} is self-adjoint. If $e$ is an {\em embedded} eigenvalue of $L_0$ then the reduced resolvent $(L_0-e)^{-1} P^\perp_e$ in \eqref{m7} is not defined as a bounded operator. However, its regularization  $(L_0-e+\i \epsilon)^{-1}P^\perp_e$ certainly is, for any $\epsilon>0$. We may hope that in a sense, the perturbation expansion \eqref{m7} stays valid also for embedded eigenvalues, upon regularizing the resolvent and taking $\epsilon\rightarrow 0_+$. But due to the regularization, the operator $(L_0-e+\i \epsilon)^{-1}P^\perp_e$ is not self-adjoint any longer! So according to \eqref{m7}, the second order corrections to the embedded eigenvalue would become complex numbers. This is, however, not compatible with $L_\lambda$ being self-adjoint. We may then intuit that for small nonzero $\lambda$, the number $m_e'$ of eigenvalues of $L_\lambda$ close to $e$ might be strictly reduced, $m_e' <  m_e$, and that this reduction is accounted for by the existence of complex eigenvalues of the so-called level shift operator
\begin{equation}
\label{34}
\Lambda_e=-P_e I P^\perp _e(L_0-e+\i 0_+)^{-1} IP_e.
\end{equation}
Here, $(L_0-e+\i 0_+)^{-1}$ is the limit of $(L_0-e+\i\epsilon)^{-1}$ as $\epsilon\rightarrow 0_+$, taken in the sense of the operator norm in the expression \eqref{34}. This mechanism has the following precise formulation. 

Let $Q$ be an orthogonal projection on $\mathcal H$ and set $Q^\perp =\bbbone-Q$. The {\em Feshbach map} applied to an operator $A$ on $\mathcal H$ is defined by
\begin{equation}
	\label{Feshbach}
	{\mathfrak F}(A;Q) = Q\big(A-A Q^\perp (Q^\perp AQ^\perp\upharpoonright_{{\rm Ran}Q^\perp})^{-1} A\big)Q,
\end{equation}
where it is assumed that $Q^\perp AQ^\perp\upharpoonright_{{\rm Ran}Q^\perp}$ is invertible. The Feshbach map satisfies the following {\em isospectrality property}: Let $a\in\mathbb C$ be in the resolvent set of the operator $Q^\perp AQ^\perp\upharpoonright_{{\rm Ran}Q^\perp}$, so that ${\mathfrak F}(A-a;Q)$ is well defined. Then the isospectrality property says that $a$ is an eigenvalue of $A$ if and only if zero is an eigenvalue of ${\mathfrak F}(A-a;Q)$. The redeeming quality of this mapping is that ${\mathfrak F}(A-a;Q)$ acts on ${\rm Ran}Q$, a space smaller than $\mathcal H$ (reduction in dimension). Consider now, for $z\in{\mathbb C}_-$, 
\begin{equation}
{\mathfrak F}(L_\lambda-z;P_e) = P_e\Big( e-z+\lambda I -\lambda^2 I R_z^{P_e}(\lambda) P^\perp_eI\Big)P_e,
\label{ff}
\end{equation}
where $R_z^{P_e}(\lambda)\equiv (P_e^\perp L_\lambda P_e^\perp-z)^{-1}\upharpoonright_{{\rm Ran} P^\perp_e}$ is the reduced resolvent. The isospectrality is not of any good use to analyze the spectrum of $L_\lambda$ directly, since $R_z^{P_e}(\lambda)$ is not defined for real $z$. However, one can show (see Theorem A1 of \cite{KM1} and also Theorem \ref{thmA1} below) that the condition (A2), together with the assumption\footnote{In applications, this a very weak assumption: according to \eqref{m7.1}, on the reservoir part, $P_e$ projects onto the stationary state $\Omega_\r$. This means that the operator $I$ only has to be applicable to vectors in ${\mathcal H}_\s\otimes {\mathbb C}\Omega_\r$. In particular, the reservoir part of the interaction $I$ can be an unbounded operator, as long as $\Omega_\r$ is in its domain.} that 
\begin{flalign}
\mbox{\bf (A3)}	&& \mbox{$IP_e$ is a bounded operator and\ \,}	& {\rm Ran} IP_e\subset {\mathcal D}, &
\label{16.3}
\end{flalign}
imply that the derivatives of order up to two  of
\begin{equation}
z\mapsto P_eIR_z^{P_e}(\lambda)P^\perp_eIP_e
\end{equation} 
are bounded uniformly in  $\{ z\in{\mathbb C}_- : |{\rm Re}z-e|\le g/2\}$. Here, 
\begin{equation}
	\label{g}
	g=\min\big\{ |e-e'| : e,e'\in{\mathcal E}_0, \ e\neq e'\big\}>0
\end{equation}  
denotes the minimal gap of the eigenvalues of $L_0$ (which is the same as that of $L_\s$). The Feshbach map \eqref{ff} is then well defined (by continuity) for $z\in\mathbb R$, $|z-e|\le g/2$. Now the isospectrality property can be extended to real values of $z$. Namely, one can show  (see Theorem \ref{weakFeshthm} below and also \cite{DJ,BFS,KM1}) that any $E\in{\mathbb R}$ is an eigenvalue of $L_\lambda$ if and only if zero is an eigenvalue of ${\mathfrak F}(L_\lambda-E;P_e)$, and that 
\begin{equation}
	{\rm dim}{\rm Ker}(L_\lambda -E) = {\rm dim}{\rm Ker}\big({\mathfrak F}(L_\lambda-E;P_e)\big).
	\label{m21}
\end{equation}
Due to \eqref{ff}, eigenvalues of ${\mathfrak F}(L_\lambda-E;P_e)$ are located in an $O(\lambda)$ neighbourhood of ${\mathcal E}_0$, and hence by isospectrality, so are those of $L_\lambda$. The multiplicity of $E$ as an eigenvalue of $L_\lambda$ is controlled by  \eqref{m21}. We assume now
\begin{flalign}
\mbox{\bf (A4)}	&& P_e IP_e & = 0, &
\label{13}
\end{flalign}
as this condition does not alter the emergence of complex eigenvalues (because $P_eIP_e$ is self-adjoint). We could dispense with the condition \eqref{13} by a simple modification of our arguments. In applications \cite{MAOP,Markov2,Mcorr}, this assumption is naturally satisfied. According to \eqref{34}, \eqref{ff}, \eqref{13},
\begin{equation}
	{\mathfrak F}(L_\lambda-E;P_e) = (e-E) P_e + \lambda^2\Lambda_e +
	O(\lambda^3).
	\label{m22}
\end{equation}
As it acts on the $m_e$-dimensional space ${\rm Ran}P_e$, the operator $\Lambda_e$ has $m_e$ generally complex eigenvalues. Since $\Lambda_e$ it is a dissipative operator, meaning that 
\begin{equation}
	{\rm Im }\Lambda_e = \lim_{\epsilon\rightarrow 0_+}  P_e I\frac{\epsilon P_e^\perp  }{(L_0-e)^2+\epsilon^2} IP_e\, \ge 0,
\end{equation}
its eigenvalues have non-negative imaginary parts.\footnote{\label{foot1} It is sometimes useful to note that the {\em real}  eigenvalues of $\Lambda_e$ are automatically semi-simple as they lie on the boundary of the numerical range of $\Lambda_e$, see {\em e.g.} Proposition 3.2 of \cite{DJ}.}
The isospectrality of the Feshbach map \eqref{m21} and the expansion \eqref{m22} show that for each eigenvalue $\Ees$ of $L_\lambda$, there is an eigenvalue $\aes\in\mathbb R$ of $\Lambda_e$ such that 
\begin{equation}
	E^{(s)}_e(\lambda) = e+\lambda^2 \aes+O(\lambda^3).
	\label{m11}
\end{equation} 
However, $\Lambda_e$ may have real eigenvalues without $L_\lambda$  having any eigenvalues close to $e$ (for $\lambda$ small, nonzero). This is so since the $O(\lambda^3)$ term in \eqref{m22} may cause the spectrum of \eqref{m22} to be non-real. In this case, $L_\lambda$ does not have any eigenvalues close to $e$, according to \eqref{m21}. To simplify the analysis, we do not consider this higher order effect. Instead, we assume that the real eigenvalues of $\Lambda_e$ are in {\em bijection} with the eigenvalues $E_e^{(s)}(\lambda)$ of $L_\lambda$ close to $e$. One way to ensure this is to impose the condition
\begin{flalign*}
\mbox{\bf (A5)}	&& \mbox{The eigenvalues of $\Lambda_e$ are simple and $\Lambda_e$ has exactly $m'_e$ real eigenvalues.} &  &
\end{flalign*}
We recall that $m'_e$, defined before \eqref{0m2}, is  the number of (distinct and simple) eigenvalues of $L_\lambda$ close to $e$, for small $\lambda\neq 0$. Assuming $\Lambda_e$ to have purely simple spectrum is done for convenience of the presentation. This restriction can be removed easily and our approach still works, as long as $\Lambda_e$ is diagonalizable. In some models, it can happen though that $\Lambda_e$ is not diagonalizable at so-called exceptional points of parameters; one then expects a qualitatively different behaviour of the dynamics (Jordan blocks of $\Lambda_e$ cause polynomial corrections to exponential decay in time). We do not further explore this interesting aspect here.

The condition (A5) without the simplicity assumption is also  called the {\em Fermi Golden Rule Condition}. It ensures that stability or instability of eigenvalues of $L_\lambda$ is detected at the lowest order ($\lambda^2$) in the perturbation.  In terms of dynamical properties, the condition (A5)  means that metastable states have life-times of $O(\lambda^{-2})$, as shown in Theorem \ref{motherthm}.

\subsection{How to link stable and unstable eigenvalues to the dynamics}
\label{sec:link}

We have seen above that the (partially) stable eigenvalues of $L_\lambda$ are the real eigenvalues of the level shift operators $\Lambda_e$ and that the $\Lambda_e$ may have eigenvalues $a_e^{(s)}$ with strictly positive imaginary part. We introduced the level shift operators $\Lambda_e$ as the second order $(\lambda^2)$ contributions in the Feshbach map,  \eqref{m22}. The isospectrality property of the Feshbach map then linked $\Lambda_e$ to the spectrum of $L_\lambda$. As it turns out, the Feshbach map is also an ingredient in  the block decomposition of an operator $H$ acting on ${\mathcal H}={\rm Ran}Q\oplus {\rm Ran}Q^\perp$, where $Q$ is an orthogonal projection. More precisely, one can readily verify that for any operator $H$ such that ${\mathfrak F}(H;Q)$ exists, the following identity holds (see Section \ref{Fdecsect}, in particular \eqref{fesh2}), 
\begin{equation}
H= 
\begin{pmatrix}
		\bbbone & QH Q^\perp R_z^Q\\
		0 & \bbbone
\end{pmatrix}
\begin{pmatrix}
		{\mathfrak F}(H;Q) & 0\\
		0 & Q^\perp H Q^\perp
\end{pmatrix}
\begin{pmatrix}
		\bbbone & 0 \\
		R^Q_zQ^\perp HQ & \bbbone
\end{pmatrix}.
\label{24.1}
\end{equation}
The first component in the $2\times 2$ decomposition of \eqref{24.1} is that of ${\rm Ran}Q$, the second one that of ${\rm Ran}Q^\perp$. Choosing in \eqref{24.1} $Q=P_e$ and $H=L_\lambda-z$ for $z\in{\mathbb C}\backslash{\mathbb R}$ gives the decomposition
\begin{equation}
(L_\lambda-z)^{-1}  = [{\mathfrak F}(L_\lambda-z;P_e)]^{-1} + {\mathfrak B}(z) + R_z^{P_e}(\lambda),
\label{25}
\end{equation}
where ${\mathcal B}(z)$ is a term of $O(\lambda)$, see \eqref{16}. The first term on the right side is the inverse of the Feshbach map, an operator acting on ${\rm Ran}P_e$ and the last term acts on ${\rm Ran}P_e^\perp$. The resolvent, \eqref{25}, is linked to the propagator via the Fourier-Laplace transform, for $w>0$,
\begin{equation}
e^{\i t L_\lambda} = \frac{-1}{2\pi\i}\int_{{\mathbb R}-\i w} e^{\i t z} (L_\lambda-z)^{-1} dz.
\label{26}
\end{equation}
Using \eqref{25} in \eqref{26} provides a link between the Feshbach map and the dynamics. We then expand the (inverse) of the first term on the right side of \eqref{25}, ${\mathfrak F}(L_\lambda-z;P_e) = e-z+\lambda^2 \Lambda_e +O(|\lambda|^3 +\lambda^2|z-e|)$. For $z$ close to $e$, this links  $[{\mathfrak F}(L_\lambda-z;P_e)]^{-1}$ to $(e-z+\lambda^2\Lambda_e)^{-1}$, which has poles $z=e+\lambda^2 a_e^{(s)}$ at the eigenvalues of $e+\lambda^2\Lambda_e$. Upon ``integration around'' those poles, in accordance with \eqref{26}, one can extract the dynamical factors $e^{\i t (e+\lambda^2 a_e^{(s)})}$. This procedure works locally, that is, for $z$ close to a fixed $e$. So we will subdivide the integration contour in \eqref{26} into regions close to $e\in{\mathcal E}_0$, for each $e$, and apply the Feshbach map with the appropriate $P_e$. On the complement, where $z$ is away from all eigenvalues of $L_0$, a similar analysis is done using the Feshbach map with projection $P_\r$. This is the outline of the idea, and we refer to Section \ref{proofsect} for the detailed analysis. 

\bigskip

{\bf Proof of \eqref{m4.1}.\ } Let $\phi,\psi\in {\rm Ran}P_\r^\perp$ and use the  resolvent representation (Cauchy formula or Fourier-Laplace transform),
\begin{equation}
\langle \phi, e^{\i t \bar L_\lambda}  \psi\rangle = \frac{-1}{2\pi\i} \int_{{\mathbb R}-\i w} e^{\i t z} \langle \phi, R^{P_\r}_z(\lambda)\psi\rangle dz,
\label{m3.1}
\end{equation}
where $w>0$ is arbitrary. We have 
\begin{eqnarray}
R_z^{P_\r}(\lambda) &=&(z+\i)^{-1}[-\bbbone +R_z^{P_\r}(\lambda)\, (\bar L_\lambda+\i)]\nonumber\\
&=& -(z+\i)^{-1} -(z+\i)^{-2}(\bar L_\lambda+\i) +(z+\i)^{-2} R_z^{P_\r}(\lambda)\, (\bar L_\lambda+\i)^2.
\label{m3.2}
\end{eqnarray}
Since $\int_{\mathbb R}\frac{e^{\i t x}}{(x+\i(1-w))^k}dx=0$ for $t>0$ and $w<1$, the relations \eqref{m3.1} and \eqref{m3.2} imply
\begin{equation}
\langle \phi, e^{\i t \bar L_\lambda}  \psi\rangle = \frac{-1}{2\pi\i} \int_{{\mathbb R}-\i w} \frac{e^{\i t z}}{(z+\i)^2} \langle (\bar L_\lambda+\i)\phi, R^{P_\r}_z(\lambda)(\bar L_\lambda+\i)\psi\rangle dz,
\label{m12.1}
\end{equation}
for $\phi,\psi\in{\rm Dom}(\bar L_\lambda)$. We now use $e^{\i t z} = (\i t)^{-1}\partial_z e^{\i t z}$ and integrate in \eqref{m12.1} by parts $k$ times,
\begin{eqnarray}
\langle \phi, e^{\i t L_\lambda}  \psi\rangle &=& \frac{1}{(\i t)^k} \frac{-1}{2\pi\i} \int_{{\mathbb R}-\i w} e^{\i t z} \partial^k_z\big\{  (z+\i)^{-2}\langle (\bar L_\lambda+\i) \phi, R_z^{P_\r}(\lambda)  (\bar L_\lambda+\i)\psi\rangle \big\}dz.\qquad 
\label{m3}
\end{eqnarray}
It now follows from \eqref{m3} and \eqref{m03}, with $0\le j\le 2$ replaced by $0\le j\le k$, that 
\begin{equation}
\big|\langle \phi, e^{\i t \bar L_\lambda} \psi\rangle\big| \le \frac{C_1\big((\bar L_\lambda+\i)\phi,(\bar L_\lambda+\i)\psi\big)}{(1+t^2)^{k/2}},
\label{m4}
\end{equation}
provided $(\bar L_\lambda+\i)\phi\in {\mathcal D}$ and $(\bar L_\lambda+\i)\phi\in{\mathcal D}$. 
In this way, smoothness of the resolvent gives rise to dissipation in the dynamics. This proves \eqref{m4.1}.

\section{Main result}

Our main result is Theorem \ref{motherthm} below. It involves a reduced system dynamics $e^{\i t M(\lambda)}$, and we explain the generator $M(\lambda)$ now. The simplicity of the spectrum assumed in (A5) implies that the level shift operators $\Lambda_e$ is diagonalizable, 
\begin{equation}
	\label{83}
	\Lambda_e = \sum_{s=1}^{m_e} \aes \Qes.
\end{equation}
The $\aes$ are the eigenvalues and the $\Qes$ are the rank-one spectral projections, satisfying the disjointness and completeness relations
$$
Q_e^{(s)}Q_{e'}^{(s')}=Q_e^{(s)}\delta_{s,s'}\delta_{e,e'}\quad \mbox{ 
and}\quad  \sum_{s=1}^{m_e} Q_e^{(s)}=P_e. 
$$
Given an eigenvalue $e\in{\mathcal E}_0$ of $L_0$, we partition the indices $s=1,\ldots,m_e$ into the oscillating and decaying classes
\begin{equation}
{\mathcal S}_e^{\rm osc} = \big\{ s : \aes\in{\mathbb R} \big\}\qquad \mbox{and}\qquad {\mathcal S}_e^{\rm dec} = \big\{ s : {\rm Im} \aes>0\big\}.
\label{27}
\end{equation}
We define the operators
\begin{equation}
	M(\lambda) = \bigoplus_{e\in{\mathcal E}_0} M_e,\qquad M_e = \sum_{s\in{\mathcal S}_e^{\rm dec}} \big(e+\lambda^2\aes\big) \Qes + \sum_{s\in{\mathcal S}^{\rm osc}_e} \Ees \Qes.
	\label{M}
\end{equation}
Since $Q_e^{(s)}=Q_e^{(s)}P_\r=P_\r Q_e^{(s)}$ we view each $M_e$ as an operator on ${\rm Ran}P_{\s,e}$, the eigenspace of $L_\s$ associated to the eigenvalue $e$. The operator $M(\lambda)$ is then also an operator acting on the system Hilbert space ${\mathcal H}_\s$ alone.  By construction, $M(\lambda)$ and $L_\s$ commute, and $M(0)=L_\s$. 
\medskip

\begin{thm}[Resonance expansion of propagator]
\label{motherthm}
There is a constant $c_0>0$ such that if  $0<|\lambda|< c_0$, then we have weakly on ${\mathcal D}$,  
\begin{equation}
e^{\i tL_\lambda} = e^{\i t M(\lambda)}\otimes P_\r + P^\perp_\r e^{\i t P^\perp_\r L_\lambda P^\perp_\r} P^\perp_\r +R(\lambda,t), 
	\label{n76}
\end{equation}
where the operator $R(\lambda,t)$ satisfies
\begin{equation}
\big|\langle \phi , R(\lambda,t) \psi\rangle \big|\le C |\lambda|^{1/4} \, K(\phi,\psi), \qquad \forall \phi,\psi\in{\mathcal D},
\label{31.1}
\end{equation}
for constants $C$ and $K$ independent of $t\ge0$. 
\end{thm}
If interested in $\langle\phi, e^{-\i tL_\lambda}\psi\rangle$, $t\ge0$, then it suffices to take the adjoint of \eqref{n76}. The choice of the sign in the exponent on the left side of \eqref{n76} is motivated by the application of our result to the reduced system dynamics, \cite{Markov2}. The generator $M(\lambda)$ is defined in its diagonalized form \eqref{M} and so $e^{\i t M(\lambda)}$ is easily obtained from the functional calculus: The equality \eqref{n76} means that for all $\phi,\psi\in{\mathcal D}$,
\begin{eqnarray}
\langle\phi, e^{\i tL_\lambda}\psi\rangle  &=& \sum_{e\in{\mathcal E}_0} \Big[ \sum_{s\in{\mathcal S}_e^{\rm dec}} e^{\i t (e+\lambda^2 \aes)} \big\langle\phi, \Qes\psi\big\rangle + \sum_{s\in{\mathcal S}_e^{\rm osc}}  \, e^{\i t \Ees}\big\langle \phi, \Qes\psi\big\rangle\Big]\nonumber\\
&&+\langle \phi, P^\perp_\r e^{\i t P^\perp_\r L_\lambda P^\perp_\r} P^\perp_\r \psi\rangle  +\langle\phi, R(\lambda,t)\psi\rangle.
\label{76.1.1.1}
\end{eqnarray}

Discussion of Theorem \ref{motherthm} 
\begin{itemize}

\item[(i)] The operator $M(\lambda)$ generally {\em contains all orders of $\lambda$}  in the oscillatory parts $E_e^{(s)}(\lambda)$, see \eqref{M}. It is clear that we need to include the {\em exact} phases (eigenvalues $E_e^{(s)}(\lambda)$ of $L_\lambda$ to all orders in $\lambda$) in the approximate dynamics on the right side of \eqref{n76}, for otherwise we have diverging phase differences, and we cannot get an approximation small in $\lambda$ for all times. However, in the decaying terms, we can truncate the exponents at the second order in $\lambda$, keeping only $e+\lambda^2\aes$, and the time decay still allows us a control uniform in $t$.

\item[(ii)] On the finite-dimensional part, the dynamics  $e^{\i t M(\lambda)}\otimes P_\r$ on the right side of \eqref{n76}, there are terms oscillating in time, $\propto e^{\i t E_e^{(s)}(\lambda)}$, and terms decaying in time, $\propto e^{\i t (e+\lambda^2 a_e^{(s)})}$. The decay is exponential, with rates  $\lambda^2 {\rm Im}a_e^{(s)}>0$, since  $|e^{\i t (e+\lambda^2 a_e^{(s)})}| =e^{-\lambda^2 t{\rm Im}a_e^{(s)}}$. On the infinite-dimensional part ({\em i.e.}, on ${\rm Ran}P_\r^\perp$), the dynamics is also decaying, but only at a polynomial rate $1/t^2$, as per \eqref{m4.1}. 

\item[(iii)] The asymptotic dynamics for times larger than all the life times $(\lambda^2 {\rm Im}a_e^{(s)})^{-1}$ and large enough so that the dissipative term $\langle \phi, P^\perp_\r e^{\i t P^\perp_\r L_\lambda P^\perp_\r} P^\perp_\r \psi\rangle\sim t^{-2}$ has decayed ({\em c.f.} \eqref{m4.1}),  is given by $e^{\i t L_\lambda} \approx e^{\i t M(\lambda)}\otimes P_\r$. If $E_e^{(s)}(\lambda)=0$ is the only eigenvalue of $L_\lambda$, then the dynamics relaxes to the final state $Q_e^{(s)}\otimes P_\r$ modulo an error $O(|\lambda|^{1/4})$. 

\item[(iv)] We may call $e^{\i t M(\lambda)}\otimes P_\r$ and  $ P^\perp_\r e^{\i t P^\perp_\r L_\lambda P^\perp_\r} P^\perp_\r$ of \eqref{n76} the quasi-static and dissipative parts, respectively. The decomposition \eqref{n76} is reminiscent of the structure of solutions of dispersive partial differential equations, which split into a dispersive wave plus a part converging towards an invariant manifold \cite{PW,SW}. In our case, after the decay of the dissipative term, the dynamics first approaches the quasi-static manifold spanned by the $Q_e^{(s)}$ for all $e, s$. The orbits stay close to the quasi-static manifold for times up to $\sim 1/\lambda^2$, after which the dynamics moves to the final,  invariant manifold spanned, modulo $O(|\lambda|^{1/4})$,  by the $Q_e^{(s)}\otimes P_\r$ with $e\in{\mathcal E}_0$ and $s\in{\mathcal S}_e^{\rm osc}$.

\item[(v)]  The approach we take to prove Theorem \ref{motherthm} is similar to \cite{KM1}. In that paper, an expansion of the dynamics was given as a sum of a main part plus a remainder, similar to \eqref{n76}. The remainder of \cite{KM1} converges to zero in the limit of large times, but it is not shown to be small in $\lambda$. In the current work, the remainder is small in $\lambda$ for all times \eqref{31.1}, but it does not converge to zero for large times. This means, incidentally, that the main term of the current work has a simpler form than the one of \cite{KM1}. Indeed, the current main term is a second order approximation along the decaying terms, with  $s\in{\mathcal S_{\rm dec}}$, of that of \cite{KM1}. The result of \cite{KM1} is not suitable for proving the validity of the Markovian approximation, but the current result here is.
\end{itemize}

\medskip

{\bf Uncorrelated initial states.\ } For vectors of the form $\psi=\psi_\s\otimes\Omega_\r$, $\phi=\phi_\s\otimes\Omega_\r$ the state-dependent constant $K$ in the remainder \eqref{31.1}, satisfies $K(\phi,\psi)=\|\phi_\s\|\, \|\psi_\s\|$ (see \eqref{137.1} below). The term $\langle \phi, P^\perp_\r e^{\i t P^\perp_\r L_\lambda P^\perp_\r} P^\perp_\r \psi\rangle $ vanishes. We then obtain the following result directly from \eqref{n76}.
\begin{cor}
	\label{corollary1}
There is a constant $c_0$  such that if $0<|\lambda|< c_0$, then for all $\psi_\s, \phi_\s\in {\mathcal H}_\s$, $t\ge 0$, 
	\begin{equation}
	\Big| \langle\phi_\s \otimes\Omega_\r, e^{\i tL_\lambda}\psi_\s\otimes\Omega_\r \rangle_{{\mathcal H}_\s\otimes{\mathcal H}_\r}  - \langle\phi_\s, e^{\i t M(\lambda)}\psi_\s\rangle_{{\mathcal H}_\s}\Big| \leq C |\lambda|^{1/4}\, \|\phi_\s\|\, \|\psi_\s\|. 
	\label{80.1}
	\end{equation}
\end{cor}
This corollary is the starting point for a proof that the Markovian approximation is valid for all times, which we give in \cite{Markov2}.  We show that $M(\lambda)$ gives rise to a completely positive trace-preserving semigroup acting on system density matrices, which coincides with the well-known Gorini-Kossakovski-Sudarshan-Lindblad semigroup \cite{AL,CP,MAOP}. We point out, though, that correlated intitial states are also treatable with our method -- indeed, the main result, Theorem \ref{motherthm}, does not make an assumption on intital $\s\r$ correlations. We explore this in \cite{Mcorr} (see also Section \ref{sect:newresults}). 
\bigskip

{\bf Parameter dependence.\ } We take some care in estimating $c_0$, $C$ and $K$ appearing in Theorem \ref{motherthm} and Corollary \ref{corollary1} in terms of various parameters, as we explain now. The analysis is based on a perturbation theory ($\lambda$ small) and the corrections of the spectrum are governed by the level shift operators $\Lambda_e$, whose spectral decomposition is given in \eqref{83}. To quantify the perturbation theory, we  introduce the quantities
\begin{eqnarray}
a&=& \min_{e,s}\big\{ {\rm Im}\aes : \aes\not\in{\mathbb R} \big\}  \qquad \mbox{(smallest nonzero imaginary part)}\label{FGR}\\
\alpha&=&  \max_{e,s} |\aes| = \max_e {\rm spr}(\Lambda_e)\qquad \mbox{(maximal spectral radius)}\label{alpha}\\
\delta&=&	\min_{e, s,s'}\big\{|a_e^{(s)} -a_e^{(s')}| : s\neq s'\big\}  \label{delta} \qquad \mbox{(gap in spectrum of the  $\Lambda_e$)}\\
\kappa&=& 	\max_{e,s} \|\Qes\|\qquad \mbox{(size of biggest spectral projection)}
\label{kappa}
\end{eqnarray}
We also define 
\begin{equation}
\label{195.1}
\varkappa_1 = \max_{m,n} C_1\big(I(\varphi_m\otimes\Omega_\r), I(\varphi_n\otimes\Omega_\r)\big),
\end{equation}
where $C_1(\cdot,\cdot)$ is given in \eqref{m03} and where $\{\varphi_m\}_{m=1}^{\dim {\mathcal H}_\s}$ is an orthonormal eigenbasis of $L_\s$. The parameter $\varkappa_1$ is well defined since due to \eqref{16.3}, the vectors $I \varphi_m\otimes\Omega_\r$ are in $\mathcal D$. We combine all these constants and the spectral gap $g$ of $L_0$ (see  \eqref{g}) into the following two effective constants, 
\begin{eqnarray}
\varkappa_0&=&
		\max\Big[1, 1/g,\kappa/a,\alpha\kappa,\varkappa_1(1+\varkappa_1)(1+g+1/g),\nonumber\\
		&&\varkappa_1(1+\varkappa_1^4)\kappa \max\big\{ 1,\tfrac{1+\alpha}{\delta}, 1/a, \tfrac{1+\kappa}{a\delta}, \kappa^2\big( \tfrac{\kappa}{a}(1+\kappa^3/\delta^3)(1+1/a)+1/\delta^2
		\big)\big\}\Big] \qquad
 \label{160.1.1}\\
 \lambda_0 &=& \frac{\min\Big[ 1,a,\delta/\kappa^2, \|I P_\r\|, g^{3/2}\Big]}{\max\Big[ 1, \varkappa_1\kappa (1+\varkappa_1\kappa/\delta), \alpha,\varkappa_1\Big]}.
 \label{lambdanot}
\end{eqnarray}
Furthermore, for vectors $\phi,\psi\in{\mathcal D}$ such that $\bar L_\lambda \phi, \bar L_\lambda\phi\in{\mathcal D}$, with $\bar L_\lambda=P^\perp_\r L_\lambda P^\perp_\r$, we define
\begin{eqnarray}
	K(\phi,\psi) &=& \|\phi\|\,\|\psi\| + \max_{j=1,2} C_j(\phi,\psi)\nonumber \\
	&& +\max_{j=1,2} \big( \max_m C_j(I\varphi_m\otimes\Omega_\r,\phi) \max_m C_j(I\varphi_m\otimes\Omega_\r,\psi)\big)\nonumber\\
	&&+  \sym \|\phi\| \max_{j,m}C_j(I\varphi_m\otimes\Omega_\r, \psi)\nonumber\\
	&&  +  \sym \Big( \|\phi\| +\max_m C_1(I\varphi_m\otimes\Omega_\r, \phi)\Big)\nonumber\\
	&&\quad \times \Big(  \|P_\r I\|  \|P^\perp_\r\psi\|+ \max_m C_1\big(I\varphi_m\otimes\Omega_\r,(\bar L_\lambda+\i)P^\perp_\r\psi\big) \Big).
	\label{137.1}
\end{eqnarray}
In \eqref{137.1} we use the notation
\begin{equation}
\sym E(\phi,\psi) = E(\phi,\psi)+E(\psi,\phi)
\label{40} 
\end{equation}
for any function $E$ of $\phi$ and $\psi$. 

\begin{prop} [Parameter dependence of $c_0,C, K$] 
	\label{prop2}
The constants $c_0, C$ of Theorem \ref{motherthm} and Corollary \ref{corollary1} can be written as $c_0=c'\lambda_0^{4/3}$ and $C=C'\varkappa_0$, where $c', C'$ do not depend on $\lambda, g, a, \alpha, \kappa, \delta$ and $\lambda_0$, $\varkappa_0$ are given in \eqref{lambdanot} and \eqref{160.1.1}, respectively. The constant $K$ in \eqref{31.1}, which is a function on ${\mathcal D}\times{\mathcal D}$, is given by \eqref{137.1}.
\end{prop}

{\em Remark.\ } The parameter $\varkappa_1$,  \eqref{195.1}, is defined in terms of the function $C_1$, \eqref{m03}. The latter describes the dissipative nature of the interacting dynamics and, {\em a priori}, depends on $\lambda$. However, the dissipation is due to the nature of the environment alone and so in applications, one can bound $\varkappa_1$ uniformly in $\lambda$, for small enough $\lambda$ (see \cite{Markov2}). See also the point (ii) in the discussion of condition (A2) above. In this sense, $\varkappa_1$ and thus $\lambda_0$, are independent of $\lambda$. 
\bigskip

\section{Proof of Theorem \ref{motherthm}}
\label{proofsect}

We are presenting the core strategy in Section \ref{sect4.0}. It involves estimates which we derive in Sections \ref{sect4.1} and \ref{sect4.2}. We then collect the estimates and implement the outlined strategy in Section \ref{proofpropersect}. As we explained after Theorem \ref{motherthm}, our approach is similar to that of \cite{KM1}. However, in order to be able to show that the remainder is small in $\lambda$ for all times, \eqref{31.1}, we need to substantially modify the analysis of \cite{KM1}. The main alteration is the introduction of two new energy scales $\eta$ and $\vartheta$ which allow for a more detailed estimation of the resolvent. See also Fig.\,2 and Fig.\,4.
\medskip

{\bf Notational convention. }  For $X,Y\ge 0$, we write  $X\prec Y$ to mean that there is a constant $C$ independent of the coupling constant $\lambda$ as well as the parameters  $g, a,\alpha,\delta,\kappa$,  such that $X\le C Y$. (Recall the definitions \eqref{g}, \eqref{FGR}-\eqref{kappa} of those parameters.)

\subsection{Strategy}
\label{sect4.0}

Given an orthogonal projection $Q$ on ${\mathcal H}$ we set $Q^\perp=\bbbone-Q$  and we denote the resolvent and reduced resolvent operators by
\begin{equation} 
	R_z (\lambda)= (L_\lambda-z)^{-1} \qquad \mbox{and}\qquad R_z^Q (\lambda)= (Q^\perp L_\lambda Q^\perp -z)^{-1}\upharpoonright_{{\rm Ran}Q^\perp}.
	\label{m12}
\end{equation}
The resolvent representation of the propagator is
\begin{equation}
	\label{15}
\langle \phi ,  \, e^{\i t L} \psi\rangle =\frac{-1}{2\pi\i} \int_{{\mathbb R}-\i w} e^{\i tz}\langle \phi \, R_z(\lambda)  \psi \rangle \, dz,
\end{equation}
valid for any $w>0$ if either of $\psi$ or $\phi$ belong to ${\rm Dom}(L)$ (see \cite{EN}). Given an orthogonal projection $Q$, the resolvent has a decomposition into a sum of three parts:\footnote{{\em Reminiscent of ``... est omnis divisa in partes tres ...''}, \cite{Cesar}} a part acting on ${\rm Ran}Q$, one acting on ${\rm Ran}Q^\perp$ and one part mixing these subspaces. This decomposition reads, for $z\in{\mathbb C}\backslash{\mathbb R}$,
\begin{equation}
\label{16}
R_z(\lambda) = {\mathfrak F}(z)^{-1} +{\mathfrak B}(z) + R^Q_z(\lambda).
\end{equation}
Here, $\mathfrak F$ is the Feshbach map \eqref{Feshbach},
\begin{equation}
 {\mathfrak F}(z) \equiv {\mathfrak F}(L-z;Q)= Q(L-z - LQ^\perp R_z^Q(\lambda)L)Q.
 \label{16.4}
\end{equation}
The operator ${\mathfrak B}(z)$ in \eqref{16} is given by
\begin{equation}
\label{17}
{\mathfrak B}(z) = -{\mathfrak F}(z)^{-1} QL R_z^Q - R_z^QLQ{\mathfrak F}(z)^{-1} +R^Q_z LQ{\mathfrak F}(z)^{-1} QL R_z^Q. 
\end{equation}
Here and in what follows, it is convenient to simply write $R_z^Q$ for $R_z^Q(\lambda)$. The existence of ${\mathfrak F}(z)^{-1}$ is automatic, this operator equals $QR_z(\lambda)Q$, see \eqref{16}. It is elementary to establish the relation \eqref{16}, see Section \ref{Fdecsect} for more detail.
\medskip

\begin{figure}[t]
	\centering
\includegraphics[width=15cm]{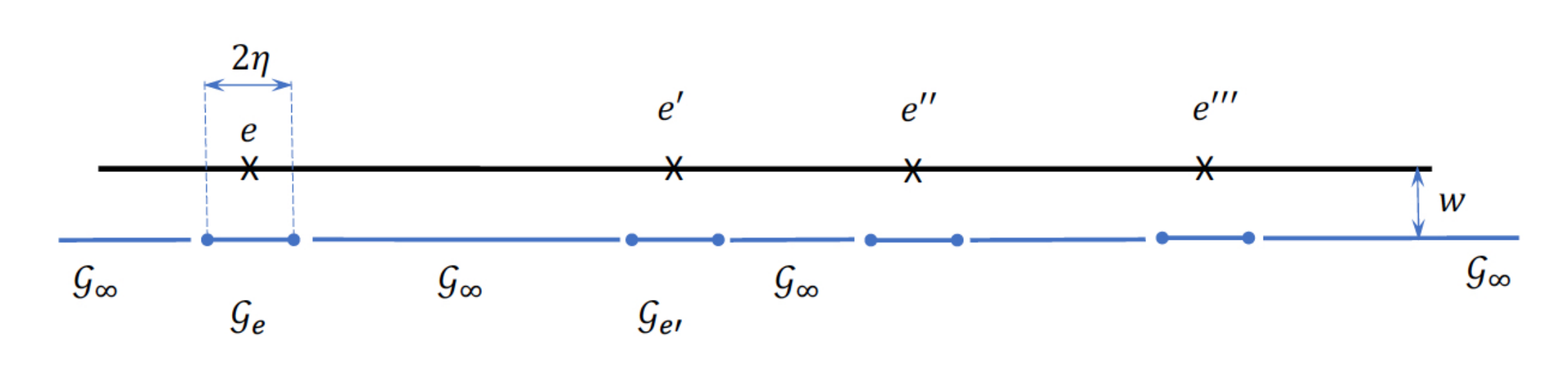}	
	\caption{Subdivision of the integration path ${\mathbb R}-\i w$.}
	\label{Fig2}
\end{figure}

Our {\bf strategy} is to partition the integration contour ${\mathbb R}-\i w$ in \eqref{15} into segments, see Fig.~\ref{Fig2}, and apply \eqref{16} with suitable projections $Q$ on each segment. To describe the partition, fix an $\eta>0$. It will be chosen as a certain positive power of  $|\lambda|$ later on, so $\eta$ is a small parameter (relative to the eigenvalue gap $g$ of $L_0$, \eqref{g}).  For every eigenvalue $e$ of $L_0$  define the segment (see Fig. \!\ref{Fig2})  
\begin{equation} 
\label{018}
{\mathcal G}_e =\big\{x-\i w\ :\ |x-e|\le \eta\big\} \quad \mbox{and set}	\quad {\mathcal G}_\infty = \big\{x-\i w\ :\ {\rm dist}(x, {\mathcal E}_0)>\eta \big\},
\end{equation}
where $w>0$ is the parameter in \eqref{15} and ${\mathcal E}_0$ is the set of all eigenvalues of $L_0$.  

We have the disjoint decomposition  ${\mathbb R}-\i w = {\mathcal G}_\infty \bigcup_{e\in {\mathcal E}_0} {\mathcal G}_e$ and \eqref{15} gives 
\begin{equation}
	\langle \phi, e^{\i t L_\lambda}\psi\rangle = \sum_{e\in{\mathcal E}_0} J_e(t) +J_\infty(t), \quad \mbox{with}\quad 
J_\# (t) =\frac{-1}{2\pi\i} \int_{{\mathcal G}_\#} e^{\i tz}\langle \phi,   \, R_z\,    \psi \rangle \, dz.
	\label{19}
\end{equation}

On ${\mathcal G}_e$ we will apply the Feshbach map with projection $Q=P_e$, \eqref{m7.1}, and on ${\mathcal G}_\infty$ we use the projection $Q=P_\r$, \eqref{POmega}. Accordingly, it is important that we have dissipative bounds on the reduced resolvents, {\em i.e.}, the following {\em local limiting absorption principles}, which follow from the global ones given in \eqref{m03}, \eqref{m03.1}.

\begin{thm}[Local limiting absorption principle]
	\label{thmA1}
	If $\lambda^2\varkappa_1\prec g$ then we have for all $\phi,\psi\in{\mathcal D}$ and all $e\in{\mathcal E}_0$
	\begin{eqnarray}
	\max_{0\le j\le 2}\sup_{ \{z\in{\mathbb C}_- \, :\,  |z-e|\le g/2 \}}\Big| \partial_z^j \langle\phi, R_z^{P_e}(\lambda) \psi\rangle\Big| &\le& C_2(\phi,\psi)<\infty,
	\label{32}\\
	\sup_{ \{z\in{\mathbb C}_- \, :\,  |z-e|\le g/2 \}}\Big| \partial_\lambda\langle\phi, R_z^{P_e}(\lambda) \psi\rangle\Big| &\le& C_2(\phi,\psi)<\infty,
	\label{034}
	\end{eqnarray}
	where $C_2(\cdot,\cdot)$ is well defined (finite) on ${\mathcal D}\times{\mathcal D}$ and satisfies
	\begin{eqnarray}
	\lefteqn{
		C_2(\phi,\psi) \prec    C_1(\phi,\psi) +\max\{1,1/g^3\} \Big[ \|\phi\|\, \|\psi\|} \label{198.1}\\
	&& +\lambda^2  \max_m C_1(\phi, I\varphi_m\otimes\Omega_\r) C_1(\psi, I\varphi_m\otimes\Omega_\r) +|\lambda| \sym \|\psi\| \max_m C_1(\phi, I \varphi_m\otimes \Omega_\r)\Big].\nonumber
	\end{eqnarray}
	Furthermore, we have
	\begin{equation}
	\max_{0\le k\le 2}			\sup_{ 	\{z\in{\mathbb C}_-\, : \, |{\rm Re}z-e|\ge g/2\  \forall e \}} \ \big|\partial_z^k \langle\phi, R_z^{P_\r}\psi\rangle\big| \le C_2(\phi,\psi).
	\label{33}
	\end{equation}
\end{thm}

We give a proof of Theorem \ref{thmA1} in Section \ref{proofthmA1section}. Similar to $\varkappa_1$ defined in \eqref{195.1}, we set
\begin{equation}
\varkappa_2 = \max_{m,n} C_2\big(I(\varphi_m\otimes\Omega_\r), I(\varphi_n\otimes\Omega_\r)\big).
\label{39.1}
\end{equation}
Using \eqref{198.1} we get the majorization 
\begin{equation}
\varkappa_2 \prec \varkappa_1 +\max\{1,1/g^3\}\big( |\lambda| \varkappa_1 +\|IP_\r\| \big)^2.
\label{40.1}
\end{equation}
Furthermore, imposing the constraints
\begin{equation} 
|\lambda|\varkappa_1\prec \|IP_\r\|\quad\mbox{and}\quad \lambda^2\varkappa_1\prec \min\{1, g^3\}
\label{constraints}
\end{equation}
gives from \eqref{40.1} the simple bound,
\begin{equation}
\varkappa_2\prec \varkappa_1.
\label{kappa12}
\end{equation}

\subsection{Cheat sheet}
\label{cheatsheetsect}

We keep track of several constants during the estimates to follow. The following cheat sheet is presented for the convenience of the reader.
\begin{table}[ht]
	\centering 
	\begin{tabular}{c c c } 
		Symbol & Meaning & Definition  \\ [.5ex] 
	\hline \\ [-2ex]
$a$& smallest nonzero imaginary part of all $\Lambda_e$ & \eqref{FGR}  \\ 
$a_e^{(s)}$ & eigenvalues of $\Lambda_e$ &  \eqref{83}\\ 
$\alpha$  & maximal spectral radius of all $\Lambda_e$ & \eqref{alpha}  \\
$\beta$ & inverse temperature & \\
$C_1(\phi,\psi)$, $C_2(\phi,\psi)$ & global, local limiting absorption constants & \eqref{m03}, \eqref{32}\\
$\delta$ & minimal gap of all $\Lambda_e$ & \eqref{delta}  \\
$e$ & eigenvalues of $L_\s$ (or $L_0$) & after \eqref{m1}  \\ 
$\eta$ & integration domain parameter &\eqref{018} \\
$g$ & gap of $L_\s$ & \eqref{g}  \\
${\mathcal G}_e$, ${\mathcal G}_\infty$ & integration domains & \eqref{018}\\
$\varkappa_1$, $\varkappa_2$ & global, local limiting absorption constants & \eqref{195.1}, \eqref{39.1}\\
$\varkappa_0$, $\varkappa_3$, $\varkappa_4$, $\varkappa_5$ & & \eqref{160.1}, \eqref{119.0}, \eqref{119.1}, \eqref{kappa5}\\
$\kappa$ & biggest norm of spectral projections of all  $\Lambda_e$& \eqref{kappa}\\
$\lambda$ & coupling constant & \eqref{m1} \\
$\Lambda_e$ & level shift operator & \eqref{34}, \eqref{83}\\
$Q_e^{(s)}$ & eigenprojections of $\Lambda_e$ & \eqref{83}\\
${\mathcal S}_e^{\rm osc}$, ${\mathcal S}_e^{\rm dec}$ & oscillating and decaying index sets & \eqref{27}\\
$\vartheta$ & integration domain parameter & beginning Section \ref{sect4.2}\\
${\mathfrak S}_{\phi\leftrightarrow\psi}$ &  symmetrizer & \eqref{40} \\
$w$ & vertical integration offset & \eqref{15} \\
$\prec$ & parameter independent majorization & beginning section \ref{proofsect}\\
[1ex] 
		\hline 
	\end{tabular}
\end{table}

\subsection{Analysis of $J_e(t)$}
\label{sect4.1}

Fix an eigenvalue $e\in{\mathcal E}_0$ of $L_\s$ and choose $Q=P_e$ in the Feshbach decompositon \eqref{16}, where $P_e$ is the eigenprojection  \eqref{m7.1}. The Feshbach operator \eqref{16.4} reads
\begin{equation}
{\mathfrak F}(z) = e-z +\lambda^2 A_e(z,\lambda), \qquad A_e(z,\lambda)=  - P_e I  R_z^{P_e}(\lambda) IP_e.
\label{23.1}
\end{equation}
For $z=e$ and $\lambda=0$, $A_e(z,\lambda)$ reduces to the level shift operator $\Lambda_e$. We show in Lemma \ref{lem2.4} below that for $z\in{\mathbb C}_-$ with $|z-e|$ not too large and $\lambda$ not too large, $A_e(z,\lambda)$ has simple eigenvalues (a property inherited from $\Lambda_e$) and we denote the spectral representation by 
\begin{equation}
	A_e(z,\lambda)  = \sum_{s=1}^{m_e} \aes(z,\lambda) \Qes(z,\lambda).
	\label{89.1}
\end{equation}
Here $a_e^{(s)}(z,\lambda)$ are the eigenvalues and $Q_e^{(s)}(z,\lambda)$ the rank one eigenprojections which extend by continuity to $z\in{\mathbb R}$ (Corollary \ref{cor1}).

\begin{figure}[t]
	\centering
 	\includegraphics[width=7.5cm]{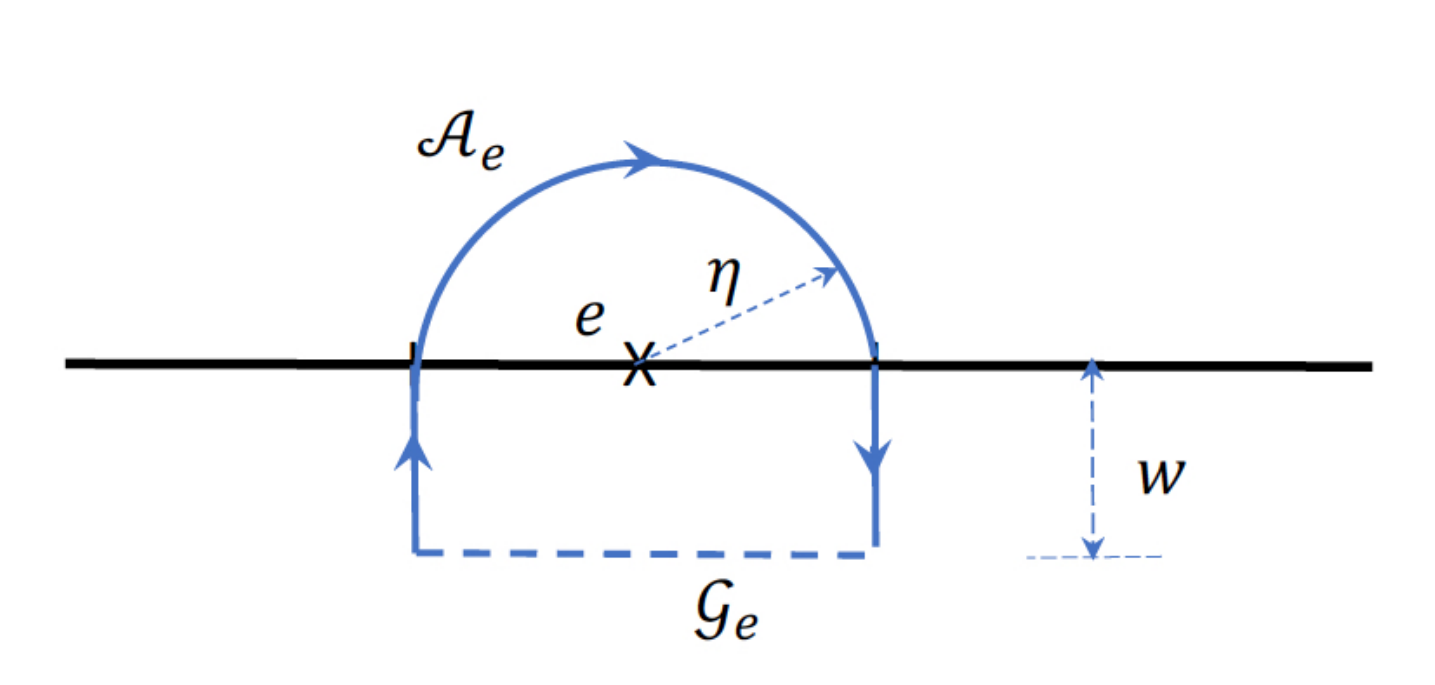}
	\caption{The contour ${\mathcal A}_e$.}
	\label{Fig3}
\end{figure}

We define the oriented contour ${\mathcal A}_e$, depicted in  Fig.~\ref{Fig3}, by
 \begin{equation}
 	\label{40-1}
 	{\mathcal A}_e = \big\{e-\eta+\i y : y\in[-w,0]\big\} \cup \big\{e+ \eta e^{ia} : a\in[\pi,2\pi]\big\} \cup \big\{e+ \eta+\i y : y\in[0,-w]\big\}.
 \end{equation}

The main result of this section is 
\begin{prop}
	\label{propm1}
Suppose that 
\begin{equation}
	\label{cond1}
|\lambda| +\eta\prec \frac{a}{\varkappa_4 },  \quad \lambda^2\varkappa_4 \prec 1\quad \mbox{and}\quad  \lambda^2 \frac{\alpha+|\lambda| \varkappa_4}{\eta} \prec 1.
\end{equation}
Then
 \begin{eqnarray}
 	\label{024}
J_e(t) &=&\sum_{s\in{\mathcal S}_e^{\rm dec}}e^{\i t (e+\lambda^2 \aes(e,\lambda))} \big\langle\phi, \Qes(e,\lambda)\psi\big\rangle +\sum_{s\in {\mathcal S}_e^{\rm osc} }  \, e^{\i t \Ees}\big\langle \phi, \Qes(\Ees,\lambda)\psi\big\rangle \nonumber\\
&&-\frac{1}{2\pi\i}\int_{{\mathcal A}_e}  \,  e^{\i t z}\big\langle\phi, (L_\s- z)^{-1}P_\r\psi\big\rangle  \, dz +S_e,
\end{eqnarray}
where
\begin{align}	
&\| S_e\| \prec \\
& \ \ \  \Big\{ e^{wt}\frac{w +\eta}{\eta} \Big( \frac{\eta}{g}+ \lambda^2+\varkappa_3 ( |\lambda|+\eta)\Big) +\lambda^2 \nonumber\\
&\qquad + \eta e^{wt}\Big( \varkappa_3+\lambda^2\varkappa_5 + \Big( \frac{\varkappa_4\kappa}{a} +\varkappa_3\Big)  \Big( 1+\frac{\eta}{a\lambda^2}\Big) \Big)\Big\} \|\phi\|\, \|\psi\|\nonumber\\
&+ \Big\{ 
 \frac{\kappa\eta}{a|\lambda|} e^{wt} + |\lambda|\big( 1 +\eta(1+\varkappa_3)\big) + \frac{(w+\eta) e^{wt}}{\eta} (|\lambda| +\lambda^2)\nonumber\\
&\qquad + e^{wt}\lambda^2\eta\varkappa_5(|\lambda|+\lambda^2)+ \lambda^2 \Big( 1+ (w+\eta) e^{wt}(1+\varkappa_3)  \Big)\Big\} \nonumber\\
& \qquad \times \Big[\max_j C_2(I\varphi_j\otimes\Omega,\phi) \max_j C_2(I\varphi_j\otimes\Omega,\psi)+  \sym \|\phi\| \max_jC_2(I\varphi_j\otimes\Omega_\r, \psi) \Big] \nonumber\\
& +\eta e^{wt} C_2(\phi,\psi)
\label{m57}
\end{align}
with the various constants  defined in Section \ref{cheatsheetsect}. 
\end{prop}

{\em Proof of Proposition \ref{propm1}.\ } We show 
\eqref{024}  by estimating separately the three contributions coming from the decomposition \eqref{16} of the resolvent $R_z$,
\begin{eqnarray}
\int_{{\mathcal G}_e} e^{\i t z} R_z dz &=& \int_{{\mathcal G}_e} e^{\i t z}{\mathfrak F}(z)^{-1} dz +\int_{{\mathcal G}_e} e^{\i t z}{\mathfrak B}(z) dz + \int_{{\mathcal G}_e} e^{\i t z}R^Q_z dz\nonumber\\
& \equiv& D_1+D_2+D_3. 
\label{16.1}
\end{eqnarray}
The corresponding bounds are presented in \eqref{43}, \eqref{82} and \eqref{53} below.
\bigskip 
 
\noindent
{\bf Estimate of $D_1$ in \eqref{16.1}. }  Our main estimate for $D_1$ is given in \eqref{43} below. All estimates in this section are controlled in the operator norm sense for operators on ${\rm Ran}P_e$. We have 
\begin{equation}
	{\mathfrak F}(z) = e-z +\lambda^2 A_e(z,\lambda),
	\label{23}
\end{equation}
where the properties of $A_e(z,\lambda)=- P_e I  R_z^{P_e} IP_e,$ are discussed in Section \ref{sectAe}.  It follows from \eqref{89} that 
\begin{equation}
	\label{f-1}
 {\mathfrak F}(z)^{-1} = \sum_{s=1}^{m_e}  \frac{\Qes(z,\lambda)}{e-z+\lambda^2 \aes(z,\lambda)} .
\end{equation}
$\bullet$ Consider \eqref{f-1} for an $s\in{\mathcal S}_e^{\rm dec}$.   The case $s\in{\mathcal S}_e^{\rm osc}$ is dealt with below. Let $z\in{\mathcal G}_e$. We show that 
\begin{equation}
	\frac{\Qes(z,\lambda)}{e-z+\lambda^2 \aes(z,\lambda)} = \frac{\Qes(e,\lambda)}{e-z+\lambda^2 \aes(e,\lambda)}  + T(z,\lambda),
	\label{32.1}
\end{equation}
where (recall the definitions \eqref{FGR}  and \eqref{alpha} of $a, \alpha$)
\begin{equation}
\label{32.2}
\| T(z,\lambda)\| \prec\Big( \frac{\varkappa_4\kappa}{a} +\varkappa_3\Big)  \Big( 1+\frac{\alpha+|\lambda|\varkappa_4}{a}\Big).
\end{equation}
 To show \eqref{32.1}, we start with the expression for $T(z,\lambda)$, 
\begin{eqnarray}
T(z,\lambda) &=& T'(z,\lambda) +T''(z,\lambda)\nonumber\\
T'(z,\lambda) &=& -\lambda^2 \Qes(z,\lambda)  \frac{\aes(z,\lambda)-\aes(e,\lambda)}{e-z} \frac{e-z}{[e-z+\lambda^2 \aes(z,\lambda)][e-z+\lambda^2 \aes(e,\lambda)]}\nonumber\\
T''(z,\lambda) &=& \frac{\Qes(z,\lambda) -\Qes(e,\lambda)}{e-z+\lambda^2 \aes(e,\lambda)}. 
	\label{0032}
\end{eqnarray}
Recall that $\aes(e,0)=\aes$ is an eigenvalue of $\Lambda_e$. We  have
\begin{eqnarray}
	\lefteqn{
|e-z+\lambda^2 \aes(z,\lambda)|}\nonumber\\
 &=& \big|e-z+\lambda^2 \aes +\lambda^2(\aes(e,\lambda)-\aes) +\lambda^2\big(\aes(z,\lambda)-\aes(e,\lambda) \big)\big|\nonumber\\
	&\ge& |e-z+\lambda^2 \aes| -C \lambda^2\varkappa_4 \big(|\lambda|+|z-e|\big),
	\label{032}
\end{eqnarray}
where we have used \eqref{119.1} (see also Corollary \ref{cor1}). Now $|e-z+\lambda^2\aes|\ge {\rm Im} ( e-z+\lambda^2\aes)
\ge \lambda^2{\rm Im }\aes \ge a \lambda^2$ (see \eqref{FGR}), so \eqref{032} yields $|e-z+\lambda^2 \aes(z,\lambda)|\ge a\lambda^2/2$, since $\varkappa_4\big(|\lambda| +|z-e|\big) \prec a$ by \eqref{cond1} (note that $z\in{\mathcal G}_e$ so we have $|z-e|\le \sqrt{w^2+\eta^2}\le \sqrt2 \,\eta$ (we will take $w\rightarrow 0$ at fixed $\eta$, so $w<\eta$ without loss of generality)). We get 
\begin{equation}
	\label{33.1}
	\frac{1}{|e-z+\lambda^2 \aes(z,\lambda)|}\prec \frac{1}{a \lambda^2}.
\end{equation} 
Moreover, we have
\begin{eqnarray} 
\Big| \frac{e-z}{e-z+\lambda^2\aes(e,\lambda)}\Big| &\le& 1 +\lambda^2\Big|
\frac{\aes(e,\lambda)}{e-z+\lambda^2\aes(e,\lambda)} \Big|\nonumber\\
&\prec& 1+\frac{|\aes(e,\lambda)|}{a} \prec 1+ \frac{\alpha +|\lambda|\varkappa_4}{a}
	\label{0033}
\end{eqnarray} 
(use \eqref{alpha} and \eqref{119.1} in the last estimate). Now we estimate $T'(z,\lambda)$ given in \eqref{0032}, using the bound $\|\Qes(z,\lambda)\|\prec \kappa$ (see \eqref{bnd}) and \eqref{33.1}, \eqref{0033} and \eqref{119.1},
\begin{equation}
\|T'(z,\lambda)\| \prec \frac{\varkappa_4\kappa}{a} \Big(1+\frac{\alpha +|\lambda|\varkappa_4}{a} \Big). 
\label{41.1}
\end{equation}
To deal with $T''(z,\lambda)$ in \eqref{0032} we write 
\begin{equation}
\label{32.3}
T''(z,\lambda) =\frac{\Qes(z,\lambda) -\Qes(e,\lambda)}{e-z} \frac{e-z}{e-z+\lambda^2 \aes(e,\lambda)} 
\end{equation}
and use the bounds \eqref{119.0} and  \eqref{0033}  to get
\begin{equation}
\|T''(z,\lambda)\| \prec \varkappa_3 \Big( 1+\frac{\alpha+|\lambda|\varkappa_4}{a}\Big).
\label{44.1}
\end{equation}
The combination of \eqref{41.1} and \eqref{44.1} gives \eqref{32.2}. 

$\bullet$ Next we consider a term in \eqref{f-1} with $s\in{\mathcal S}_e^{\rm osc}$. We have to modify the above argument, since  it used that the imaginary part of $\aes$ was strictly positive, which is not the case any more now. By the isospectrality of the Feshbach map, there is exactly one  eigenvalue $E_e^{(s)}(\lambda)$ of $L_\lambda$ with 
\begin{equation}
E_e^{(s)}(\lambda) = e+\lambda^2\aes\big(E_e^{(s)}(\lambda), \lambda\big).
\label{m69} 
\end{equation}
Moreover, the projection $\Qes(\Ees,\lambda)$ is {\em orthogonal} and so it has norm one (\cite{DJ}, Theorem 3.8). 
We decompose
\begin{eqnarray}
\frac{\Qes(z,\lambda)}{e-z+\lambda^2 \aes(z,\lambda)} &=& \frac{\Qes(E_e^{(s)}(\lambda),\lambda)}{E_e^{(s)}(\lambda)-z}
+\frac{\Qes(z,\lambda) - \Qes(E_e^{(s)}(\lambda),\lambda )}{e-z+\lambda^2\aes(z,\lambda)}\nonumber\\
&&+\Qes(E_e^{(s)}(\lambda),\lambda) \frac{E_e^{(s)}(\lambda)-e-\lambda^2\aes(z,\lambda))}{(E_e^{(s)}(\lambda)-z)(e-z+\lambda^2\aes(z,\lambda))}.
\label{m71}
\end{eqnarray}
Using \eqref{m69}, the second term on the right side of \eqref{m71} becomes
\begin{eqnarray}
	\lefteqn{
-\frac{ \Qes(E_e^{(s)}(\lambda),\lambda)- \Qes(z,\lambda) }{E_e^{(s)}(\lambda)-z-\lambda^2[\aes(E_e^{(s)}(\lambda),\lambda)-  \aes(z,\lambda)]} } \nonumber\\
&& \qquad = -\frac{ \Qes(E_e^{(s)}(\lambda),\lambda) - \Qes(z,\lambda) }{E_e^{(s)}(\lambda)-z} \frac{1}{1-\lambda^2\, \frac{\aes(E_e^{(s)}(\lambda),\lambda)-  \aes(z,\lambda)}{E_e^{(s)}(\lambda)-z}}.
\label{m72}
\end{eqnarray}
We estimate the last fraction in \eqref{m72} by  using \eqref{119.1} as   
\begin{equation}
\Big|\Big[ 1-\lambda^2\, \frac{\aes(E_e^{(s)}(\lambda),\lambda)-  \aes(z,\lambda)}{E_e^{(s)}(\lambda)-z}\Big]^{-1} \Big| \le  \frac{1}{1-C\lambda^2\varkappa_4} \prec 1,
\label{47.1}
\end{equation}
where the last relations holds since $\lambda^2\varkappa_4\prec 1$ (see \eqref{cond1}). Estimates \eqref{47.1} and \eqref{119.0} show
\begin{equation}
\Big\| \eqref{m72}\Big\| \prec \varkappa_3. 
\label{25.1}
\end{equation}
Consider now the last term on the right side of \eqref{m71}. Using \eqref{m69}, the fraction reads
\begin{eqnarray}
\lambda^2 \ \frac{\aes(E_e^{(s)}(\lambda),\lambda)- \aes(z,\lambda)}{(E_e^{(s)}(\lambda)-z)^2} \frac{1}{1-\lambda^2\, \frac{\aes(E_e^{(s)}(\lambda),\lambda) -\aes(z,\lambda)}{E_e^{(s)}(\lambda)-z}}.
\label{m75} 
\end{eqnarray}
The second fraction in \eqref{m75} is $\prec 1$ by  \eqref{47.1}. Next, we obtain from point 3. of Lemma  \ref{lem2.4}
\begin{eqnarray}
	\lefteqn{
\Big| \frac{\aes(E_e^{(s)}(\lambda),\lambda) - \aes(z,\lambda)}{E_e^{(s)}(\lambda)-z} -\partial_z \aes(\Ees,\lambda) \Big| }\nonumber\\
&=&\Big| \int_z^{\Ees} \big( \partial_\zeta \aes(\zeta,\lambda) - \partial_\zeta \aes(\Ees,\lambda)\big) \frac{d\zeta}{\Ees-z}\Big|\nonumber\\
&=&\Big| \int_z^{\Ees} \frac{d\zeta}{\Ees-z} \int_\zeta^{\Ees} dw \ \partial_w^2  \aes(w,\lambda)  \Big| \prec \varkappa_5 |\Ees-z|.
\label{49.1}
\end{eqnarray}
It follows that 
\begin{equation}
\Big| \frac{\aes(E_e^{(s)}(\lambda),\lambda) - \aes(z,\lambda)}{(E_e^{(s)}(\lambda)-z)^2} -\frac{\partial_z \aes(\Ees,\lambda)}{\Ees-z} \Big| \prec  \varkappa_5.
\label{50.1}
\end{equation}
Now
\begin{equation}
 \eqref{m75} = \lambda^2 \frac{\partial_z\aes(\Ees,\lambda)}{\Ees-z}\  \frac{1}{1-\lambda^2\partial_z
  \aes(\Ees,\lambda)} +T''',
 \label{50.2}
\end{equation}
where 
\begin{eqnarray}
T''' &=&\lambda^2 \Big( \frac{\aes(z,\lambda)-\aes(z,\lambda)}{(\Ees-z)^2}  - \frac{\partial_z \aes(\Ees,\lambda)}{\Ees-z}\Big)\frac{1}{1-\lambda^2 \frac{\aes(\Ees,\lambda)-
\aes(z,\lambda)}{\Ees-z}} \label{51.1}\\
&&+ \lambda^2\frac{\partial_z \aes(\Ees,\lambda)}{\Ees -z} \Big(\frac{1}{1-\lambda^2 \frac{\aes(\Ees,\lambda)-
		\aes(z,\lambda)}{\Ees-z}}- \frac{1}{1-\lambda^2 \partial_z\aes(\Ees,\lambda)}\Big).
\nonumber
\end{eqnarray}
By \eqref{119.1}, $|\partial_z\aes(\Ees,\lambda)|\prec \varkappa_4$  and so the second fraction in \eqref{50.2} is bounded $\prec 1$ since $\lambda^2 \varkappa_4\prec 1$ (see \eqref{cond1}). 

The estimates \eqref{50.1} and \eqref{47.1} show that the first summand on the right side of  \eqref{51.1} is $\prec\lambda^2 \varkappa_5$. Next,
\begin{eqnarray}
\lefteqn{
\frac{1}{\Ees-z}\Big( \Big[ 1-\lambda^2 \frac{\aes(\Ees,\lambda)-
		\aes(z,\lambda)}{\Ees-z}
	\Big]^{-1}- \Big[ 1-\lambda^2 \partial_z\aes(\Ees,\lambda)\Big]^{-1}\Big)
 }\nonumber\\
&=&\lambda^2  \frac{\partial_z\aes(\Ees,\lambda) - \frac{\aes(\Ees,\lambda)-\aes(z,\lambda)}{\Ees-z}}{\Ees-z}\nonumber\\
&&\times \big[1-\lambda^2  \frac{\aes(\Ees,\lambda)-\aes(z,\lambda)}{\Ees-z}\big]^{-1} \big[1-\lambda^2 \partial_z \aes(\Ees,\lambda)\big]^{-1}
\end{eqnarray}
which has modulus  $\prec \lambda^2 \varkappa_5$ (use \eqref{47.1}, \eqref{50.1}, \eqref{119.1}). In conclusion, we obtain 
\begin{equation}
| T''' |\prec \lambda^2 \varkappa_5.
\label{52}
\end{equation}
Combining \eqref{m71}, \eqref{25.1}, \eqref{50.2} and \eqref{52}, and also using that $\|\Qes(\Ees,\lambda)\|=1$ ({\em c.f.} after \eqref{m69}), gives the bound for $s\in{\mathcal S}_e^{\rm osc}$, 
\begin{align}
\Big\| & \frac{\Qes(z,\lambda)}{e-z+\lambda^2 \aes(z,\lambda)} -  \frac{\Qes(\Ees,\lambda)}{\Ees-z} \Big[ 1+\lambda^2  \frac{\partial_z\aes(\Ees,\lambda)}{1-\lambda^2\partial_z
	\aes(\Ees,\lambda)} \Big]\Big\| \nonumber\\
&\prec \varkappa_3+\lambda^2\varkappa_5.
\label{35}
\end{align}

$\bullet$ Combining \eqref{f-1}, \eqref{32.1} and \eqref{35} shows that for every $e\in{\mathcal E}_0$,
\begin{eqnarray}
{\mathfrak F}(z)^{-1} &=& \sum_{s\in{\mathcal S}_e^{\rm dec}}  \frac{\Qes(e,\lambda)}{e-z+\lambda^2 \aes(e,\lambda) }  \label{037}\\
&&  + \sum_{s\in{\mathcal S}_e^{\rm osc}} \frac{\Qes(\Ees,\lambda)}{\Ees-z}\Big[ 1+\lambda^2  \  \frac{\partial_z\aes(\Ees,\lambda)}{1-\lambda^2\partial_z
	\aes(\Ees,\lambda)}\Big] +{\mathcal T}(z,\lambda),
\nonumber
\end{eqnarray}
where the remainder satisfies
\begin{equation}
\|{\mathcal T}(z,\lambda) \|
 \prec \varkappa_3+\lambda^2\varkappa_5 + \Big( \frac{\varkappa_4\kappa}{a} +\varkappa_3\Big)  \Big( 1+\frac{\alpha+|\lambda|\varkappa_4}{a}\Big).
\label{57.1}
\end{equation}
We now analyze $D_1$, see \eqref{16.1}. From \eqref{037},  
\begin{eqnarray}
	\lefteqn{
D_1\equiv \int_{{\mathcal G}_e} e^{\i t z} \, {\mathfrak F}(z)^{-1} dz =\Qes(e,\lambda) \sum_{s\in{\mathcal S}_e^{\rm dec}}  \int_{{\mathcal G}_e}  \,  \frac{e^{\i t z}}{e-z+\lambda^2 \aes(e,\lambda)} \, dz \ }\label{24}\\
&+& \sum_{s\in{\mathcal S}_e^{\rm osc}} \Qes(\Ees,\lambda) \Big[1+\lambda^2  \frac{\partial_z\aes(\Ees,\lambda)}{1-\lambda^2\partial_z
	\aes(\Ees,\lambda)}\Big] \int_{{\mathcal G}_e} \frac{e^{\i tz}}{\Ees-z} dz	 + S_1, \nonumber
\end{eqnarray}
where the remainder $S_1$ is due to the integration of ${\mathcal T}(z,\lambda)$, estimated by (take into account that $|{\mathcal G}_e| =2\eta$)
\begin{eqnarray}
\|S_1\|  &\prec& \eta e^{wt}\Big( \varkappa_3+\lambda^2\varkappa_5 + \Big( \frac{\varkappa_4\kappa}{a} +\varkappa_3\Big)  \Big( 1+\frac{\alpha+|\lambda|\varkappa_4}{a}\Big) \Big)\nonumber\\
&\prec& \eta e^{wt}\Big( \varkappa_3+\lambda^2\varkappa_5 + \Big( \frac{\varkappa_4\kappa}{a} +\varkappa_3\Big)  \Big( 1+\frac{\eta}{a\lambda^2}\Big) \Big).\nonumber
\end{eqnarray}
We used \eqref{cond1} in the second estimate. Now
\begin{equation}
\int_{{\mathcal G}_e} \frac{e^{\i t z} }{\Ees-z}  d z  = -2\pi\i e^{\i t\Ees}  + \int_{{\mathcal A}_e}  \frac{e^{\i t z}}{\Ees-z}  d z,
\label{m88}
\end{equation}
where ${\mathcal A}_e$ is given in \eqref{40-1}.  Note that by \eqref{m69} and \eqref{119.1}, $|\Ees-e| = \lambda^2 |\aes(\Ees,\lambda)| \prec \lambda^2|\aes| +\lambda^2 \varkappa_4 (|\Ees-e| +|\lambda|)$, which, since $\lambda^2\varkappa_4 \prec 1$ (see \eqref{cond1}),  we can solve for (see \eqref{alpha}):
\begin{eqnarray}
\label{alphar}
|\Ees-e| \prec \lambda^2 (\alpha+|\lambda| \varkappa_4 ) \prec\eta.
\end{eqnarray}
In the last estimate, we used that 
\begin{equation}
	\label{38.1}
\lambda^2	\frac{\alpha+|\lambda| \varkappa_4}{\eta}\prec 1
\end{equation}
(see \eqref{cond1}). Thus \eqref{m88} gives
\begin{equation}
\label{37.1}
\Big|\int_{{\mathcal G}_e} \frac{e^{\i t z} }{\Ees-z}  d z\Big| \prec   1+ \frac{(w+\eta) e^{wt}}{\eta}. 
\end{equation} 
Using this bound in \eqref{24} shows that 
\begin{eqnarray}
\Big\| D_1 &-& \sum_{s\in{\mathcal S}_e^{\rm dec}} \int_{{\mathcal G}_e}  \,  \frac{e^{\i t z}\Qes(e,\lambda) }{e-z+\lambda^2 \aes(e,\lambda)} \, dz -\sum_{s\in{\mathcal S}_e^{\rm osc}} \int_{{\mathcal G}_e}  \,  \frac{e^{\i t z}\Qes(\Ees,\lambda) }{\Ees-z} \, dz \ \Big\| \nonumber\\
&\prec&  \|S_1\| + \lambda^2\Big(1+\frac{(w+\eta) e^{wt}}{\eta}\Big).
\label{38}
\end{eqnarray}
Similarly to \eqref{m88}, we also write the contour integrals in the first sum in \eqref{38} as
\begin{equation}
\int_{{\mathcal G}_e} \frac{ e^{\i t z}  }{e-z+\lambda^2 \aes(e,\lambda)}dz  =- 2\pi\i e^{\i t (e+\lambda^2 \aes(e,\lambda))} + \int_{{\mathcal A}_e}  \frac{e^{\i t z} }{e-z+\lambda^2 \aes(e,\lambda)}dz,
\label{39}
\end{equation}
where ${\mathcal A}_e$ is the contour \eqref{40-1}. For $z\in{\mathcal A}_e$ we have 
$|z-e|\ge\eta$ and so $\lambda^2|a_e^{(s)}(e,\lambda)/(e-z)|<1$ (because we have  $\lambda^2(\alpha+|\lambda|\varkappa_4)/\eta <1$). Thus the geometric series converges, 
\begin{eqnarray}
\frac{1}{e-z+\lambda^2 \aes(e,\lambda)} =\frac{1}{e-z} \sum_{n\ge 0}\big(-\frac{\lambda^2 a_e^{(s)}(e,\lambda)}{e-z}\big)^n,
\end{eqnarray}
and gives the bound
\begin{equation}
\Big|\frac{1}{e-z+\lambda^2 \aes(e,\lambda)} -\frac{1}{e-z} \Big| \prec \frac{\lambda^2}{\eta}(\alpha+|\lambda|\varkappa_4). 
\end{equation}
With $|{\mathcal A}_e| = 2w +\pi\eta$, we obtain
\begin{equation}
	\label{41}
 \Big| \int_{{\mathcal A}_e}  \frac{e^{\i t z}  }{e-z+\lambda^2 
 	\aes(e,\lambda)}dz -  \int_{{\mathcal A}_e} \frac{e^{\i t z}  }{e-z} dz \Big| \prec e^{wt}\frac{w +\eta}{\eta} \lambda^2(\alpha+|\lambda|\varkappa_4).
\end{equation}
In the same way we obtain
\begin{equation}
	\label{41.2}
	\Big| \int_{{\mathcal A}_e}  \frac{e^{\i t z}  }{\Ees -z}dz -  \int_{{\mathcal A}_e} \frac{e^{\i t z}  }{e-z} dz \Big| \prec e^{wt}\frac{w +\eta}{\eta} \lambda^2(\alpha+|\lambda|\varkappa_4).
\end{equation}
Then using \eqref{39}  and \eqref{41} in  \eqref{38} (and similarly for the integrals over the singularity $(\Ees-z)^{-1}$)  gives
\begin{eqnarray}
	\Big\| D_1 &+&\sum_{s\in{\mathcal S}_e^{\rm dec}} \Big[ 2\pi\i  \, e^{\i t(e+\lambda^2\aes(e,\lambda))} \Qes(e,\lambda) -\int_{{\mathcal A}_e} \frac{e^{\i t z}\Qes(e,\lambda)}{e-z}dz \Big]\nonumber\\
&&+\sum_{s\in{\mathcal S}_e^{\rm osc}} \Big[ 2\pi\i \, e^{\i t \Ees}\Qes(\Ees,\lambda) -\int_{{\mathcal A}_e} \frac{e^{\i t z}\Qes(\Ees,\lambda)}{e-z}dz\Big] \Big\| \nonumber\\
	&\prec& \|S_1\| + \lambda^2\Big(1 + \frac{(w+\eta) e^{wt}}{\eta}\big(1+\kappa(\alpha+|\lambda|\varkappa_4)\big)\Big) .
	\label{38.3}
\end{eqnarray}
Next we replace the projections in the remaining integrals in \eqref{38.3} by their values at $\lambda=0$: By \eqref{119.0} we have  $\|\Qes(e,\lambda)-\Qes(e,0)\|\prec |\lambda|\varkappa_3$ and $\| \Qes(\Ees,\lambda)-\Qes(e,0)\|\prec \varkappa_3 (|\Ees-e|+|\lambda|)\prec  |\lambda|\varkappa_3  [1+|\lambda|(\alpha+|\lambda| \varkappa_4)] \prec \varkappa_3(|\lambda|+\eta)$, see also \eqref{alphar}, \eqref{cond1}. Upon replacing the projections in the integrals of \eqref{38.3} by their values at $\lambda=0$, we thus make an error $\prec   \frac{(w+\eta)e^{wt}}{\eta} \varkappa_3 (|\lambda|+\eta)$. Once this replacement is made, we  use the fact that $\sum_{s=1}^{m_e} \Qes(e,0) = P_e$ is the spectral projection of $L_0$ associated to the eigenvalue $e$ and so we get from \eqref{38.3}
\begin{eqnarray}
D_1 &=&-2\pi\i \sum_{s\in{\mathcal S}_e^{\rm dec}}  \, e^{\i t(e+\lambda^2\aes(e,\lambda))} \Qes(e,\lambda) - 2\pi\i\sum_{s\in{\mathcal S}_e^{\rm osc}}  \, e^{\i t \Ees}\Qes(\Ees,\lambda)  \nonumber\\
&&+\int_{{\mathcal A}_e} \frac{e^{\i t z}P_e}{e-z}dz +S_2,	
\label{38.4}
\end{eqnarray}
where
\begin{equation}
\| S_2\| \prec  \|S_1\|+ \lambda^2+ \frac{e^{wt}(w+\eta)}{\eta}\Big( \lambda^2\big(1+\kappa(\alpha+|\lambda|\varkappa_4)\big)+\varkappa_3( |\lambda|+\eta)  \Big).
\label{m99}
\end{equation}
We now further analyze the integral in \eqref{38.4}. 
Denoting $P^\perp_{\s,e} =\bbbone_\s-P_{\s,e}$, where $P_{\s,e}$ is the spectral projection of $L_\s$ onto the eigenspace associated to $e$ (see \eqref{m7.1}), we have (as $\eta<g/2$)
\begin{equation}
\label{66}
\Big \|\int_{{\mathcal A}_e} \frac{e^{\i tz}}{L_\s-z}dz \, P^\perp_{\s,e}\Big\| \prec  \frac{e^{wt}(w+\eta)}{g},
\end{equation}
where $g$ is the spectral gap of $L_\s$, \eqref{g}.  So we can replace $P_e=P_{\s,e}\otimes P_\R$ in \eqref{38.4} by simply $P_\r$, incurring an error of size of the right hand side of \eqref{66},
\begin{eqnarray}
D_1 &=&-2\pi\i \sum_{s\in{\mathcal S}_e^{\rm dec}}  \, e^{\i t(e+\lambda^2\aes(e,\lambda))} \Qes(e,\lambda) - 2\pi\i\sum_{s\in{\mathcal S}_e^{\rm osc}}  \, e^{\i t \Ees}\Qes(\Ees,\lambda) \nonumber\\
&&  +\int_{{\mathcal A}_e}  \,  \frac{e^{\i t z} P_\r}{L_\s- z}  \, dz  +S_3,	
\label{43}
\end{eqnarray}
with 
\begin{equation}
	\| S_3\| \prec \|S_1\| + \lambda^2+ \frac{e^{wt}(w+\eta)}{\eta}\Big(\frac{\eta}{g}+ \lambda^2\big(1+\kappa(\alpha+|\lambda|\varkappa_4)\big)+\varkappa_3( |\lambda|+\eta)  \Big).
	\label{m99.1}
\end{equation}

\medskip

\noindent
{\bf Estimate of $D_2$ in \eqref{16.1}. }  Our main estimate is given in \eqref{82} below. 
We have from \eqref{17} with $Q=P_e$, 
\begin{eqnarray}
	\label{44}
\lefteqn{ \langle \phi, D_2\psi\rangle =\int_{{\mathcal G}_e} e^{\i t z} \langle\phi, {\mathcal B}(z)\psi\rangle dz}\\
&=&\int_{{\mathcal G}_e} e^{\i tz}\langle \phi,   \, \{-\lambda\, {\mathfrak F}(z)^{-1}  P_e I R_z^{P_e} - \lambda R_z^{P_e} I P_e {\mathfrak F}(z)^{-1} +\lambda^2 R_z^{P_e} I P_e{\mathfrak F}(z)^{-1} P_e I R_z^{P_e} \}\,    \psi \rangle \, dz.
\nonumber
\end{eqnarray}
We begin by analyzing the first term on the right side of \eqref{44}. Using \eqref{89},
\begin{equation}
\int_{{\mathcal G}_e} e^{\i tz}\langle \phi,   \, {\mathfrak F}(z)^{-1} \lambda P_e I R_z^{P_e} \psi\rangle dz = \lambda\sum_{s=1}^{m_e} \int_{{\mathcal G}_e}e^{\i tz} \frac{ \langle\phi, \Qes(z,\lambda)P_e IR^{P_e}_z \psi\rangle  }{e-z+\lambda^2 \aes(z,\lambda)} dz.
\label{71.1}
\end{equation}
$\bullet$ For $s\in{\mathcal S}_e^{\rm dec}$ we have from \eqref{33.1}
\begin{eqnarray}
\Big| \lambda\sum_{s=1}^{m_e} \int_{{\mathcal G}_e}e^{\i tz}\frac{ \langle\phi, \Qes(z,\lambda)P_e IR^{P_e}_z \psi\rangle  }{e-z+\lambda^2 \aes(z,\lambda)} dz\Big| &\prec& \frac{|{\mathcal G}_e|}{a|\lambda|}e^{w t} \max_{z\in{\mathcal G}_e} | \langle\phi, \Qes(z,\lambda)P_e IR^{P_e}_z \psi\rangle |\nonumber\\  
&\prec& \frac{\kappa\eta}{a|\lambda|}e^{w t}\|\phi\|  \max_j C_2(I\varphi_j\otimes\Omega, \psi).\quad
\label{m101}
\end{eqnarray}
To get the last bound, we took into account that  $P_e=P_e P_\Omega$, that $IP_e\subset {\mathcal D}$,  so that from  \eqref{bnd} and \eqref{32}, we obtain
\begin{eqnarray}
| \langle\phi, \Qes(z,\lambda)P_e IR^{P_e}_z \psi\rangle | &\le& \sum_j |\langle \phi, \Qes(z,\lambda)\varphi_j\otimes\Omega\rangle |\ |\langle\varphi_j\otimes\Omega, IR_z^{P_e}\psi\rangle| \nonumber\\
&\prec&\kappa \|\phi\|  \max_j C_2(I\varphi_j\otimes\Omega, \psi).
\label{m102.1}
\end{eqnarray}
$\bullet$ For $s\in{\mathcal S}_e^{\rm osc}$ we use \eqref{m69} to get
\begin{equation}
\frac{\Qes(z,\lambda)P_eIR_z^{P_e}}{e-z+\lambda^2\aes(z,\lambda)} =\frac{\Qes(z,\lambda)P_eIR_z^{P_e}}{\Ees-z} \frac{1}{1-\lambda^2\frac{\aes(\Ees,\lambda) - \aes(z,\lambda)}{\Ees -z}}.
\label{m102}
\end{equation}
Now 
\begin{eqnarray}
\lefteqn{
\frac{\Qes(z,\lambda)P_eIR_z^{P_e}}{\Ees-z} = \frac{\Qes(z,\lambda)-\Qes(\Ees,\lambda)}{\Ees-z} P_eIR_z^{P_e} } \nonumber\\
&&- \Qes(\Ees,\lambda) P_eI \frac{R^{P_e}_{\Ees} - R_z^{P_e}}{\Ees-z} +\frac{\Qes(\Ees,\lambda)P_eIR_{\Ees}^{P_e}}{\Ees-z}.
\label{m103}
\end{eqnarray}
The first two terms of the right side of \eqref{m103}, when used in \eqref{m102}, are estimated by taking into account that the second factor on the right side of \eqref{m102} is $\prec 1$ due to \eqref{47.1}. However, the last term on the right side of \eqref{m103} has a singularity in $z$ which must be removed by integrating over ${\mathcal G}_e$. To do this, we write
\begin{eqnarray*}
	\lefteqn{
\frac{1}{1-\lambda^2\frac{\aes(\Ees,\lambda) - \aes(z,\lambda)}{\Ees -z}} =\frac{1}{1-\lambda^2\partial_z\aes(\Ees,\lambda)} } \nonumber\\
&&-\lambda^2\frac{\partial_z\aes(\Ees,\lambda) - \frac{\aes(\Ees,\lambda) - \aes(z,\lambda)}{\Ees -z}}{ [1-\lambda^2\partial_z\aes(\Ees,\lambda)  ] [ 1-\lambda^2\frac{\aes(\Ees,\lambda) - \aes(z,\lambda)}{\Ees -z}] }.
\end{eqnarray*}
The denominator of the second summand on the right side is $\succ 1$ and its numerator is $\prec\varkappa_5 |\Ees-z|$ by \eqref{49.1}. Thus 
\begin{eqnarray}
	\lefteqn{
\Big| \int_{{\mathcal G}_e} \frac{e^{\i t z}}{\Ees-z}\frac{1}{1-\lambda^2\frac{\aes(\Ees,\lambda) - \aes(z,\lambda)}{\Ees -z}} dz}\label{m106.1}\\
&&  - \frac{1}{1-\lambda^2\partial_z\aes(\Ees,\lambda)}\int_{{\mathcal G}_e} \frac{e^{\i t z}}{\Ees-z}dz\Big| \prec e^{wt}\lambda^2|{\mathcal G}_e| \varkappa_5 \prec e^{wt}\lambda^2\eta\varkappa_5.\qquad 
\nonumber
\end{eqnarray}
Now $|1-\lambda^2\partial_z\aes(\Ees,\lambda)|^{-1}\prec 1$ (see \eqref{cond1}) and so by \eqref{m106.1} and \eqref{37.1}, we have 
\begin{equation}
\Big| \int_{{\mathcal G}_e} \frac{e^{\i t z}}{\Ees-z}\frac{1}{1-\lambda^2\frac{\aes(\Ees,\lambda) - \aes(z,\lambda)}{\Ees -z}} dz\Big| \prec  1+ \frac{(w+\eta) e^{wt}}{\eta}+ e^{wt}\lambda^2\eta\varkappa_5.
\label{m107.1}
\end{equation}
We combine \eqref{m107.1} with \eqref{m102}, \eqref{m103} to get
\begin{eqnarray}
\lefteqn{
\Big|\lambda  \int_{{\mathcal G}_e} e^{\i tz}\frac{ \langle\phi, \Qes(z,\lambda)P_e IR^{P_e}_z \psi\rangle  }{e-z+\lambda^2 \aes(z,\lambda)} dz\Big|}\nonumber\\
&\prec& |\lambda| |  \langle \phi, \Qes(\Ees,\lambda)P_eIR_{\Ees}^{P_e}\psi\rangle| \Big( 1+ \frac{(w+\eta) e^{wt}}{\eta}+ e^{wt}\lambda^2\eta\varkappa_5\Big) \nonumber\\
&& + |\lambda| |{\mathcal G}_e| \max_{z\in{\mathcal G}_e} \Big| \langle \phi, \frac{\Qes(z,\lambda)-\Qes(\Ees,\lambda)}{\Ees-z} P_eIR_z^{P_e} \psi\rangle \Big|  \nonumber\\
&& + |\lambda| |{\mathcal G}_e| \max_{z\in{\mathcal G}_e} \Big|\langle \phi, \Qes(\Ees,\lambda) P_eI \frac{R^{P_e}_{\Ees} - R_z^{P_e}}{\Ees-z} \psi\rangle \Big|.
\label{m105}
\end{eqnarray}
Using \eqref{37.1} and the argument of \eqref{m102.1} (with  $\|\Qes(\Ees,\lambda)\| =1$ as this one is an orthogonal projection) we get $|  \langle \phi, \Qes(\Ees,\lambda)P_eIR_{\Ees}^{P_e}\psi\rangle | 
\prec  \|\phi\| \max_j C_2(I\varphi_j\otimes\Omega,\psi)$.
Next, by \eqref{119.0}, \eqref{32} and again by the argument of \eqref{m102.1}, the last two terms on the right hand side of \eqref{m105} has an upper bound
$\prec |\lambda| \eta  \|\phi\|  (1+ \varkappa_3) \max_j C_2(I\varphi_j\otimes\Omega, \psi)$. We thus obtain from \eqref{m105} that for $s\in{\mathcal S}_e^{\rm osc}$,
\begin{eqnarray}
	\lefteqn{
		\Big|\lambda  \int_{{\mathcal G}_e} e^{\i tz}\frac{ \langle\phi, \Qes(z,\lambda)P_e IR^{P_e}_z \psi\rangle  }{e-z+\lambda^2 \aes(z,\lambda)} dz\Big|
	}\label{m107}\\
&\prec& |\lambda| \|\phi\| \max_j C_2(I\varphi_j\otimes\Omega,\psi)\Big(1+ \frac{(w+\eta) e^{wt}}{\eta} +e^{wt}\lambda^2\eta\varkappa_5 +\eta(1+\varkappa_3) \Big).
	\nonumber
\end{eqnarray}

$\bullet$  The relations \eqref{71.1}, \eqref{m101} and \eqref{m107}  give 
\begin{eqnarray}
\lefteqn{\Big|\int_{{\mathcal G}_e} e^{\i tz}\langle \phi,   \, {\mathfrak F}(z)^{-1} \lambda P_e I R_z^{P_e} \psi\rangle dz \Big|\prec \|\phi\|  \max_j C_2(I\varphi_j\otimes\Omega,\psi)}\nonumber\\
&& \times \Big( \frac{\kappa\eta}{a|\lambda|} e^{wt} + |\lambda|\big( 1+ \frac{(w+\eta) e^{wt}}{\eta} +e^{wt}\lambda^2\eta\varkappa_5 +\eta(1+\varkappa_3)\big)\Big).\qquad
\label{76.1.1}
\end{eqnarray}
The integral $\int_{{\mathcal G}_e} e^{\i tz}\langle \phi,  \lambda R_z^{P_e} I P_e {\mathfrak F}(z)^{-1} \psi\rangle dz $  in \eqref{44} has the same upper bound \eqref{76.1.1}, but with $\phi$ and $\psi$ interchanged.
\medskip

$\bullet$ We now estimate the term in \eqref{44} involving $ \lambda^2 R_z^{P_e} I{\mathfrak F}(z)^{-1} I R_z^{P_e}$. We use again \eqref{f-1}. Proceeding as above, we obtain from \eqref{33.1} that for $s\in{\mathcal S}_e^{\rm dec}$,
\begin{equation}
\lambda^2 \Big| \int_{{\mathcal G}_e} e^{\i tz}\frac{\langle \phi,   R^{P_e}_z I P_e \Qes(z,\lambda) I R_z^{P_e} \,    \psi \rangle}{e-z+\lambda^2\aes(z,\lambda)} \, dz\Big|
 \prec \frac{\eta e^{wt}}{a}   \max_{z\in{\mathcal G}_e} |\langle \phi, R_z^{P_e} IP_e  \Qes(z,\lambda) IR_z^{P_e} \psi\rangle|.	
\label{78}
\end{equation}
Now 
\begin{equation}
|\langle \phi,R_z^{P_e} IP_e  \Qes(z,\lambda) IR_z^{P_e} \psi\rangle| \prec \|\Qes(z,\lambda)\| \|P_eI (R_z^{P_e})^*\phi\| \|P_eI R_z^{P_e}\psi\|
\label{m111}
\end{equation} 
and 
\begin{eqnarray}
\|P_eI (R_z^{P_e})^*\phi\| 
&\prec&\max_j |\langle\varphi_j\otimes\Omega, I (R_z^{P_e})^* \phi\rangle| = \max_j |\langle\phi,  R_z^{P_e} I \varphi_j\otimes\Omega \rangle| \nonumber\\
&\prec&  \max_j C_2(\phi, I\varphi_j\otimes\Omega)
\nonumber
\end{eqnarray}
and similarly, $\|P_eIR_z^{P_e}\psi\|\prec \max_j C_2( I\varphi_j\otimes\Omega,\psi)$. Furthermore by \eqref{bnd}, $\|\Qes(z,\lambda)\|\prec \kappa$. Combining these estimates with \eqref{78} yields that for all $s\in{\mathcal S}_e^{\rm dec}$,
\begin{align}
\lambda^2 \Big| \int_{{\mathcal G}_e} e^{\i tz} & \frac{\langle \phi,   R^{P_e}_z I P_e \Qes(z,\lambda) I R_z^{P_e} \,    \psi \rangle}{e-z+\lambda^2\aes(z,\lambda)} \, dz\Big| \nonumber\\
& \prec \frac{\eta e^{wt}}{a} \kappa  \max_j C_2(I\varphi_j\otimes\Omega, \phi) \max_j C_2(I\varphi_j\otimes\Omega, \psi).
\end{align}

Next we treat the case $s\in{\mathcal S}_e^{\rm osc}$, for which we show the bound (recall that  $e=E_e^{(s)}(\lambda)+\lambda^2a_e^{(s)}(E_e^{(s)}(\lambda),\lambda)$, \eqref{m69})
\begin{align}
\lambda^2 \Big| \int_{{\mathcal G}_e} e^{\i tz} & \frac{\langle \phi,   R^{P_e}_z I P_e\Qes(z,\lambda) I R_z^{P_e} \,    \psi \rangle}{E_e^{(s)}(\lambda)-z-\lambda^2[\aes(\Ees,\lambda) -\aes(z,\lambda)]} \, dz\Big|\nonumber\\
& \prec  \lambda^2 \Big( 1+ (w+\eta) e^{wt}(1+\tfrac1\eta +\varkappa_3) + e^{wt}\lambda^2\eta\varkappa_5\Big) \nonumber\\
&\quad \times \max_j C_2(I\varphi_j\otimes\Omega,\phi) \max_j C_2(I\varphi_j\otimes\Omega,\psi)
\qquad
\label{79.1}
\end{align}
as follows. We proceed  as in \eqref{m102} and \eqref{m103} and get terms with $[\Qes(\Ees,\lambda)-\Qes(z,\lambda)][\Ees-z]^{-1}$ and $[R_{\Ees}^{P_e}-R_z^{P_e}][\Ees-z]^{-1}$, all of which are controlled by bounds on the derivatives. We find that they are $\prec \lambda^2(\eta+w)e^{wt}(1+\varkappa_3)\max_j C_2(I\varphi_j\otimes\Omega,\phi) \max_j C_2(I\varphi_j\otimes\Omega,\psi)$. The only remaining term to consider is 
\begin{align}
\lambda^2 |\langle \phi,  &  R^{P_e}_{\Ees} I P_e \Qes(\Ees, \lambda) I R_{\Ees}^{P_e} \,    \psi \rangle| \nonumber\\
\times  &\Big| \int_{{\mathcal G}_e} \frac{e^{\i t z}}{\Ees-z}\frac{1}{1-\lambda^2\frac{\aes(\Ees,\lambda) - \aes(z,\lambda)}{\Ees -z}} dz\Big| \nonumber\\
&\prec \lambda^2 \Big( 1+ \frac{(w+\eta) e^{wt}}{\eta}+ e^{wt}\lambda^2\eta\varkappa_5\Big)\max_j C_2(I\varphi_j\otimes\Omega,\phi) \max_j C_2(I\varphi_j\otimes\Omega,\psi),
\end{align}
where we have used \eqref{m107.1}. This yields \eqref{79.1}. 
\medskip

We collect the estimates \eqref{76.1.1}, \eqref{79.1} and use them in \eqref{44} to get 
\begin{align}
|\langle \phi, D_2 & \psi\rangle| \prec  \Big\{ 
 \frac{\kappa\eta}{a|\lambda|} e^{wt} + |\lambda|\big( 1 +\eta(1+\varkappa_3)\big)+ \frac{(w+\eta) e^{wt}}{\eta} ( |\lambda| +\lambda^2)\nonumber\\
& +e^{wt}\lambda^2\eta\varkappa_5(|\lambda|+\lambda^2)+ \lambda^2 \Big( 1+ (w+\eta) e^{wt}(1+\varkappa_3)  \Big)\Big\} \nonumber\\
& \times\Big[\max_j C_2(I\varphi_j\otimes\Omega,\phi) \max_j C_2(I\varphi_j\otimes\Omega,\psi)+  \sym \|\phi\| \max_jC_2(I\varphi_j\otimes\Omega_\r, \psi) \Big]. 
\label{82}
\end{align}

\medskip

\noindent
{\bf Estimate of $D_3$ in \eqref{16.1}. }  The bound \eqref{32}  gives immediately 
\begin{equation}
	\label{53}
\big|\langle \phi, D_3\psi\rangle\big| = \Big| \int_{{\mathcal G}_e}e^{\i tz} \langle\phi,R_z^{P_e}\psi\rangle dz\Big|  \prec \eta e^{wt} C_2(\phi,\psi).
\end{equation}
The three bounds \eqref{43}, \eqref{82} and \eqref{53}, together with \eqref{16.1}, show \eqref{024} and \eqref{m57} and this completes the proof of Proposition \ref{propm1}.\hfill \qed

\subsection{Analysis of  $J_\infty(t)$}
\label{sect4.2}

\begin{figure}[b]
	\centering
	\includegraphics[width=12cm]{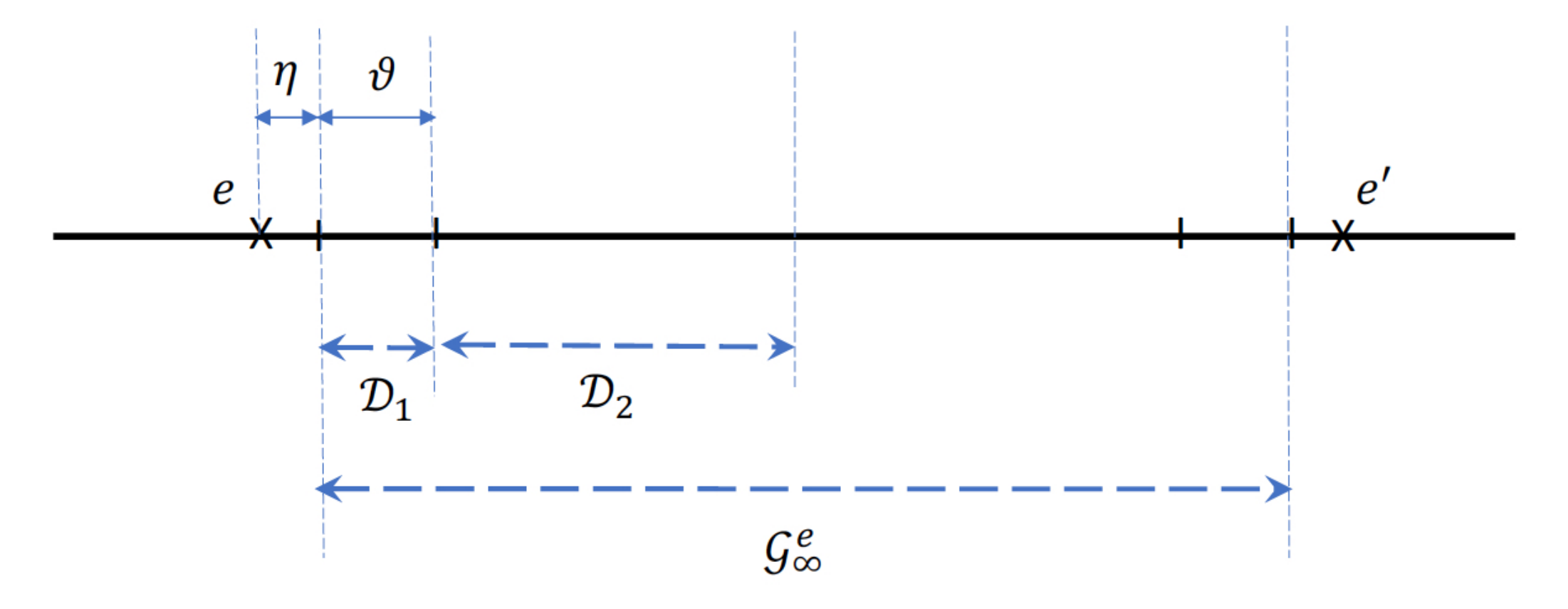}
	\caption{ Different regions for the real part of $z$, for  $z\in {\mathcal G}_\infty^e$.}
	\label{Fig4}
\end{figure}

We introduce a new parameter $\vartheta>0$ which will be chosen suitably small (a power of $|\lambda|$) below, and which is depicted in Fig.~\ref{Fig4}. The goal of this section is to show the following result.

\begin{prop}
\label{prop3}
If $\varkappa_1\lambda^2/\eta\prec 1$, then we have 
\begin{align}
\Big|J_\infty(t) & +\frac{1}{2\pi\i} \int_{{\mathcal G}_\infty} e^{\i t z}\langle \phi, (L_\s-z)^{-1}\psi\rangle dz - \langle \phi, P^\perp_\r e^{\i t P^\perp_\r L_\lambda P^\perp_\r} P^\perp_\r \psi\rangle \Big| \nonumber\\
\prec  &\ 
\lambda^2 e^{wt}\Big[ \frac{\varkappa_1}{\eta}+ \frac{g\varkappa_1}{(\eta+\vartheta)^2} + \frac{g\varkappa^2_1\lambda^2}{\eta^3} \nonumber\\
& \qquad \quad+\vartheta \Big(\frac{\varkappa_1(\eta+\vartheta+|\lambda|)+\alpha\kappa}{\eta^2} +\frac{\varkappa_1}{g}\big(1/\eta +1/g\big)\Big)
\Big] \|\phi\|\, \|\psi\|	\nonumber\\
&+\eta e^{wt}C_1(\phi,\psi)\nonumber\\
& +  e^{wt} \Big( \frac{|\lambda|\vartheta}{\eta} + \frac{|\lambda|}{\eta+\vartheta} +\frac{|\lambda|^3\varkappa_1}{\eta^2} \Big[ 1+\frac{\lambda^2\varkappa_1}{\eta}\Big] \Big)\sym \|\phi\|  \max_j C_1(I\varphi_j\otimes\Omega_\r,\psi)  \nonumber \\
&  +e^{wt}\frac{\lambda^2}{\eta} \Big[1+\lambda^2\varkappa_1(1+1/\eta) \Big]\max_j  C_1(I\varphi_j\otimes\Omega_\r, \phi)  \max_j C_1(I\varphi_j\otimes \Omega_\r,\psi) \nonumber\\
&  + e^{wt}\frac{|\lambda| }{\eta+\vartheta} \sym \|\phi\| \Big(  \|P_\r I\|  \|P^\perp_\r\psi\|+ \max_j C_1\big(I\varphi_j\otimes\Omega_\r,(\bar L_\lambda+\i)\psi\big) \Big)\nonumber\\
&  + e^{wt}\frac{\lambda^2}{\eta}  \sym \max_j C_1(I\varphi_j\otimes\Omega_\r, \phi)\Big( \|P_\r I\| \|P^\perp_\r\psi\| + \max_j C_1\big(I\varphi_j\otimes\Omega_\r,(\bar L_\lambda+\i)\psi\big)\Big).
\label{69}
\end{align}
In the last line, we use the notation $\bar L_\lambda\equiv P^\perp_\r L_\lambda P^\perp_\r\upharpoonright_{{\rm Ran}P^\perp_\r}$.
\end{prop}

\medskip
\noindent
{\em Proof of Proposition \ref{prop3}.\ } Consider two adjacent eigenvalues $e<e'$ of $L_\s$ and set
\begin{equation} 
{\mathcal G}^e_\infty=\{x-\i w : e+\eta \le x\le e'-\eta\}.
\label{116} 
\end{equation}
We introduce a new small parameter $\vartheta>0$ and set ${\mathcal D}_1=\{x-\i w : e+\eta \le x\le e+\eta+\vartheta\}$ and ${\mathcal D}_2 = \{x-\i w : e+\eta+\vartheta\le x\le e+\tfrac{e'-e}{2} \}$, as depicted in Fig.~\ref{Fig4}.

We use the Feshbach decomposition \eqref{16} with $Q=P_\r = |\Omega_\r\rangle\langle \Omega_\r|$ and accordingly, we get three contributions to the resolvent as in \eqref{16.1},
\begin{eqnarray}
	\label{16.2}
	\int_{{\mathcal G}_\infty^e} e^{\i t z} R_z dz &=& \int_{{\mathcal G}_\infty^e} e^{\i t z}{\mathfrak F}(z)^{-1} dz +\int_{{\mathcal G}_\infty^e} e^{\i t z}{\mathfrak B}(z) dz + \int_{{\mathcal G}_\infty^e} e^{\i t z}R^{P_\r}_z dz\nonumber\\
& \equiv& D'_1+D'_2+D'_3. 
\end{eqnarray}
\bigskip

\noindent
{\bf Estimating $D'_1$ of \eqref{16.2}. } The main bound we prove is given in \eqref{D_1'}. We have, in the sense of operators acting on ${\rm Ran}P_\r$,
\begin{equation} 
{\mathfrak F}(z) = P_\r\big( L_\s -z-\lambda^2 I R_z^{P_\r}I \big) P_\r =\big(\bbbone-\lambda^2 P_\r I R_z^{P_\r}I P_\r(L_\s-z)^{-1} \big) (L_\s-z).
\label{54}
\end{equation}
For $z\in{\mathcal G}_\infty^e$, $\|(L_\s-z)^{-1}\|\le 1/\eta$ and by \eqref{90} we get $\|\lambda^2  P_\r IR_z^{P_\r} IP_\r(L_\s-z)^{-1}\|\prec  \frac{\varkappa_1\lambda^2}{\eta}$ 
and so we get from \eqref{54}, since  $\frac{\varkappa_1\lambda^2}{\eta}\prec 1$, 
\begin{equation} 
	\label{55}
	{\mathfrak F}(z)^{-1} = (L_\s-z)^{-1} \big[ \bbbone +\lambda^2 P_\r I R_z^{P_\r}I P_\r(L_\s-z)^{-1} + S_4\big],\qquad 
\|S_4\| \prec \frac{\varkappa^2_1\lambda^4}{\eta^2}.  
\end{equation}
We consider now $z\in{\mathcal D}_1\cup{\mathcal D}_2$  (the region  closer to $e$ than to $e'$, see Fig.\,4). By inserting $\bbbone_\s = P_{\s,e} +P_{\s,e}^\perp$, using that $P_e=P_{\s,e}\otimes P_\r$ and that  $\|P_{\s,e}^\perp (L_\s-z)^{-1}\|\le 2/g$ (see \eqref{g} for the gap $g$), we estimate the term in \eqref{55} involving the resolvent $R_z^{P_\r}$ as follows,
\begin{equation}
	\label{57}
\Big\| \lambda^2 (L_\s-z)^{-1}  P_\r I R_z^{P_\r}I P_\r(L_\s-z)^{-1}- \frac{\lambda^2}{(e-z)^2} P_e I R_z^{P_\r}I P_e\Big\|\prec  \frac{\varkappa_1\lambda^2}{g}\big( 1/\eta +1/g\big).
\end{equation}
By \eqref{90}  we have $\| P_eI R_z^{P_\r}IP_e - P_eI (R_e^{P_\r}|_{\lambda=0} )IP_e \|\prec \varkappa_1(|z-e|+|\lambda|)$. Thus, taking into account the definition \eqref{34} of the level shift operator $\Lambda_e$ and using that $P^\perp_\r IP_e = P^\perp_e IP_e$,  we have
\begin{equation}
\label{59}
\| P_eI R_z^{P_\r}IP_e - \Lambda_e \| \prec \varkappa_1\big(|z-e|+|\lambda|\big).
\end{equation} 
The reason for introducing the new small parameter $\vartheta$ (see Fig.\,4), is that for $z\in{\mathcal D}_1$, the right hand side of \eqref{59} is bounded above by $\varkappa_1(\eta +\vartheta +|\lambda|)$. It then follows from \eqref{59} and \eqref{57} that $\forall z\in{\mathcal D}_1$, 
\begin{equation*}
\| \lambda^2 (L_\s-z)^{-1}  P_\r I R_z^{P_\r}I P_\r(L_\s-z)^{-1} \| \prec   \frac{\lambda^2 \|\Lambda_e\|}{\eta^2}  + \varkappa_1 \lambda^2 \Big[\frac{\eta+\vartheta+|\lambda|}{\eta^2} +\frac{1}{g}\big(1/\eta +1/g\big)\Big]
\end{equation*}
and hence, as $\|\Lambda_e\|\prec \alpha \kappa$, we get
\begin{eqnarray} 
		\label{62.1.1}
\lefteqn{	\Big\| \int_{{\mathcal D}_1} e^{\i tz} \lambda^2 (L_\s-z)^{-1}  P_\r I R_z^{P_\r}I P_\r(L_\s-z)^{-1} dz \Big\|}\\
& \prec& \lambda^2\vartheta e^{wt} \Big[\frac{\varkappa_1(\eta+\vartheta+|\lambda|)+\alpha\kappa}{\eta^2} +\frac{\varkappa_1}{g}\big(1/\eta +1/g\big)\Big]. 
\nonumber
\end{eqnarray}
Next we find a bound for the integral on the left side of \eqref{62.1.1} with ${\mathcal D}_1$ replaced by ${\mathcal D}_2$. For $z\in{\mathcal D}_2$, we have $\|(L_\s-z)^{-1}\|\le (\eta+\vartheta)^{-1}$ and so it follows from  \eqref{90} that $\|(L_\s-z)^{-1}  P_\r I R_z^{P_\r}I P_\r(L_\s-z)^{-1}\|\prec \varkappa_1(\eta+\vartheta)^{-2}$. Consequently, since $|{\mathcal D}_2|<g/2$, we get
\begin{equation}
\Big\|	\int_{{\mathcal D}_2} e^{\i tz} \lambda^2 (L_\s-z)^{-1}  P_\r I R_z^{P_\r}I P_\r(L_\s-z)^{-1}  dz\Big\|  \prec \frac{\lambda^2 \varkappa_1 g e^{wt}}{(\eta+\vartheta)^2}.
	\label{62.2}
\end{equation}
We now combine \eqref{62.1.1} and \eqref{62.2} with \eqref{55},
\begin{eqnarray}
\lefteqn{\Big\| \int_{{\mathcal D}_1\cup{\mathcal D}_1}  e^{\i tz} \, {\mathfrak F}(z)^{-1} dz  - \int_{{\mathcal D}_1\cup{\mathcal D}_2}  \frac{e^{\i tz}}{L_\s-z} dz \Big\|}	\label{62}\\
&\prec& \lambda^2 e^{wt}\Big[ \frac{g\varkappa_1}{(\eta+\vartheta)^2} + \frac{g\varkappa^2_1\lambda^2}{\eta^3} +\vartheta \Big(\frac{\varkappa_1(\eta+\vartheta+|\lambda|)+\alpha\kappa}{\eta^2} +\frac{\varkappa_1}{g}\big(1/\eta +1/g\big)\Big)
\Big].
\nonumber
\end{eqnarray}
For $z\in{\mathcal G}_\infty^e$ lying closer to $e'$ than $e$ ({\em i.e.}, $z\in {\mathcal G}_\infty^e\backslash( {\mathcal D}_1\cup{\mathcal D}_2)$) the analysis is the same, as only the distance from $x$ to the nearest eigenvalue of $L_\s$ plays a role in the estimates. We conclude that the bound \eqref{62} holds with ${\mathcal D}_1\cup{\mathcal D}_2$ replaced by ${\mathcal G}_\infty^e$ in the integrals on the left side:
\begin{eqnarray}
	\lefteqn{
\Big\| D'_1- \int_{{\mathcal G}_\infty^e} \frac{e^{\i tz}}{L_\s-z} dz \Big\| 
}
\label{D_1'}\\
&\prec & 
\lambda^2 e^{wt}\Big[ \frac{g\varkappa_1}{(\eta+\vartheta)^2} + \frac{g\varkappa^2_1\lambda^2}{\eta^3} +\vartheta \Big(\frac{\varkappa_1(\eta+\vartheta+|\lambda|)+\alpha\kappa}{\eta^2} +\frac{\varkappa_1}{g}\big(1/\eta +1/g\big)\Big)
\Big].\nonumber
\end{eqnarray}
\medskip

\noindent
{\bf Estimating $D'_2$ of \eqref{16.2}. } Our main etimate is given in \eqref{132.1}. According to \eqref{17}, the term ${\mathcal B}(z)$ gives three contributions. The ones involving only one resolvent $R_z^{P_\r}$ are all estimated in the same way as follows. We have for all $z\in{\mathbb C}_-$ (use \eqref{m03}),
\begin{equation}
\big| \langle \phi, {\mathfrak F}(z)^{-1} P_\r I R_z^{P_\r}\psi\rangle\big| \prec \max_j\big| \langle \phi, {\mathfrak F}(z)^{-1}\varphi_j\otimes\Omega_\r\rangle\big|  \, C_1(I\varphi_j\otimes\Omega_\r, \psi).
\label{127.3}
\end{equation}
According to \eqref{55}, the main term of ${\mathfrak F}(z)^{-1}$ is $(L_\s-z)^{-1}$ the norm of which is bounded above by $1/\eta$ for $z\in{\mathcal D}_1$  and by $1/(\eta+\vartheta)$ for $z\in{\mathcal D}_2$. Then,
\begin{equation}
\Big| \int_{{\mathcal D}_1 \cup {\mathcal D}_2} e^{\i tz}\langle \phi, (L_\s-z)^{-1}\lambda I R_z^{P_\r}\psi\rangle dz \Big| \prec e^{wt} \Big( \frac{|\lambda|\vartheta}{\eta} + \frac{|\lambda|}{\eta+\vartheta} \Big) \|\phi\|  \max_j C_1(I\varphi_j\otimes\Omega_\r,\psi). 
\label{128}
\end{equation}
The terms in \eqref{55}, which are of order two and higher in $\lambda$, are estimated as
\begin{equation}
\Big\| (L_\s-z)^{-1} \big[ \lambda^2 P_\r I R_z^{P_\r}I P_\r(L_\s-z)^{-1} + S_4\big] \Big\| \prec \frac{\lambda^2\varkappa_1}{\eta^2} \Big[1+  \frac{\lambda^2\varkappa_1}{\eta}\Big]
\label{129.1}
\end{equation}
for $z\in{\mathcal G}_\infty^e$. It follows from \eqref{55}, \eqref{128} and  \eqref{129.1} that 
\begin{align}
\Big| \int_{{\mathcal D}_1 \cup {\mathcal D}_2} e^{\i tz}\langle \phi, {\mathfrak F}(z)^{-1}\lambda I R_z^{P_\r}\psi\rangle dz \Big|  \prec & 
 \ e^{wt} \Big( \frac{|\lambda|\vartheta}{\eta} + \frac{|\lambda|}{\eta+\vartheta} +\frac{|\lambda|^3\varkappa_1}{\eta^2} \Big[ 1+\frac{\lambda^2\varkappa_1}{\eta}\Big]\Big)\nonumber\\
&\times  \|\phi\|  \max_j C_1(I\varphi_j\otimes\Omega_\r,\psi).
\label{130.1}
\end{align}
The same upper bound is achieved for the term involving $R_z^{P_\r}\lambda I P_\r{\mathfrak F}(z)^{-1}$ in \eqref{17}. To deal with the term in \eqref{17} involving the resolvent twice, we use the bound 
\begin{equation}
\lambda^2 \big|\langle\phi, R_z^{P_\r}I {\mathfrak F}(z)^{-1} IP_\r R_z^{P_\r}\psi\rangle\big| \prec \lambda^2 \|{\mathfrak F}(z)^{-1}\| \max_j  C_1(I\varphi_j\otimes\Omega_\r, \phi)  C_1(I\varphi_j\otimes \Omega_\r,\psi).
\label{131.1}
\end{equation}
From \eqref{55}, $\|{\mathfrak F}(z)^{-1}\|\prec \frac{1}{\eta}[1+\lambda^2\varkappa_1/\eta]$ for $z\in{\mathcal G}_\infty^e$, which gives, combined with \eqref{131.1} and  \eqref{130.1} the result
\begin{align}
\big| \langle \phi, D'_2\psi\rangle \big| & = \Big| \int_{{\mathcal G}_\infty^e} e^{\i tz}\langle \phi, {\mathfrak B}(z)\psi\rangle dz \Big|\\
  & \prec  
	\ e^{wt} \Big( \frac{|\lambda|\vartheta}{\eta} + \frac{|\lambda|}{\eta+\vartheta} +\frac{|\lambda|^3\varkappa_1}{\eta^2} \Big[ 1+\frac{\lambda^2\varkappa_1}{\eta}\Big]\Big)  \sym  \|\phi\|  \max_j C_1(I\varphi_j\otimes\Omega_\r,\psi)  \nonumber \\
& \quad +e^{wt}\frac{\lambda^2}{\eta} \Big[1+\frac{\lambda^2\varkappa_1}{\eta}\Big]\max_j  C_1(I\varphi_j\otimes\Omega_\r, \phi)  C_1(I\varphi_j\otimes \Omega_\r,\psi).
	\label{132.1}
\end{align}
\medskip

\noindent
{\bf Estimating $D'_3$ of \eqref{16.2}. } Recalling the definition \eqref{116} of ${\mathcal G}_\infty^e$ and using \eqref{m03}, we extend the integration by an amount of $2\eta$ to $e\le x\le e'$, 
\begin{equation}
\langle \varphi, D'_3 \psi\rangle = \int_{{\mathcal G}_\infty^e} e^{\i t z} \langle \phi, R_z^{P_\r} \psi\rangle dz = \int_e^{e'} e^{\i t (x-\i w)} \langle \phi, R_{x-\i w}^{P_\r}\psi\rangle dx  +S_5,
\label{126.2}
\end{equation}
with 
\begin{equation}
\label{127.2}
\|S_5\| \prec  \eta e^{wt} C_1(\phi, \psi).
\end{equation}

\bigskip

\noindent
{\bf Estimates on the infinite parts of ${\mathcal G}_\infty$.} So far in this section, we have dealt with all $z$ between any two eigenvalues of $L_\s$, see Figs. f1 and f3. Let $e_+$  and $e_-$ be the largest and smallest eigenvalues of $L_\s$, respectively and set 
\begin{equation}
{\mathcal G}_\infty^+ = \{x-\i w : x\ge e_++\eta \}\qquad\mbox{and}\qquad {\mathcal G}_\infty^- = \{x-\i w : x\le e_--\eta\}.
\end{equation} 
We choose $Q=P_\r$ for the projection in the Feshbach decomposition \eqref{16}. We use \eqref{54} to expand 
\begin{equation}
	\label{65}
	{\mathfrak F}(z)^{-1}= (L_\s-z)^{-1} \sum_{n\ge 0}\lambda^{2n} \big[P_\r IR_z^{P_\r}I P_\r(L_\s-z)^{-1}\big]^n,
\end{equation}
which converges for $\lambda^2\varkappa_1\|(L_\s-z)^{-1}\|\prec 1$. For $z\in{\mathcal G}_\infty^+$, we have $\|(L_\s-z)^{-1}\| \le 1/\eta$ and so, since $\lambda^2\varkappa_1/\eta\prec 1$, the series \eqref{65} converges uniformly in $z\in{\mathcal G}_\infty^+$ an uniformly in $w$. Splitting off the first term in the series \eqref{65} gives 
\begin{equation}
\label{64}
\int_{{\mathcal G}_\infty^+}e^{\i t z} {\mathfrak F}(z)^{-1} dz  =\int_{{\mathcal G}_\infty^+}\frac{e^{\i t z}}{L_\s-z}dz +{\mathcal I},
\end{equation}
where ($z=x-\i w$)
\begin{eqnarray}
\|{\mathcal I}\| &\le& e^{wt} \int_{e_++\eta}^\infty \big\| (L_\s-z)^{-1} \sum_{n\ge 1}  [\lambda^2 P_\r IR_z^{P_\r}IP_\r(L_\s-z)^{-1}]^n \big\| dx \nonumber\\
&\le & e^{wt} \int_{e_++\eta}^\infty \frac{1}{x-e_+}  \sum_{n\ge 1}
\frac{(C \varkappa_1\lambda^2)^n}{(x-e_+)^n}  dx \nonumber\\
&=& e^{wt}\sum_{n\ge 1} (C \varkappa_1\lambda^2)^n \int_\eta^\infty \frac{dx}{x^{n+1}} = e^{wt} \sum_{n\ge1}\frac{1}{n}\Big( \frac{C\varkappa_1\lambda^2}{\eta}\Big)^n \nonumber\\
&\le& e^{wt} \frac{C\varkappa_1\lambda^2}{\eta}\sum_{n\ge0} \Big( \frac{C\varkappa_1\lambda^2}{\eta}\Big)^n\prec e^{wt} \frac{\varkappa_1\lambda^2}{\eta}, 
\label{139}
\end{eqnarray}
since $\frac{\varkappa_1\lambda^2}{\eta}\prec 1$. The analysis and estimates for $z\in{\mathcal G}_\infty^-$  are the same. 
We conclude that 
\begin{equation}
	\Big\|
\int_{{\mathcal G}_\infty^+\cup {\mathcal G}_\infty^-} e^{\i t z} {\mathfrak F}(z)^{-1} dz  -\int_{{\mathcal G}_\infty^+\cup {\mathcal G}_\infty^-}\frac{e^{\i t z}}{L_\s-z}dz \Big\| \prec
 e^{wt} \frac{\varkappa_1\lambda^2}{\eta}. 
\label{67}
\end{equation}
Next we need to estimate $\int_{{\mathcal G}_\infty^+\cup {\mathcal G}_\infty^-} e^{\i tz} \langle \phi, {\mathfrak B}(z) \psi\rangle  dz$, where ${\mathfrak B}(z)$ is the sum of three terms according to \eqref{17}. In each term, we use \eqref{65} to split off the main part. Accordingly, we have
\begin{align}
{\mathfrak B}(z)  = & -\lambda (L_\s-z)^{-1} P_\r I R_z^{P_\r} - \lambda R_z^{P_\r} I P_\r(L_\s-z)^{-1} \nonumber\\
 & +\lambda^2 R_z^{P_\r} IP_\r (L_\s-z)^{-1} IR_z^{P_\r} +{\mathfrak B}_2(z), 
 \label{141}
\end{align} 
which defines the quantity ${\mathfrak B}_2(z)$.  Proceeding as in the derivation of \eqref{139}, we get
\begin{align}
\Big| \int_{{\mathcal G}_\infty^+\cup {\mathcal G}_\infty^-} & e^{\i tz} \langle \phi, {\mathfrak B}_2(z) \psi\rangle  dz\Big| \nonumber\\
 \prec &  \quad e^{wt} \frac{\varkappa_1|\lambda|^3}{\eta} \sym \|\phi\|  \max_j C_1(I\varphi_j\otimes\Omega_\r,\psi) \nonumber\\
& \quad + e^{wt} \frac{\varkappa_1\lambda^4}{\eta}  \max_j C_1(I\varphi_j\otimes\Omega,\psi) \max_j C_1(I\varphi_j\otimes\Omega,\phi) .
	\label{63.1}
\end{align}
Now we analyze the integral associated to the first term in \eqref{141}, which is given by $-\lambda \int_{{\mathcal G}_\infty^+}e^{\i tz}\langle\phi, (L_\s-z)^{-1} P_\r IR_z^{P_\r}\psi\rangle dz$. We split the integration domain into  $e_++\eta\le x\le e_++\eta+\vartheta$ and $x\ge e_++\eta+\vartheta$. On the compact domain, the integral has the bound $\prec e^{wt} \frac{|\lambda|\vartheta}{\eta} \|\phi\| \max_j C_1(I\varphi_j\otimes\Omega_\r,\psi)$, where the factor $1/\eta$ is due to the resolvent $(L_\s-z)^{-1}$ and $\vartheta$ is the length of the interval of integration. On the infinite integration domain, we need to control the integrability for large $x$. Denoting $\bar L_\lambda\equiv P^\perp_\r L_\lambda P^\perp_\r\upharpoonright_{{\rm Ran}P^\perp_\r}$, we have
\begin{equation}
R_z^{P_\r} = (z+\i)^{-1}\big[-\bbbone + R_z^{P_\r}(\bar L_\lambda+\i)\big]
\label{144}
\end{equation}
so that ($z=x-\i w$)
\begin{align}
\Big| \int_{e_++\eta+\vartheta}^\infty e^{\i tz}  \langle\phi, & (L_\s-z)^{-1} P_\r  IR_z^{P_\r}\psi\rangle dz \Big| \nonumber\\
\prec &  \ e^{wt}  \|\phi\| \Big( \|P_\r I\| \|P^\perp_\r\psi\| + \max_j C_1\big(I\varphi_j\otimes\Omega_\r,(\bar L_\lambda+\i)\psi\big)\Big) \nonumber\\
&  \times \int_{e_++\eta+\vartheta}^\infty\  \frac{dx }{|e_+-z| |z+\i|}.
  \label{144.1}
\end{align}
The last integral is $\prec (\eta+\vartheta)^{-1}$, thus 
\begin{align}
\Big| \lambda \int_{{\mathcal G}_\infty^+}e^{\i tz}\langle\phi, & (L_\s-z)^{-1} P_\r IR_z^{P_\r}\psi\rangle dz\Big|\nonumber\\
\prec\  &  e^{wt} \frac{|\lambda|\vartheta}{\eta} \|\phi\| \max_j C_1(I\varphi_j\otimes\Omega_\r,\psi)\nonumber\\
&  + e^{wt}\frac{|\lambda| }{\eta+\vartheta}\|\phi\|\Big( \|P_\r I\| \|P^\perp_\r\psi\| + \max_j C_1\big(I\varphi_j\otimes\Omega_\r,(\bar L_\lambda+\i)\psi\big)\Big).
\label{145}
\end{align}
Of course, we get the same upper bound for $-\lambda \int_{{\mathcal G}_\infty^+}e^{\i tz}\langle\phi, R_z^{P_\r} I P_\r (L_\s-z)^{-1} \psi\rangle dz$ (see \eqref{141}), with $\phi$ and $\psi$ exchanged on the right side of \eqref{145}. Finally, we estimate 
\begin{align}
\Big| \int_{{\mathcal G}_\infty^+} e^{\i tz}\langle \phi, \, & R_z^{P_\r} IP_\r (L_\s-z)^{-1} IR_z^{P_\r} \psi\rangle d z\Big| \nonumber\\
\prec & \ e^{wt} \max_j C_1(I\varphi_j\otimes\Omega_\r, \phi)\Big( \|P_\r I\| \|P^\perp_\r\psi\| + \max_j C_1\big(I\varphi_j\otimes\Omega_\r,(\bar L_\lambda+\i)\psi\big)\Big)\nonumber\\
& \times \int_{e_++\eta}^\infty \frac{dx }{|e_+-z| |z+\i|}.
\label{146}
\end{align}
The last integral is $\prec 1/\eta$.  Collecting the bounds \eqref{63.1}, \eqref{145} and \eqref{146} and using them in \eqref{141}, we obtain
\begin{align}
\Big| & \int_{{\mathcal G}_\infty^+\cup {\mathcal G}_\infty^-} \langle \phi, {\mathfrak B}(z)\psi\rangle dz\Big|\nonumber\\
& \prec   e^{wt} \frac{\varkappa_1|\lambda|^3+|\lambda|\vartheta }{\eta} 
 \sym \|\phi\|  \max_j C_1(I\varphi_j\otimes\Omega_\r,\psi) \nonumber\\
&  \quad + e^{wt} \frac{\varkappa_1\lambda^4}{\eta}  \max_j C_1(I\varphi_j\otimes\Omega,\psi) \max_j C_1(I\varphi_j\otimes\Omega,\phi) \nonumber\\
&  \quad + e^{wt}\frac{|\lambda| }{\eta+\vartheta} \sym \|\phi\| \Big(  \|P_\r I\|  \|P^\perp_\r\psi\|+ \max_j C_1\big(I\varphi_j\otimes\Omega_\r,(\bar L_\lambda+\i)\psi\big) \Big)\nonumber\\
& \quad + e^{wt}\frac{\lambda^2}{\eta}  \sym \max_j C_1(I\varphi_j\otimes\Omega_\r, \phi)\Big( \|P_\r I\| \|P^\perp_\r\psi\| + \max_j C_1\big(I\varphi_j\otimes\Omega_\r,(\bar L_\lambda+\i)\psi\big)\Big).
\label{148}
\end{align}
Next, extending the integration domain by $2\eta$ similar to \eqref{126.2},   we estimate
\begin{equation}
\int_{{\mathcal G}_\infty^+ \cup {\mathcal G}_\infty^-} e^{\i t z} \langle\phi, R_z^{P_\r}\psi\rangle dz = \int_{(-\infty,e_-]\cup [e_+,\infty)}   e^{\i t (x-\i w)} \langle\phi, R_{x-\i w}^{P_\r}\psi\rangle dx +S_6,
\label{133.1}
\end{equation}
where
\begin{equation}
\| S_6\| \prec \eta e^{wt} C_1(\phi,\psi). 
\end{equation}

\bigskip

Finally we note that ${\mathcal G}_\infty = \cup_{e<e_+} {\mathcal G}_\infty^e \cup {\mathcal G}_\infty^+ \cup{\mathcal G}_\infty^-$ and that
\begin{eqnarray}
\frac{-1}{2\pi\i} \int_{{\mathbb R}-\i w} e^{\i t z}\langle \phi, R_z^{P_\r} \psi\rangle dz =  \langle \phi, P^\perp_\r e^{\i t P^\perp_\r L_\lambda P^\perp_\r} P^\perp_\r \psi\rangle.
\label{120}
\end{eqnarray} 
Combining the estimates \eqref{D_1'},  \eqref{132.1}, \eqref{126.2},  \eqref{67}, \eqref{148}  and \eqref{133.1} yields \eqref{69}. This completes the proof of Proposition \ref{prop3}. \hfill \qed

\subsection{Combining the estimates: end of the proof of Theorem \ref{motherthm}. } 
\label{proofpropersect}

\begin{figure}[t]
	\centering
	\includegraphics[width=15cm]{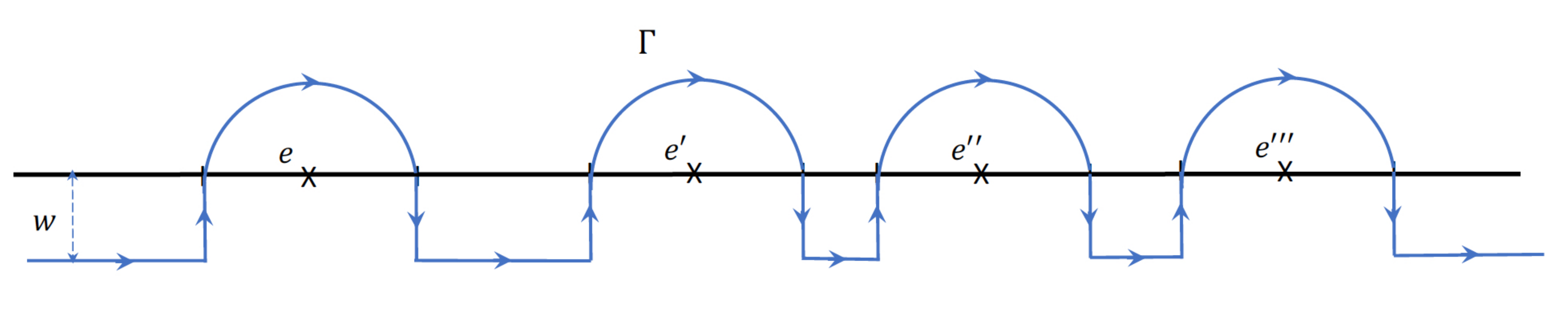}
	\caption{The contour $\Gamma$ appearing in \eqref{134.1}.}
	\label{Fig5}
\end{figure}

We combine the estimates \eqref{024} and \eqref{69}  (and \eqref{120}) into
\begin{eqnarray}
\lefteqn{
\Big| \sum_{e\in{\mathcal E}_0} J_e(t) +J_\infty(t) - 
\sum_{e\in{\mathcal E}_0}\sum_{s\in{\mathcal S}_e^{\rm dec}}e^{\i t (e+\lambda^2 \aes(e,\lambda))} \big\langle\phi, \Qes(e,\lambda)\psi\big\rangle}\nonumber\\ &&-\sum_{e\in{\mathcal E}_0}\sum_{s\in{\mathcal S}_e^{\rm osc}}  \, e^{\i t \Ees}\big\langle \phi, \Qes(\Ees,\lambda)\psi\big\rangle  -\langle \phi, P^\perp_\r e^{\i t P^\perp_\r L_\lambda P^\perp_\r} P^\perp_\r \psi\rangle \nonumber \\
&&+ \frac{1}{2\pi\i}  \int_{\Gamma}  \,  e^{\i t z} \langle\phi, (L_\s- z)^{-1}P_\r\psi\rangle  \, dz \Big|  \prec S(w,\phi,\psi),
\label{134.1}
\end{eqnarray}
where the contour $\Gamma$ appearing in \eqref{134.1} is given by $\Gamma =  \bigcup_e{\mathcal A}_e \, \cup\, {\mathcal G}_\infty$. It is represented in Fig.~\ref{Fig5}.

The error term $S(w,\phi,\psi)$  in \eqref{134.1} is the sum of the error terms in \eqref{024} and \eqref{69}, satisfying
\begin{align}
& \!\!\!\!\!\! S(w,\phi,\psi) \prec  \\
& \Big\{ e^{wt}\frac{w +\eta}{\eta} \big( \eta/g+ \varkappa_3 (|\lambda|+\eta)\big)  +  \eta e^{wt}\Big( \varkappa_3+\lambda^2\varkappa_5 + \Big( \frac{\varkappa_4\kappa}{a} +\varkappa_3\Big)  \Big( 1+\frac{\eta}{a\lambda^2}\Big) \Big) \nonumber\\
& +\lambda^2 e^{wt}\Big( \frac{\varkappa_1}{\eta}+ \frac{g\varkappa_1}{(\eta+\vartheta)^2} + \frac{g\varkappa^2_1\lambda^2}{\eta^3} +\vartheta \big[\frac{\varkappa_1(\eta+\vartheta+|\lambda|)+\alpha\kappa}{\eta^2} +\frac{\varkappa_1}{g}(1/\eta +1/g)\big] \Big)
\Big\} \|\phi\|\, \|\psi\|	\nonumber\\
&+ \Big\{ 
 \frac{\kappa\eta}{a|\lambda|} e^{wt} + |\lambda|\big( 1 +\eta(1+\varkappa_3)\big)+ \frac{(w+\eta) e^{wt}}{\eta} ( |\lambda| +\lambda^2)\nonumber\\
&\qquad + e^{wt}\lambda^2\eta\varkappa_5(|\lambda|+\lambda^2)+ \lambda^2 \big( 1+ (w+\eta) e^{wt}(1+\varkappa_3)  \big)\Big\} \nonumber\\
& \qquad \times \Big[\max_m C_2(I\varphi_m\otimes\Omega_\r,\phi) \max_m C_2(I\varphi_m\otimes\Omega_\r,\psi)+  \sym \|\phi\| \max_mC_2(I\varphi_m\otimes\Omega_\r, \psi) \Big] \nonumber\\
& +\eta e^{wt} \big\{ C_1(\phi,\psi) + C_2(\phi,\psi)\big\}\nonumber \\
& +  e^{wt} \Big( \frac{|\lambda|\vartheta}{\eta} + \frac{|\lambda|}{\eta+\vartheta} +\frac{|\lambda|^3\varkappa_1}{\eta^2} \Big[ 1+\frac{\lambda^2\varkappa_1}{\eta}\Big] \Big)\sym \|\phi\|  \max_m C_1(I\varphi_m\otimes\Omega_\r,\psi)  \nonumber \\
&  +e^{wt}\frac{\lambda^2}{\eta} \Big[1+\lambda^2\varkappa_1(1+1/\eta) \Big]\max_m C_1(I\varphi_m\otimes\Omega_\r, \phi)  \max_mC_1(I\varphi_m\otimes \Omega_\r,\psi).\nonumber\\
&  + e^{wt}\frac{|\lambda| }{\eta+\vartheta} \sym \|\phi\| \Big(  \|P_\r I\|  \|P^\perp_\r\psi\|+ \max_m C_1\big(I\varphi_m\otimes\Omega_\r,(\bar L_\lambda+\i)P^\perp_\r\psi\big) \Big)\nonumber\\
&  + e^{wt}\frac{\lambda^2}{\eta}  \sym \max_m C_1(I\varphi_m\otimes\Omega_\r, \phi)\Big( \|P_\r I\| \|P^\perp_\r\psi\| + \max_m C_1\big(I\varphi_m\otimes\Omega_\r,(\bar L_\lambda+\i)P^\perp_\r\psi\big)\Big).\nonumber
\end{align}

Since $\Gamma$ does not enclose any of the eigenvalues of $L_\s$, we have $\int_{\Gamma} \frac{e^{\i tz}}{L_\s-z}dz =0$, that is, the last term on the left side of \eqref{134.1} vanishes. Upon taking $w\rightarrow 0$, we obtain from \eqref{134.1}, \eqref{19}
\begin{eqnarray}
	\lefteqn{
\Big| \langle\phi, e^{\i tL_\lambda}\psi\rangle  - \sum_{e\in{\mathcal E}_0}\sum_{s\in{\mathcal S}_e^{\rm dec}}e^{\i t (e+\lambda^2 \aes(e,\lambda))} \big\langle\phi, \Qes(e,\lambda)\psi\big\rangle}	\label{134.2}\\ &&-\sum_{e\in{\mathcal E}_0}\sum_{s\in{\mathcal S}_e^{\rm osc}}  \, e^{\i t \Ees}\big\langle \phi, \Qes(\Ees,\lambda)\psi\big\rangle -\langle \phi, P^\perp_\r e^{\i t P^\perp_\r L_\lambda P^\perp_\r} P^\perp_\r \psi\rangle \Big| \prec \varkappa(0)K(\phi,\psi),
\nonumber
\end{eqnarray}
with
\begin{eqnarray}
K(\phi,\psi) &=& \|\phi\|\,\|\psi\| + \max_{j=1,2} C_j(\phi,\psi)\nonumber \\
&& +\max_{j=1,2} \big( \max_m C_j(I\varphi_m\otimes\Omega_\r,\phi) \max_m C_j(I\varphi_m\otimes\Omega_\r,\psi)\big)\nonumber\\
&&+  \sym \|\phi\| \max_{j,m}C_j(I\varphi_m\otimes\Omega_\r, \psi)\nonumber\\
&&  +  \sym \Big( \|\phi\| +\max_m C_1(I\varphi_m\otimes\Omega_\r, \phi)\Big)\nonumber\\
&&\quad \times \Big(  \|P_\r I\|  \|P^\perp_\r\psi\|+ \max_m C_1\big(I\varphi_m\otimes\Omega_\r,(\bar L_\lambda+\i)P^\perp_\r\psi\big) \Big)
\label{137}
\end{eqnarray}
and
\begin{eqnarray}
\lefteqn{
\varkappa(0) =	
\eta/g+ \varkappa_3 (|\lambda|+\eta) +\varkappa_5\eta(\lambda^2+\eta/a)(\varkappa_4\kappa/a+\varkappa_3)  }  \nonumber\\
&&+\lambda^2 \Big[ 1+\frac{\varkappa_1}{\eta}+ \frac{g\varkappa_1}{(\eta+\vartheta)^2} + \frac{g\varkappa^2_1\lambda^2}{\eta^3} +\vartheta \Big(\frac{\varkappa_1(\eta+\vartheta+|\lambda|)+\alpha\kappa}{\eta^2} +\frac{\varkappa_1}{g}\big(1/\eta +1/g\big)\Big)
\Big] \nonumber\\
&&+ \frac{\kappa\eta}{a|\lambda|} +|\lambda|\big(1+\eta(1+\varkappa_3)\big) +\lambda^2\eta\varkappa_5 (1+\lambda^2)  +\lambda^2\eta(1+\varkappa_3)\nonumber\\
&&+ \frac{\lambda^2}{\eta} \big( 1+|\lambda|\varkappa_1/\eta\big) \big(1+ \lambda^2\varkappa_1/\eta\big) +|\lambda|\vartheta/\eta + \frac{|\lambda|}{\eta+\vartheta}.
\label{117.1}
\end{eqnarray}

Next we use \eqref{119.0} and \eqref{119.1} to get 
\begin{equation}
\|\Qes(e,\lambda) -\Qes\|\prec \varkappa_3|\lambda| \qquad \mbox{and}\qquad |\aes(e,\lambda) - \aes|\prec\varkappa_4|\lambda|, 
\label{74.1}
\end{equation} 
where $\aes$ and $\Qes$ are the spectral data of $\Lambda_e$, \eqref{83}. For $(e,s)$ such that $\aes\notin{\mathbb R}$  we have
\begin{eqnarray}
\Big| e^{\i t(e+\lambda^2 \aes(e,\lambda))} - e^{\i t(e+\lambda^2 \aes)}\Big| &=& e^{-t\lambda^2{\rm Im}\aes} \Big| e^{ \i t\lambda^2 (\aes(e,\lambda)- \aes)} -1\Big|\nonumber\\
&=& e^{-t\lambda^2{\rm Im}\aes} \Big| \int_0^{t\lambda^2(\aes(e,\lambda)- \aes)} e^{\i  z} dz \Big|,
\label{74}
\end{eqnarray}
where the integration path is the straight line linking the endpoints. On this path, $|{\rm Im}z|\le t\lambda^2|\aes(e,\lambda)-\aes|\le \tfrac12 t\lambda^2 {\rm Im}\aes$, as $|\lambda|\varkappa_4\prec a$ (see \eqref{cond1}).  Thus the integral is bounded above by $t\lambda^2|\aes(e,\lambda)-\aes| e^{\frac12 t\lambda^2 {\rm Im}\aes}$, which gives
\begin{eqnarray}
\Big| e^{\i t(e+\lambda^2 \aes(e,\lambda))} - e^{\i t(e+\lambda^2 \aes)}\Big| &\prec & \frac{|\lambda|\varkappa_4}{a} (\tfrac12 t \lambda^2{\rm Im}\aes) e^{-\frac12 t\lambda^2{\rm Im}\aes} \prec \frac{|\lambda|\varkappa_4}{a},
\label{75.1}
\end{eqnarray}
uniformly in $t\ge 0$. For $(e,s)$ such that $\aes\in\mathbb R$, we cannot replace $e^{\i t\Ees}$ by $e^{\i t (e+\lambda^2\aes)}$ in a manner uniform in time, as $\sup_{t\ge 0}|e^{i t\Ees}- e^{\i t (e+\lambda^2\aes)}|=2$, even though $\Ees-\aes$ is of order $\lambda^2$. We conclude that we can replace, for the {\em decaying} terms in \eqref{134.2}, the $\aes(e,\lambda)$ by $\aes$ and the  $\Qes(e,\lambda)$ by $\Qes$ and by doing this, we incur an error $\prec |\lambda| (\varkappa_3+\varkappa_4/a)$, uniformly in time $t\ge 0$. Furthermore, denoting the eigenprojection of $L_\lambda$ associated to the eigenvalue $\Ees$ by $\Pi_{e,\lambda}$, we have (in the strong sense) $\Pi_{e,\lambda} = \lim_{\varepsilon\rightarrow 0_+}\i \varepsilon (L_\lambda -E_e^{(s)}(\lambda)+\i\varepsilon)^{-1}$ and by \eqref{16}
\begin{equation}
P_e \Pi_{e,\lambda}P_e = \lim_{\varepsilon\rightarrow 0_+}\i \varepsilon \big[{\mathfrak F}(L_\lambda -E_e^{(s)}(\lambda)+\i\varepsilon; P_e)\big]^{-1}.
\label{163}
\end{equation}
Taking into account \eqref{23.1} and \eqref{89.1},
\begin{equation}
	P_e \Pi_{e,\lambda}P_e = \lim_{\varepsilon\rightarrow 0_+}\i \varepsilon \sum_{s'=1}^{m_e} \frac{Q_e^{(s')}(E_e^{(s)}(\lambda) -\i\varepsilon, \lambda)}{e-E_e^{(s)}(\lambda) +\i\varepsilon+\lambda^2 a_e^{(s')}(E_e^{(s)}(\lambda)-\i\varepsilon, \lambda)}.
	\label{164}
\end{equation}
According to \eqref{m69} we have $e-E_e^{(s)}(\lambda) = -\lambda^2 a_e^{(s)}(E_e^{(s)}(\lambda), \lambda)$ and it is readily seen that the limit $\varepsilon\rightarrow 0_+$ vanishes for the terms $s'\neq s$ in the sum, because the denominator stays bounded. We conclude that 
\begin{equation}
P_e \Pi_{e,\lambda}P_e = Q_e^{(s)}(E_e^{(s)}(\lambda), \lambda).
\label{165}
\end{equation}
Combining \eqref{165} with \eqref{119.0} gives
\begin{equation}
\|P_e \Pi_{e,\lambda}P_e - \Qes\|\prec \varkappa_3\big( |E_e^{(s)}(\lambda)-e|+|\lambda|\big)\prec |\lambda| \varkappa_3 \big(1+|\lambda|\alpha\big).
\label{166.1} 
\end{equation}
To arrive at the last bound in \eqref{166.1}, we notice that by \eqref{m69} and \eqref{119.1},  $|E_e^{(s)}(\lambda)-e|\prec \lambda^2(\alpha+\varkappa_4(|E_e^{(s)}(\lambda)-e|+|\lambda| ))$, which we can solve since $\lambda^2\varkappa_4\prec 1$ (see \eqref{cond1}) to yield $|E_e^{(s)}(\lambda)-e|\prec \lambda^2(\alpha+|\lambda|\varkappa_4)$. 

Making the replacements in the decaying terms as discussed after \eqref{75.1} and using \eqref{166.1}, we then obtain  from \eqref{134.2} that  
\begin{eqnarray}
\lefteqn{
	\Big| \langle\phi, e^{\i tL_\lambda}\psi\rangle  - \sum_{e\in{\mathcal E}_0}\sum_{s\in{\mathcal S}_e^{\rm dec}} e^{\i t (e+\lambda^2 \aes)} \big\langle\phi, \Qes\psi\big\rangle}\nonumber	\\
&& -\sum_{e\in{\mathcal E}_0}\sum_{s\in{\mathcal S}_e^{\rm osc}}  \, e^{\i t \Ees}\big\langle \phi, \Qes\psi\big\rangle
 -\langle \phi, P^\perp_\r e^{\i t P^\perp_\r L_\lambda P^\perp_\r} P^\perp_\r \psi\rangle \Big| \prec \varkappa'(0) K(\phi,\psi),
 \label{76}
\end{eqnarray}
where
\begin{equation}
\varkappa'(0) = \varkappa(0) + |\lambda|\big( \varkappa_3(1+|\lambda|\alpha)+ \varkappa_4/a\big).
\label{166}
\end{equation}
We choose
\begin{equation}
	\eta = |\lambda|^{1+\epsilon}, \quad \vartheta = |\lambda|^{1-\epsilon'}
	\label{73}
\end{equation}
for $\epsilon',\epsilon>0$ to be determined below. Then it follows from \eqref{117.1}, \eqref{166} that
\begin{eqnarray}
\varkappa'(0) &\prec&  |\lambda|^{3+\epsilon} +|\lambda|^{1-\epsilon-\epsilon'} +|\lambda|^{\epsilon'} + |\lambda|^{1+\epsilon}/g +|\lambda|^\epsilon \kappa/a +  |\lambda|^{1-2\epsilon-\epsilon'} \alpha\kappa \nonumber\\
&&+ \varkappa_1\Big(  |\lambda|^{1-\epsilon}+|\lambda|^{1-2\epsilon} + |\lambda|^{2-2\epsilon-2\epsilon'}+ |\lambda|^{2-\epsilon-\epsilon'} +g( |\lambda|^{2\epsilon'}+|\lambda|^{1-3\epsilon}\varkappa_1)  +\nonumber\\
&& \qquad +|\lambda|^{2-3\epsilon}\varkappa_1+ |\lambda|^{2-\epsilon-\epsilon'}/g +|\lambda|^{3-\epsilon'}/g^2 \Big)\nonumber\\
&& + \varkappa_3 \Big( |\lambda|+|\lambda|^{2+\epsilon} +\varkappa_5|\lambda|^{3+\epsilon} +|\lambda|^{2+2\epsilon}\varkappa_5/a+\lambda^2\alpha
\Big)\nonumber\\
&& + \varkappa_4\Big(\lambda^{3+\epsilon} \varkappa_5\kappa/a + \lambda^{2+2\epsilon}\varkappa_5\kappa/a^2 +|\lambda|/a\Big) \nonumber\\
&&+ \varkappa_5\Big( |\lambda|^{3+\epsilon} + |\lambda|^{5+\epsilon}\Big). 
\label{159}
\end{eqnarray}
Taking $|\lambda|\le1$ and $\epsilon=\epsilon'=1/4$, we obtain from \eqref{159}, 
\begin{eqnarray}
\label{160}
\varkappa'(0) &\prec& |\lambda|^{1/4} \max\Big\{1, 1/g,\kappa/a,\alpha\kappa,\varkappa^2_1,g\varkappa_1(1+\varkappa_1) , \varkappa_1/g^2, \varkappa_3(1+\alpha+\varkappa_5+\varkappa_5/a),\nonumber\\
&&\qquad \ \ \varkappa_4\big(1/a+\varkappa_5\kappa(1+1/a)/a\big), \varkappa_5\Big\}.
\label{171}
\end{eqnarray}
We further bound the maximum in \eqref{171}. Namely, using \eqref{kappa3,4} and \eqref{kappa5} we  get
\begin{eqnarray}
\varkappa_3(1+\varkappa_5+\varkappa_5/a) &\prec& \varkappa_2(1+\varkappa_2^2) \frac{\kappa}{\delta}\big(1+\alpha+1/a\big) \label{169}\\
\varkappa_4/a &\prec& \varkappa_2(1+\varkappa_2) \frac{\kappa}{a}(1+\kappa/\delta) \label{173}\\
\varkappa_4\varkappa_5\frac{\kappa}{a} (1+1/a) &\prec& \varkappa_2^2(1+\varkappa_2^3) \frac{\kappa^4}{a}(1+\kappa^3/\delta^3)(1+1/a) \label{174}\\
\varkappa_5 &\prec& \varkappa_2(1+\varkappa_2)(1+\kappa^2/\delta^2)\kappa.
\label{172}
\end{eqnarray}
According to \eqref{kappa12} we have $\varkappa_2\prec \varkappa_1$ and the sum of the right hand sides of \eqref{169} - \eqref{172} is  bounded above by $\varkappa_1(1+\varkappa_1^4)\kappa \max\big\{ 1,\tfrac{1+\alpha}{\delta}, 1/a, \tfrac{1+\kappa}{a\delta}, \kappa^2\big( \tfrac{\kappa}{a}(1+\kappa^3/\delta^3)(1+1/a)+1/\delta^2
\big)\big\}$. We use this latter bound in \eqref{160} to obtain,
\begin{equation}
\varkappa'(0) \prec |\lambda|^{1/4} \varkappa_0
\end{equation}
with
\begin{eqnarray}
\varkappa_0&=&
\max\Big\{1, 1/g,\kappa/a,\alpha\kappa,\varkappa_1(1+\varkappa_1)(1+g+1/g),\nonumber\\
&&\varkappa_1(1+\varkappa_1^4)\kappa \max\big\{ 1,\tfrac{1+\alpha}{\delta}, 1/a, \tfrac{1+\kappa}{a\delta}, \kappa^2\big( \tfrac{\kappa}{a}(1+\kappa^3/\delta^3)(1+1/a)+1/\delta^2
\big)\big\}\Big\}. \qquad
\label{160.1}
\end{eqnarray}
We combine \eqref{160.1} and  \eqref{76} into 
\begin{eqnarray}
\lefteqn{
	\Big| \langle\phi, e^{\i tL_\lambda}\psi\rangle  - \sum_{e\in{\mathcal E}_0}\sum_{s\in{\mathcal S}_e^{\rm dec}} e^{\i t (e+\lambda^2 \aes)} \big\langle\phi, \Qes\psi\big\rangle} \label{76.1}\\
&&-\sum_{e\in{\mathcal E}_0}\sum_{s\in{\mathcal S}_e^{\rm osc}}  \, e^{\i t \Ees}\big\langle \phi, \Qes\psi\big\rangle -\langle \phi, P^\perp_\r e^{\i t P^\perp_\r L_\lambda P^\perp_\r} P^\perp_\r \psi\rangle \Big| \prec|\lambda|^{1/4}\varkappa_0\, K(\phi,\psi).
\nonumber
\end{eqnarray}
Along the way, in getting to \eqref{76.1}, we have made several smallness conditions on $\lambda$. They come from \eqref{cond1}: $|\lambda|\varkappa_4\prec a$, $|\lambda|^{3/4}(\alpha+|\lambda|\varkappa_4)\prec 1$, from Proposition \ref{prop3}: $|\lambda|^{3/4}\varkappa_1\prec 1$, from Theorem \ref{thmA1}: $\lambda^2\varkappa_1\prec g$,  from Lemma \ref{lem2.4}: $|\lambda|\varkappa_1\kappa^2\prec \delta$ and from estimate \eqref{constraints}: $|\lambda|\varkappa_1\prec \|I P_\r\|$, $\lambda^2\varkappa_1\prec\min\{1,g^4\}$.  They are  summarized as follows (see also \eqref{kappa3,4})
\begin{eqnarray}
|\lambda| \varkappa_1\kappa (1+\varkappa_1\kappa/\delta)&\prec& \min\{1,a\} \label{180}\\
|\lambda|^{3/4}(\alpha+\varkappa_1) &\prec& 1\\
|\lambda|\varkappa_1 &\prec& \min\{\delta/\kappa^2, \|I P_\r\|\}\\
\lambda^2\varkappa_1&\prec& g^3.
\label{182}
\end{eqnarray}
A sufficient condition for \eqref{180}-\eqref{182} to hold is
\begin{eqnarray}
|\lambda|^{3/4} \max\Big[ \varkappa_1\kappa (1+\varkappa_1\kappa/\delta), \alpha,\varkappa_1, \sqrt{\varkappa_1}\Big]\prec \min\Big[ 1,a,\delta/\kappa^2, \|I P_\r\|, g^{3/2}\Big].
\label{185.1}
\end{eqnarray}
We may replace $\sqrt{\varkappa_1}$ by $1$ on the left side of \eqref{185.1} and still have a sufficient condition for \eqref{180}-\eqref{182} to hold. 
This concludes the proof of Theorem \ref{motherthm}, including that of Proposition \ref{prop2}.\qed

\section{Proof of Theorem \ref{thmA1} and properties of $A_e(z,\lambda)$}

\subsection{Regularity of the resolvent, proof of Theorem \ref{thmA1}}

In this section, we simply write $R_z^Q$ for $R_z^Q(\lambda)$. We also identify the range of $P_\r$ with ${\mathcal H}_\s$, {\em c.f.} \eqref{POmega}, so that an operator $P_\r A P_\r$ is viewed as an operator on ${\mathcal H}_\s$. For any $\phi,\psi\in{\mathcal H}_\s$, $z\in{\mathbb C}_-$,  $k=0,\ldots,3$ we have (recall that the $\varphi_m$ are an orthonormal eigenbasis of $L_\s$, see after \eqref{195.1})
\begin{eqnarray}
\big| \langle\phi, \partial^k_z P_\r IR_z^{P_\r}IP_\r\psi\rangle\big| &=&\Big| \sum_{m,n} \langle \phi,\varphi_m\rangle \ \langle \varphi_n,\psi \rangle  \partial^k_z  \langle\varphi_m\otimes\Omega_\r,  (IR_z^{P_\r}I)\varphi_n\otimes\Omega_\r\rangle
\Big| \nonumber \\
&\le& \varkappa_1\sum_{m,n} |\langle \phi,\varphi_m\rangle| \ |\langle \psi,\varphi_n\rangle|  \prec \varkappa_1 \|\phi\|\ \|\psi\|,
\label{162}
\end{eqnarray}
where $\varkappa_1$ is defined in \eqref{195.1}.  We conclude that 
\begin{equation}
	\max_{0\le k\le 2}	\sup_{z\in{\mathbb C}_-}\| \partial_z^k P_\r I R_z^{P_\r} I P_\r\|\prec \varkappa_1 \quad \mbox{and similarly}\quad \sup_{z\in{\mathbb C}_-}\| \partial_\lambda P_\r I R_z^{P_\r} I P_\r\|\prec\varkappa_1.
	\label{90}
\end{equation}

\subsubsection{Proof of Theorem \ref{thmA1}.} 
\label{proofthmA1section}

The proof follows \cite{KM1}, where it is shown that the suprema in \eqref{32}, \eqref{034} and \eqref{33} are finite without giving a specific bound \eqref{198.1}. The key idea is to relate $R_z^{P_e}$ and $R_z^{P_\r}$. Introducing a new operator $K=P_e^\perp LP^\perp_e+\i P_e$ we have
\begin{equation} 
R_z^{P_e} = P^\perp_e(K-z)^{-1}P_e^\perp.
\label{190}
\end{equation} 
An application of the Feshbach map \eqref{fesh1} with projection $Q=P_\r$ yields
\begin{equation}
(K-z)^{-1} = 
\begin{pmatrix}
\bbbone & 0\\
-\lambda R^{P_\r}_z P_\r^\perp I \bar P_e & \bbbone
\end{pmatrix}
\begin{pmatrix}
{\mathfrak F}_z^{-1} & 0\\
0 & R_z^{P_\r}
\end{pmatrix}
\begin{pmatrix}
\bbbone & -\lambda \bar P_e I P_\r^\perp R_z^{P_\r} \\
0 & \bbbone
\end{pmatrix},
\label{fesh1.1}
\end{equation}
where $\bar P_e = P^\perp_eP_\r =\bbbone[L_\s\neq e]\otimes P_\r$. The operator ${\mathfrak F}_z$ and its inverse are given by (use that $P^\perp_e P^\perp_\r=P^\perp_\r$ and $P_\r^\perp P_e=0$, so $R_z^{P_\r}=(P^\perp_\r K P^\perp_\r -z)^{-1}\upharpoonright_{{\rm Ran} P^\perp_\r}$), 
\begin{eqnarray}
{\mathfrak F}_z &=& (i-z)P_e\oplus \bar P_e (L_\s-z-\lambda^2 P_\r IR_z^{P_\r} IP_\r)\bar P_e \label{191}\\
{\mathfrak F}_z^{-1} &=& (i-z)^{-1}P_e\oplus \bar P_e (L_\s-z)^{-1}\big(\bbbone-\lambda^2 P_\r IR_z^{P_\r} IP_\r (L_\s-z)^{-1}\big)^{-1}\bar P_e.
\end{eqnarray}
Combining \eqref{190} with \eqref{fesh1.1} yields four terms when estimating $\langle\phi, R_z^{P_e}\psi\rangle$. Those terms arise when we multiply out the matrices in \eqref{fesh1.1}. One of the terms is $\langle \phi, {\mathfrak F}_z^{-1} \psi\rangle$. For $|z-e|\le g/2$ and since $\|P_\r IR_z^{P_\r} IP_\r\|\prec \varkappa_1$ by \eqref{90}, we obtain from \eqref{191} that 
\begin{equation}
\| {\mathfrak F}_z^{-1} \| \prec \max\{1,1/g\},
\label{194.1}
\end{equation}
provided $\lambda^2\varkappa_1/g\prec 1$. Thus 
\begin{equation}
|\langle \phi, {\mathfrak F}_z^{-1} \psi\rangle| \prec \max\{1,1/g\} \|\phi\|\, \|\psi\|.
\label{194}
\end{equation} 
Another term we have to estimate is 
\begin{eqnarray}
|\langle \phi, \lambda R_z^{P_\r} I \bar P_e {\mathfrak F}_z^{-1}\psi\rangle| &\prec& |\lambda|\, \|{\mathfrak F}_z^{-1}\|\, \|\psi\| \max_m C_1(\phi, I \varphi_m\otimes \Omega_\r)\nonumber\\
&\prec& |\lambda| \max\{1,1/g\} \sym \|\psi\| \max_m C_1(\phi, I \varphi_m\otimes \Omega_\r).
\label{195}
\end{eqnarray}
A third term is of similar form and has the same upper bound \eqref{195}. The fourth term to estimate is 
\begin{eqnarray}
\lefteqn{
|\langle \phi, R_z^{P_\r}\psi\rangle |+ \lambda^2 |\langle\phi, R_z^{P_\r} I\bar P_e {\mathfrak F}_z^{-1} \bar P_e I R_z^{P_\r}\psi\rangle | }\nonumber\\
 &\prec& C_1(\phi,\psi)+ \lambda^2 \max\{1,1/g\} \max_m C_1(\phi, I\varphi_m\otimes\Omega_\r) C_1(\psi, I\varphi_m\otimes\Omega_\r). 
 \label{196}
\end{eqnarray}
Collecting the estimates \eqref{194}, \eqref{195}, \eqref{196} shows the bound \eqref{32} for $j=0$. To get a bound for the derivatives $|\partial_z^j\langle\phi, R_z^{P_e}\psi\rangle|$ we again use the representation \eqref{190} and \eqref{fesh1.1}. The $z$ derivatives are affecting the terms ${\mathfrak F}_z^{-1}$  and the reduced resolvents $R_z^{P_\r}$ in \eqref{fesh1.1}. The derivatives of $R_z^{P_\r}$ are controlled using \eqref{m03}. The derivatives of ${\mathfrak F}_z^{-1}$ are dealt with a repeated application of the formula $\partial_z {\mathfrak F}_z^{-1} = -{\mathfrak F}_z^{-1} (\partial _z {\mathfrak F}_z){\mathfrak F}_z^{-1}$, then using \eqref{194.1} and $\|\partial_z^j{\mathfrak F}_z\|\prec 1+\lambda^2\varkappa_1\prec 1$.  We get $\|\partial_z^j{\mathfrak F}_z^{-1}\|\prec \max\{1,1/g^{j+1}\}$. The bound \eqref{32} for all $j$ then readily follows. 

To prove the bound \eqref{034} we proceed in the same manner, using \eqref{190} and \eqref{fesh1.1}. The $\lambda$ derivative of the resolvent $R_z^{P_\r}$ is controlled by \eqref{m03.1} and we use ({\em c.f.} \eqref{191}) $\|\partial_\lambda {\mathfrak F}_z \|= \|\lambda \bar P_eIR_z^{P_\r}I\bar P_e -\lambda^2 \bar P_eI(\partial_\lambda R_z^{P_\r})I\bar P_e\| \prec (|\lambda| +\lambda^2) \varkappa_1\prec 1$. The bound \eqref{034}  follows.

Finally, to show \eqref{33}, we use the Feshbach representation \eqref{fesh1} with $Q=P_\r$. We proceed in the same way as above in this proof to get the result. \hfill \qed

\subsection{The operators $A_e(z,\lambda)$}
\label{sectAe}

For every $e\in{\mathcal E}_0$  and $z\in{\mathbb C}_- $ we set ({\em c.f.} \eqref{23.1}) 
\begin{equation}
	\label{31}
	A_e(z,\lambda) = - P_e I  R_z^{P_e}(\lambda) IP_e \equiv - P_e I  R_z^{P_e} IP_e.
\end{equation}
Starting from \eqref{32}, \eqref{034}, the bounds 
\begin{equation}
	\max_{0\le k\le 2}	\sup_{\{z\in{\mathbb C}_- : |z-e|\le g/2\}}\| \partial_z^k A_e(z,\lambda) \|\prec \varkappa_2, \quad \sup_{\{z\in{\mathbb C}_- : |z-e|\le g/2\}}\| \partial_\lambda A_e(z,\lambda)\|\prec \varkappa_2,
	\label{100}
\end{equation}
where $\varkappa_2$ is given in \eqref{39.1},  are derived just as in \eqref{90} above. Next, $\forall z,\zeta\in {\mathbb C}_-$ with $|z-e|\le g/2$, 
\begin{equation}
\| A_e(z,\lambda) - A_e(\zeta,\lambda) \| = \Big\| \int_z^\zeta \partial_w A_e(w,\lambda) dw\Big\| \prec |z-\zeta|\varkappa_2,
\label{102}
\end{equation}
where the integral is over the straight line linking $z$ and $\zeta$. Let  $z_n$ be a sequence in ${\mathbb C}_-$ with $|z_n-e|\le g/2$, converging to some $x\in\mathbb R$ (so $|x-e|\le g/2$). Then by \eqref{102}, $A_e(z_n,\lambda)$ is Cauchy and thus converges to a limit which we call $A_e(x,\lambda)$. Using \eqref{102} it is easy to see that the limit $A_e(x,\lambda)$ is independent of the sequence $z_n$.  We note that the level shift operator \eqref{34} equals $\Lambda_e = A_e(e,0)$.  Similarly to \eqref{102} we derive  $\|A_e(z,\lambda)-A_e(z,0)\|\prec \lambda^2\varkappa_1 \varkappa_2$ and hence
\begin{equation}
	\label{36}
	\| A_e(z,\lambda)-\Lambda_e\| \prec\varkappa_2 \big(|z-e|+|\lambda|\big).
\end{equation} 
The bound \eqref{36} is the starting point for conventional perturbation theory. Recall the definitions of the spectral gap of the level shift operators $\delta$ and the maximal operator norm $\kappa$ given in \eqref{delta} and \eqref{kappa}, respectively. 
\begin{lem} 
\label{lem2.4}
Suppose that $z\in{\mathbb C}_-$ and $|z-e|, |\lambda| \prec \frac{\delta}{\varkappa_2\kappa} \min\{1,1/\kappa\}$. Then 
\begin{itemize}
\item[{\rm 1.}] All eigenvalues of $A_e(z,\lambda)$ are simple. Call them $\aes(z,\lambda)$, $s=1,\ldots, m_e$. Each $a_{e,s}(z,\lambda)$ satisfies $|a_{e,s}(z,\lambda)-\aes|<\delta/2$ for exactly one eigenvalue $\aes$ of $\Lambda_e$. In particular, we have the diagonal form 
\begin{equation}
A_e(z,\lambda)  = \sum_{s=1}^{m_e} \aes(z,\lambda) \Qes(z,\lambda),
\label{89}
\end{equation}
		where the eigenvalues $\aes(z,\lambda)$ are simple and the (Riesz, rank one) eigenprojections are denoted by $\Qes(z,\lambda)$.
		
\item[2.] On the domain of $z$, $\lambda$ determined by the constraint stated at the beginning of the lemma, the functions, $\aes(z,\lambda)$ and $\Qes(z,\lambda)$ are analytic in $z$ and differentiable in $\lambda$. Moreover, we have 
\begin{eqnarray}
\big\| \Qes(z,\lambda)-\Qes(z',\lambda')\big\| &\prec&  \varkappa_3 \big(|z-z'|+|\lambda-\lambda'| \big) \label{119.0}\\
|\aes(z,\lambda) -\aes(z',\lambda')|
&\prec&  \varkappa_4 \big(|z-z'| +|\lambda-\lambda'| \big),
\label{119.1}
\end{eqnarray}
where 
\begin{equation}
\varkappa_3=\varkappa_2\kappa^2/ \delta,\qquad \varkappa_4=\varkappa_2\kappa (1+\varkappa_2\kappa/\delta).
\label{kappa3,4} 
\end{equation}

\item[3.] On the domain of $z$, $\lambda$ determined by the constraint stated at the beginning of the lemma, we have 
\begin{equation}
| \partial_z ^2\aes(z,\lambda)|\prec \varkappa_5\equiv \varkappa_2\kappa[1+ \frac{\varkappa_2\kappa}{\delta} (1+\varkappa_2\kappa/\delta)].
\label{kappa5}
\end{equation} 
\end{itemize}
\end{lem}

The previous result leads readily to the fact that one can extend the eigenvalues and eigenprojections as functions of $z$ continuously to the real axis: 

\begin{cor}
	\label{cor1}
The maps $z\mapsto \Qes(z,\lambda)$ and $z\mapsto \aes(z,\lambda)$ extend by continuity to $z=x\in{\mathbb R}$ provided $|x-e|$, $|\lambda|\prec \frac{\delta}{\varkappa_2\kappa} \min\{1,1/\kappa\}$. Moreover, the estimates \eqref{119.0} and \eqref{119.1} are valid if either or both of $z$, $z'$ are real. 
\end{cor}

{\em Proof of Corollary \ref{cor1}. } Let $z_n$ be a sequence in ${\mathbb C}_-$ with $|z_n-e|< \frac{\delta}{\varkappa_2\kappa}\min\{1,1/\kappa\}$ and $z_n\rightarrow x\in\mathbb R$. Then \eqref{119.0} shows that $\Qes(z_n,\lambda)$ is a Cauchy sequence, so it converges. Next, again by \eqref{119.0}, $\|\Qes(z,\lambda) - \Qes(x,\lambda) \|= \lim_n \|\Qes(z,\lambda) - \Qes(z_n,\lambda)\|$ exists and is $\prec\varkappa_3 |z-x|$. 
If $z=x'\in\mathbb R$ then approximate it by $z'_n\in{\mathbb C}_-$ and the above argument works the same. The argument is also the same for the $\aes(x,\lambda)$. \hfill \qed

\medskip

{\em Proof of the Lemma \ref{lem2.4}.\ } Proof of 1. It follows from \eqref{83} that 
\begin{equation}
	\big\| (\Lambda_e-\zeta)^{-1}\big\| \prec\frac{\kappa}{d(\zeta,{\rm spec}(\Lambda_e))},
	\label{112}
\end{equation}
where $d(\cdot,\cdot)$ denotes the distance function. For $|z-e|+ |\lambda| \prec \frac{d(\zeta,{\rm spec}(\Lambda_e))}{\varkappa_2\kappa}$ the Neumann series 
\begin{equation}
	(A_e(z,\lambda) -\zeta)^{-1} = (\Lambda_e-\zeta)^{-1}\sum_{n\ge 0} \big[ \big(\Lambda_e-A_e(z,\lambda)\big) (\Lambda_e-\zeta)^{-1}\big]^n
	\label{113}
\end{equation}
converges, so the $\zeta$ satisfying $d(\zeta,{\rm spec}(\Lambda_e)) \prec \varkappa_2\kappa(|z-e|+ |\lambda|)$ belong to the resolvent set of $A_e(z,\lambda)$. Moreover, \eqref{113}  gives the bounds
\begin{eqnarray}
\big\|  (A_e(z,\lambda) -\zeta)^{-1}\big\|  &\prec&  \frac{\kappa}{d(\zeta,{\rm spec}(\Lambda_e))},
	\label{113.1}	\\
\big\| (A_e(z,\lambda) -\zeta)^{-1} - (\Lambda_e-\zeta)^{-1} \big\| &\prec& \big(|z-e|+|\lambda|\big) \frac{\varkappa_2 \kappa^2}{[d(\zeta,{\rm spec}(\Lambda_e))]^2}\label{113.2}.\quad 
\end{eqnarray}
Let ${\mathcal C}_e^{(s)}$ be the circle centered at $\aes$ with radius $\delta/2$, then $d(\zeta,{\rm spec}(\Lambda_e))=\delta/2$ for $\zeta\in{\mathcal C}_e^{(s)}$ and so ${\mathcal C}_e^{(s)}$ belongs to the resolvent set of $A_e(z,\lambda)$. Thus  the following integral is well defined and equals the spectral projection, 
\begin{equation}
	\Qes(z,\lambda) = \frac{-1}{2\pi\i}\oint_{{\mathcal C}_e^{(s)}} (A_e(z,\lambda) - \zeta)^{-1} d\zeta.
	\label{115}
\end{equation}
We have 
\begin{eqnarray}
	\|\Qes(z,\lambda)-\Qes\| &=&\frac{1}{2\pi}\big\| \oint_{{\mathcal C}_e^{(s)}} \big[(A_e(z,\lambda) -\zeta)^{-1} - (\Lambda_e-\zeta)^{-1}\big] d\zeta\big\| \nonumber\\
	&\le&\frac\delta2 \max_{\zeta \in {\mathcal C}_e^{(s)}} \big\| (A_e(z,\lambda) -\zeta)^{-1} - (\Lambda_e-\zeta)^{-1}\big\| < 1,
\end{eqnarray}
since $|z-e|+|\lambda| \prec \delta/(\varkappa_2\kappa^2)$. By standard perturbation theory \cite{Kato}, the ranks of $\Qes(z,\lambda)$ and $\Qes$ are the same (both $=1$ by assumption (A5)) and hence $A_e(z,\lambda)$ has exactly one eigenvalue inside ${\mathcal C}_e^{(s)}$. This shows point 1. 

We now give a proof of 2.  Using \eqref{100} and \eqref{113.1} we obtain 
\begin{eqnarray}
	\big\| \partial_z \Qes(z,\lambda) \big\| &=& \frac{1}{2\pi}\big\| \oint_{{\mathcal C}_e^{(s)}} (A_e(z,\lambda) - \zeta)^{-1}\{ \partial_zA_e(z,\lambda)\} (A_e(z,\lambda) - \zeta)^{-1} d\zeta\big\|\nonumber\\
	&\le& \delta   \varkappa_2  \max_{\zeta\in{\mathcal C}_e^{(s)}} \|(A_e(z,\lambda) - \zeta)^{-1}\|^2 \prec \frac{\varkappa_2\kappa^2}{\delta}. 
	\label{115.1}
\end{eqnarray}
Taking the $z$ derivative twice (or the $\lambda$ derivative) and proceeding as in \eqref{115.1} yields the estimates
\begin{equation}
	\big\| \partial_z^2\Qes(z,\lambda) \big\| 
	\prec \frac{\varkappa_2 \kappa^2}{\delta} (1+\varkappa_2\kappa/\delta),\qquad 	\big\| \partial_\lambda \Qes(z,\lambda) \big\| \prec \frac{\varkappa_2\kappa^2}{\delta}. 
	\label{118}
\end{equation}
We combine \eqref{115.1} and \eqref{118} to obtain
\begin{eqnarray}
	\big\| \Qes(z,\lambda)-\Qes(z',\lambda')\big\| &\le& \big\| \Qes(z,\lambda)-\Qes(z',\lambda)\big\| + \big\| \Qes(z',\lambda)-\Qes(z',\lambda')\big\| \nonumber\\
	&=& \big\| \int_z^{z'} \partial_\zeta \Qes(\zeta,\lambda) d\zeta\big\| +\big\| \int_\lambda^{\lambda'} \partial_\mu \Qes(z',\mu) d\mu\big\| \nonumber\\
	&\prec& \frac{\varkappa_2 \kappa^2 }{\delta}\big(|z-z'| +|\lambda-\lambda'|\big),
	\label{130}
\end{eqnarray}
which shows \eqref{119.0}.  To show \eqref{119.1}, we note that \begin{equation}
	\aes(z,\lambda) ={\rm tr}(A_e(z,\lambda)\Qes(z,\lambda))
	\label{trace}
\end{equation} 
and so, using \eqref{100}, \eqref{115.1} and (see \eqref{115}  and \eqref{113.1})
\begin{equation}
	\label{bnd}
	\|\Qes(z,\lambda)\|\prec \kappa,
\end{equation} 
we obtain $| \partial_z\aes(z,\lambda)| = \big|{\rm tr}\big( \{\partial_zA_e(z,\lambda)\}\Qes(z,\lambda) + A_e(z,\lambda)\{\partial_z \Qes(z,\lambda)\}\big)\big|\prec \varkappa_2 \kappa (1+\varkappa_2\kappa/\delta)$. 
Proceeding in the same way we find $| \partial_\lambda\aes(z,\lambda)| \prec\varkappa_2 \kappa (1+\varkappa_2\kappa/\delta)$. By 
 integrating these bounds similarly to what we did in \eqref{130},  we get \eqref{119.1}. 
 
 We finally prove point 3.  The relation \eqref{trace} yields 
\begin{eqnarray*}
	\partial^2_z\aes(z,\lambda) &=& {\rm tr}\Big( \{\partial_z^2A_e(z,\lambda)\}\Qes(z,\lambda) + 2\{\partial_zA_e(z,\lambda)\}\{\partial_z \Qes(z,\lambda)\}\nonumber\\
	&& +  A_e(z,\lambda)\{\partial_z^2 \Qes(z,\lambda)\}\Big).
\end{eqnarray*}
The bound on the second derivative given in point 3. now follows from \eqref{100}, \eqref{bnd}, \eqref{115.1} and \eqref{118}. This completes the proof of  Lemma \ref{lem2.4}. \hfill \qed

\section{Feshbach decomposition of the resolvent}
\label{Fdecsect}

Let $Q$ be an orthogonal projection on a Hilbert space $\mathcal H$. In the decomposition ${\mathcal H}={\rm Ran}Q \oplus {\rm Ran}Q^\perp$ an operator ${\mathcal O}$ has the block decomposition
\begin{equation}
{\mathcal O}= 
\begin{pmatrix}
A & B\\
C & D
\end{pmatrix},
\end{equation}
where $A=Q{\mathcal O}Q\upharpoonright_{{\rm Ran}Q}$, $B=Q A Q^\perp\upharpoonright_{{\rm Ran}Q^\perp}$ and so on. We want to find the block decomposition of ${\mathcal O}^{-1}$, 
\begin{equation}
	{\mathcal O}^{-1}= 
	\begin{pmatrix}
		X & Y\\
		Z & W
	\end{pmatrix},
\end{equation}
assuming that $D^{-1}$ exists. Multiplying blockwise the equation ${\mathcal O}{\mathcal O}^{-1}=\bbbone$ yields the four equations
\begin{equation}
AX +BZ = \bbbone_Q, \quad AY = -BW, \quad CX = -DZ,\quad CY+DW = \bbbone_{Q^\perp}.
\label{f1}
\end{equation}
Thus $Z=-D^{-1}CX$ and $W = D^{-1}-D^{-1}CY$. Then $AX+BZ=AX-BD^{-1}CX=\bbbone_Q$, so $X=(A-BD^{-1}C)^{-1}$. Also, $AY=-BW=-BD^{-1} +BD^{-1}CY$ so $(A-BD^{-1}C)Y=-BD^{-1}$ and hence $Y=-XBD^{-1}$. In conclusion,
\begin{eqnarray*}
X &=& (A-BD^{-1}C)^{-1} \\
Y &=&-XBD^{-1}\\
Z &=& -D^{-1}CX\\
W &=& D^{-1}  -D^{-1} CXBD^{-1}.
\end{eqnarray*}
In the case ${\mathcal O}= H-z$,  for a self-adjoint $H$ (like  $H=L_\lambda$) and  $z\not\in\mathbb R$, we have indeed that $D=Q^\perp(H-z)Q^\perp\upharpoonright_{{\rm Ran Q}^\perp}$ is invertible. Also, $A-BD^{-1}C={\mathfrak F}(H-z;Q)$ and hence the sum of $X,Y,Z$ and $W$ give the right hand side of \eqref{16}.

\begin{thm}[Weak isospectrality of the Feshbach map]
	\label{weakFeshthm}
Let $H$ be a self-adjoint operator and let $E\in\mathbb R$ and suppose that the function $z\mapsto QHQ^\perp R_z^Q Q^\perp HQ$ is continuously differentiable on $z\in{\mathbb C}_-\cup\{E\}$. Denote the value of ${\mathfrak F}(H-z;Q)$ at $z=E$ by ${\mathfrak F}(H-E;Q)$. Then we have the following:
\smallskip

(1) $E$ is an eigenvalue of $H$ if and only if zero is an eigenvalue of ${\mathfrak F}(H-E;Q)$. 
\smallskip

(2) If  $H\Phi=E\Phi$, then $\varphi=Q\Phi$ satisfies ${\mathfrak F}(H-E;Q)\varphi=0$.  
\smallskip

(3) If ${\mathfrak F}(H-E;Q)\varphi=0$ then the limit of $R_z^QQ^\perp HQ\varphi$ as $z \in{\mathbb C}_-$, $z\rightarrow E$, exists. Denote it by $R_E^QQ^\perp HQ\varphi$. Then $\Phi=\varphi -R^Q_{E}Q^\perp HQ\varphi$ satisfies $H\Phi=E\Phi$.
\end{thm}

Note: It is easy to see that the correspondence $\Phi\leftrightarrow\varphi$ in Theorem \ref{weakFeshthm} is a bijection between the kernels of $H-E$ and of ${\mathfrak F}(H-E;Q)$. 
\medskip

{\em Proof of Theorem \ref{weakFeshthm}.} The implication $\Rightarrow$ in (1) together with (2) is not hard to prove, see Proposition B2 of \cite{KM1}. We give now a proof of $\Leftarrow$ in (1) and (3). One may write the resolvent using the above components in matrix form as
\begin{equation}
(H-z)^{-1} = 
\begin{pmatrix}
\bbbone & 0\\
-R^Q_zQ^\perp HQ & \bbbone
\end{pmatrix}
\begin{pmatrix}
\big[{\mathfrak F}(H-z;Q)\big]^{-1} & 0\\
0 & R_z^Q
\end{pmatrix}
\begin{pmatrix}
\bbbone & -QH Q^\perp R_z^Q \\
0 & \bbbone
\end{pmatrix}
\label{fesh1}
\end{equation}
where $R_z^Q=(Q^\perp HQ^\perp-z)^{-1}\upharpoonright_{{\rm Ran}Q^\perp}$. This relation can actually be verified simply by multiplying out the matrices. As the two outside matrices in \eqref{fesh1} are both invertible, we obtain for $z\in\mathbb C$ such that $R_z^Q$ exists:
\begin{eqnarray}
H-z &=& 
\begin{pmatrix}
\bbbone & QH Q^\perp R_z^Q\\
0 & \bbbone
\end{pmatrix}
\begin{pmatrix}
{\mathfrak F}(H-z;Q) & 0\\
0 & Q^\perp (H-z) Q^\perp
\end{pmatrix}
\begin{pmatrix}
\bbbone & 0 \\
R^Q_zQ^\perp HQ & \bbbone
\end{pmatrix}\nonumber\\
&=& 
\begin{pmatrix}
{\mathfrak F}(H-z;Q)  +QHP^\perp R^Q_z P^\perp HQ \quad &  QH Q^\perp \\	
Q^\perp HQ & Q^\perp (H-z) Q^\perp
\end{pmatrix}. 
\label{fesh2}
\end{eqnarray}
To discuss domain questions, assume for simplicity that $\dim Q<\infty$ and that  $Q^\perp HQ$ is bounded. Then the relation \eqref{fesh2} holds when applied to vectors of the form $\begin{pmatrix} \varphi \\ \chi\end{pmatrix}$ with $\chi\in{\rm Dom}(Q^\perp H Q^\perp)$. Moreover, each single product operation of the matrices in the threefold matrix product in \eqref{fesh2} makes sense individually (is well defined, when the right side is applied to vectors of the indicated form). 

Due to the assumption of the theorem, $QHQ^\perp R_z^Q Q^\perp HQ$ has a limit when $z=E-\i \epsilon$ with $\epsilon \rightarrow 0_+$. Call this limit $QHQ^\perp (\bar H-E)^{-1} Q^\perp HQ$ (the overlined $\bar H$ indicates that $H$ is restricted to the range of $Q^\perp$).  Then $\lim_{\epsilon\rightarrow 0_+}{\mathfrak F}(H-E-\i\epsilon;Q) = Q(H-E-HQ^\perp(\bar H-E)^{-1}Q^\perp H)Q \equiv {\mathfrak F}(H-E;Q)$, where the last symbol is again a definition.  Suppose that $\varphi=Q\varphi$ is such that ${\mathfrak F}(H-E;Q)\varphi=0$. For $\epsilon>0$, define $\chi_\epsilon =-Q^\perp R^Q_{E-\i\epsilon} Q^\perp HQ\varphi$. Then applying equation \eqref{fesh2} with $z=E-\i\epsilon$ gives
\begin{equation}
(H-E+\i\epsilon) (\varphi+\chi_\epsilon) = {\mathfrak F}(H-E+\i\epsilon;Q)\varphi\longrightarrow 0 \quad \mbox{as $\epsilon\rightarrow 0_+$.}
\label{fesh3}
\end{equation}
We show below that $\chi_\epsilon$ has a limit $\chi$ as $\epsilon\rightarrow 0_+$. It then follows from \eqref{fesh3}  that $\lim_{\epsilon\rightarrow 0_+}(H-E)(\varphi+ \chi_\epsilon)=0$. Since $H$ is a closed operator, the vector $\varphi+\chi$ is in the domain of $H$ and $(H-E)(\varphi+\chi)=0$. 

We now show that $\chi_\epsilon$ is Cauchy. This part of the analysis is inspired by [DJ], Theorem 3.8. Denote by $p$ the spectral projection of ${\mathfrak F}(H-E;Q)$ associated to the eigenvalue $0$. In particular, $p\varphi=\varphi$. Since ${\mathfrak F}(H-E;Q)$ is a dissipative operator and $0$ is on the boundary of its numerical range, $0$ is a semisimple eigenvalue and $p$ is the orthogonal projection onto the kernel of ${\mathfrak F}(H-E;Q)$, in particular, $p^*=p$. 
 First we show that $z\mapsto pH R_z^QHp$ is a continuously differentiable function of $z\in {\mathbb C}_-\cup{\mathbb C}_+\cup \{E\}$.  (Note: $QHR_z^QHQ$, without the restriction to the range of $p$, is not even continuous at $E$, because the limits coming from ${\mathbb C}_+$ and ${\mathbb C}_-$ do not coincide.) We point out that $z\mapsto QHQ^\perp R_z^Q Q^\perp HQ$ is continuously differentiable on $z\in{\mathbb C}_+\cup\{E\}$, which follows by taking the adjoint in the assumption of the theorem. Next, we have $p {\mathfrak F}(H-E;Q) p=0$ and hence $p(H-E)p = pHR^Q_{E-\i 0_+}Hp$, and taking the adjoint gives $p(H-E)p = pHR^Q_{E+\i 0_+}Hp$.  So we have
\begin{equation}
	pHR^Q_{E-\i 0_+}Hp = pHR^Q_{E-\i 0_-}Hp = p(H-E)p,
	\label{fesh6}
\end{equation}
which shows that $z\mapsto pHR^Q_zHp$ is continuous at $z=E$ and its value at $z=E$ is self-adjoint. Next, let $\gamma(\tau)$, $\tau\in[-1,1]$, be a smooth curve in ${\mathbb C}_-\cup\{E\}$ which is tangent to the point $E\in{\mathbb C}$, satisfying  $\gamma(0)=E$ and $\gamma'(0)=1$. Then $\tau\mapsto pHR^Q_{\gamma(\tau)}Hp$ is continuously differentiable at $\tau=0$ and $\frac{d}{d\tau}|_{\tau=0}\, {\rm Im}\, pHR_{\gamma(\tau)}^QHp = {\rm Im} \, p H (R^Q_{E-\i 0_+})^2 Hp$. On the other hand, this last derivative has to be equal to zero for the following reason.  For all $\tau$ we have ${\rm Im}\, pHR_{\gamma(\tau)}^QHp\le 0$ and from \eqref{fesh6}, ${\rm Im}\, pHR_{\gamma(\tau)}^QHp= 0$ at $\tau=0$. If $\frac{d}{d\tau}|_{\tau=0}\, {\rm Im}\, pHR_{\gamma(\tau)}^QHp$ was $>0$ or $<0$ this would contradict the fact that ${\rm Im}\, pHR_{\gamma(\tau)}^QHp\le 0$. Since this derivative vanishes, we get  ${\rm Im} \, p H (R^Q_{E-\i 0_+})^2 Hp=0$, so $p H (R^Q_{E-\i 0_+})^2 Hp$ is self-adjoint, which means that $p H (R^Q_{E-\i 0_+})^2 Hp = p H (R^Q_{E-\i 0_-})^2 Hp$. So $z\mapsto pHR^Q_zHp$ is continuously differentiable on ${\mathbb C}_-\cup{\mathbb C}_+\cup\{E\}$. 

We now use this regularity property to show that $\chi_\epsilon$ converges. Let $\epsilon, \epsilon'>0$. We have 
\begin{equation}
\big\| (R_{E-\i\epsilon}^Q - R^Q_{E-\i\epsilon'}) HQ\varphi\big\|^2 =\scalprod{\varphi}{ QH(R_{E+\i\epsilon}^Q - R^Q_{E+\i\epsilon'})(R_{E-\i\epsilon}^Q - R^Q_{E-\i\epsilon'})HQ\varphi}.
\label{fesh4}
\end{equation}
From the resolvent identity,  $R^Q_{E-\i\epsilon}R_{E+\i\epsilon'}^Q =  \frac{1}{\i (\epsilon'+\epsilon)} (R^Q_{E-\i\epsilon}-R_{E+\i\epsilon'}^Q)$ and  similarly for the other three terms on the right side of \eqref{fesh4}. Therefore, (also using that $pQ=Qp=p$),  
\begin{eqnarray}
\lefteqn{
\big\| (R_{E+\i\epsilon}^Q - R^Q_{E+\i\epsilon'}) HQ\varphi\big\|^2}\\
&=&
	\scalprod{\varphi}{ pH \tfrac{R^Q_{E-\i\epsilon}-R_{E+\i\epsilon}^Q}{2\i \epsilon} Hp\varphi} - 
	\scalprod{\varphi}{ pH \tfrac{R^Q_{E-\i\epsilon}-R_{E+\i\epsilon'}^Q}{\i(\epsilon'+\epsilon)} Hp\varphi}\nonumber\\
	&& - \scalprod{\varphi}{ pH \tfrac{R^Q_{E-\i\epsilon'}-R_{E+\i\epsilon}^Q}{\i(\epsilon+\epsilon')} Hp\varphi}+\scalprod{\varphi}{ pH \tfrac{R^Q_{E-\i\epsilon'}-R_{E+\i\epsilon'}^Q}{2\i\epsilon'} Hp\varphi}.
	\label{fesh5}
\end{eqnarray}
Since $z\mapsto pH R_z^QHp$ is a continuously differentiable function of $z\in {\mathbb C}_-\cup{\mathbb C}_+\cup \{E\}$, the right hand side of \eqref{fesh5} converges to zero as $\epsilon,\epsilon'\rightarrow 0$. Thus $\chi_\epsilon$ converges. \hfill \qed

\bigskip

{\bf Acknowledgements. } The author is really grateful to two referees who took the time to examine this work and make valuable suggestions. Their encouragement to improve the readability of the manuscript, for researchers in the quantum sciences who might not be familiar with some of the mathematical intricacies presented, was particularly beneficial.  The author was supported by a {\em Discovery Grant} from the {\em National Sciences and Engineering Research Council of Canada } (NSERC).

\bigskip
\bigskip
\medskip

\noindent
{\bf \Large Conclusion}
\bigskip

In this paper we consider the evolution of a finite-dimensional system coupled to a reservoir. The coupling is small, but fixed. The reservoir is assumed to have a unique stationary state and otherwise dissipative dynamics. We develop a method to separate the total, unitary evolution of a $\s\r$ complex into a Markovian part, a dissipative part, plus a remainder. The Markovian part describes the system dynamics with the reservoir in its stationary state and contains explicit oscillating (in time) and exponentially decaying terms. The dissipative term governs the dynamics on $\s\r$ states which do not have any overlap with the system stationary state. This term decays polynomially in time. The remainder term is small in the coupling, {\em uniformly for all $t\ge 0$}. The strengths of our results are:
\begin{itemize}
	\item[-] The decomposition is derived mathematically rigorously, with a controlled remainder which is small in the $\s\r$ coupling, for all times $t\ge0$.
	\item[-] We can treat $\s\r$  interactions with significantly less regularity than previously needed. In terms of the reservoir correlation function, this means that only polynomial decay is required (versus exponential decay in previous rigorous works).
	\item[-] We analyze the whole $\s\r$ dynamics, not only the reduced system dynamics and moreover, the initial $\s\r$ states do not need to be of product form (they can be classically correlated or entangled).
	\item[-] Our method applies to a widely used class of open systems: An $N$-level system coupled linearly to a spatially infinitely extended reservoir of thermal, non-interacting Bose particles. Detailed results on this model are presented in \cite{Markov2,Mcorr}. In particular, it is shown there that the system dynamics is well approximated by the Markovian master equation for all times $t\ge0$, even for initially correlated $\s\r$ states. It is also shown there that the Born approximation is valid for all times.
\end{itemize}

\end{document}